\documentclass[%
 reprint,
 amsmath,amssymb,
 aps,
 prx,
]{revtex4-2}

\usepackage{silence}
\usepackage{graphicx}
\usepackage{dcolumn}
\usepackage{bm}
\usepackage{subfigure}
\usepackage{threeparttable}

\usepackage{amsmath}
\usepackage{amssymb}
\usepackage{mathtools}
\usepackage{amsthm}
\usepackage{listingsutf8}
\usepackage{xcolor}
\usepackage{tcolorbox}
\usepackage{caption}

\usepackage{tabularx}
\usepackage{makecell}

\usepackage{quantikz}
\usepackage{adjustbox}

\usepackage{hyperref}

\usepackage{xcolor}
\usepackage{listings}

\lstdefinestyle{promptFormat}{
    language=Python,
    inputencoding=utf8,
    basicstyle=\ttfamily\small,            
    keywordstyle=\color{red}\bfseries,     
    commentstyle=\color{green!50!black},   
    stringstyle=\color{blue},              
    backgroundcolor=\color{gray!10},       
    frame=single,                           
    showstringspaces=false,                 
    breaklines=true,                         
    numbers=left,                            
    numberstyle=\tiny\color{gray},           
    numbersep=5pt,                           
    xleftmargin=10pt,                        
    literate={\{max\_trial\_num\}}{{\textcolor{red}{\{max\_trial\_num\}}}}{14}
             {\{current\_trial\_num\}}{{\textcolor{red}{\{current\_trial\_num\}}}}{19}
             {\{device\_n\_qubit\}}{{\textcolor{red}{\{device\_n\_qubit\}}}}{16}
             {\{idea\_num\}}{{\textcolor{red}{\{idea\_num\}}}}{10}
             {\{previous\_trial\}}{{\textcolor{red}{\{previous\_trial\}}}}{16}
             {\{review\_comment\}}{{\textcolor{red}{\{review\_comment\}}}}{16}
             {\{few\_shot\_examples\}}{{\textcolor{red}{\{few\_shot\_examples\}}}}{19}
             {\{current\_round\}}{{\textcolor{red}{\{current\_round\}}}}{15}
             {\{max\_scoring\_round\}}{{\textcolor{red}{\{max\_scoring\_round\}}}}{19}
             {\{max\_reflection\_round\}}{{\textcolor{red}{\{max\_reflection\_round\}}}}{22}
             {\{previous\_idea\}}{{\textcolor{red}{\{previous\_idea\}}}}{15}
             {\{previous\_score\}}{{\textcolor{red}{\{previous\_score\}}}}{16}
             {\{related\_work\}}{{\textcolor{red}{\{related\_work\}}}}{14}
             {\{max\_summary\_words\}}{{\textcolor{red}{\{max\_summary\_words\}}}}{19}
             {\{raw\_content\}}{{\textcolor{red}{\{raw\_content\}}}}{13}
             {\{code\}}{{\textcolor{red}{\{code\}}}}{6}
             {\{idea\}}{{\textcolor{red}{\{idea\}}}}{6}
             {\{pennylane\_operations\}}{{\textcolor{red}{\{pennylane\_operations\}}}}{22}
             {\{methods\}}{{\textcolor{red}{\{methods\}}}}{9}
             {\{pennylane\_reference\}}{{\textcolor{red}{\{pennylane\_reference\}}}}{21}
             {\{error\_messages\_string\}}{{\textcolor{red}{\{error\_messages\_string\}}}}{23}
             {\{warning\_messages\_string\}}{{\textcolor{red}{\{warning\_messages\_string\}}}}{25}
             {\{max\_suggestion\_num\}}{{\textcolor{red}{\{max\_suggestion\_num\}}}}{20}
             {\{last\_trial\_num\}}{{\textcolor{red}{\{last\_trial\_num\}}}}{16}
             {\{last\_trial\_results\}}{{\textcolor{red}{\{last\_trial\_results\}}}}{20}
             {\{quantum\_gate\_list\}}{{\textcolor{red}{\{quantum\_gate\_list\}}}}{19}
             {\{feature\_map\_constraints\}}{{\textcolor{red}{\{feature\_map\_constraints\}}}}{25}
             {\{ansatz\_constraints\}}{{\textcolor{red}{\{ansatz\_constraints\}}}}{20}
             {\{input\_data\_constraints\}}{{\textcolor{red}{\{input\_data\_constraints\}}}}{24}
             {\{research\_report\}}{{\textcolor{red}{\{research\_report\}}}}{17}
             {\{experiment\_report\}}{{\textcolor{red}{\{experiment\_report\}}}}{19}
             {\{not\_provided\_information\}}{{\textcolor{red}{\{not\_provided\_information\}}}}{26}
             {\{reference\}}{{\textcolor{red}{\{reference\}}}}{11}
             {\{current\_reflection\_round\}}{{\textcolor{red}{\{current\_reflection\_round\}}}}{26}
             {\{advocate\_response\}}{{\textcolor{red}{\{advocate\_response\}}}}{19}
             {\{additional\_information\}}{{\textcolor{red}{\{additional\_information\}}}}{24}
             {\{additional\_information\_query\}}{{\textcolor{red}{\{additional\_information\_query\}}}}{30}
             {\{expert\_role\}}{{\textcolor{red}{\{expert\_role\}}}}{13}
             {\{question\}}{{\textcolor{red}{\{question\}}}}{10}
             {\{original\_idea\}}{{\textcolor{red}{\{original\_idea\}}}}{15}
             {\{critical\_feedback\}}{{\textcolor{red}{\{critical\_feedback\}}}}{19}
             {\{molecule\_hamiltonian\_definition\}}{{\textcolor{red}{\{molecule\_hamiltonian\_definition\}}}}{33}
             {\{cost\_function\_definition\}}{{\textcolor{red}{\{cost\_function\_definition\}}}}{26}
             {\{optimizer\_settings\}}{{\textcolor{red}{\{optimizer\_settings\}}}}{20}
             {\{constraints\}}{{\textcolor{red}{\{constraints\}}}}{13}
             {\{max\_feedback\_num\}}{{\textcolor{red}{\{max\_feedback\_num\}}}}{18}
             {α}{{$\alpha$}}1
             {β}{{$\beta$}}1
             {γ}{{$\gamma$}}1
             {∑}{{$\sum$}}1
             {θ}{{$\theta$}}1
             {φ}{{$\phi$}}1
             {Φ}{{$\Phi$}}1
             {ψ}{{$\psi$}}1
             {ᵧ}{{$^\gamma$}}1
             {ₓ}{{$_x$}}1
             {ᵢ}{{$^i$}}1
             {ⱼ}{{$^j$}}1
             {₀}{{$_0$}}1
             {₁}{{$_1$}}1
             {₂}{{$_2$}}1
             {₃}{{$_3$}}1
             {₄}{{$_4$}}1
             {₅}{{$_5$}}1
             {₆}{{$_6$}}1
             {₇}{{$_7$}}1
             {₈}{{$_8$}}1
             {ⱼ}{{$_j$}}1
             {π}{{$\pi$}}1
             {…}{{$\ldots$}}1
             {×}{{$\times$}}1
             {²}{{$^2$}}1
             {√}{{$\sqrt{}$}}1
             {≤}{{$\le$}}1
             {∝}{{$\propto$}}1
             {“}{{"}}1
             {”}{{"}}1
             {‘}{{'}}1
             {’}{{'}}1
             {→}{{$\rightarrow$}}1
             {⟩}{{$\rangle$}}1
             {⟨}{{$\langle$}}1
             {≥}{{$\ge$}}1
             {†}{{$\dagger$}}1
}

\lstdefinestyle{codeFormat}{
    language=Python,
    inputencoding=utf8,
    basicstyle=\ttfamily\small,            
    keywordstyle=\color{red}\bfseries,     
    commentstyle=\color{green!50!black},   
    stringstyle=\color{blue},              
    backgroundcolor=\color{gray!10},       
    frame=single,                           
    showstringspaces=false,                 
    breaklines=true,                         
    numbers=left,                            
    numberstyle=\tiny\color{gray},           
    numbersep=5pt,                           
    xleftmargin=10pt,                        
    literate={π}{{$\pi$}}1
             {₁}{{$_1$}}1
             {λ}{{$\lambda$}}1
             {ₗ}{{$_\ell$}}1
             {γ}{{$\gamma$}}1
             {Σ}{{$\sum$}}1
             {∇}{{$\nabla$}}1
             {Δ}{{$\Delta$}}1
             {→}{{$\rightarrow$}}1
             {・}{{$\cdot$}}1
             {⊗}{{$\otimes$}}1
             {≈}{{$\approx$}}1
             {±}{{$\pm$}}1
}

\lstdefinestyle{txtFormat}{
    language=,                               
    inputencoding=utf8,
    basicstyle=\ttfamily\small,              
    keywordstyle=,                           
    commentstyle=,                           
    stringstyle=,                            
    backgroundcolor=\color{gray!10},         
    frame=single,                            
    showstringspaces=false,                  
    breaklines=true,                         
    numbersep=5pt,                           
    xleftmargin=10pt,                        
    literate={π}{{$\pi$}}1
             {σ}{{$\sigma$}}1
             {ρ}{{$\rho$}}1
             {γ}{{$\gamma$}}1
             {ℓ}{{$\ell$}}1
             {α}{{$\alpha$}}1
             {Σ}{{$\Sigma$}}1
             {Δ}{{$\Delta$}}1
             {ξ}{{$\xi$}}1
             {ε}{{$\varepsilon$}}1
             {φ}{{$\varphi$}}1
             {τ}{{$\tau$}}1
             {‖}{{$\lVert$}}1
             {→}{{$\rightarrow$}}1
             {≤}{{$\leq$}}1
             {≥}{{$\geq$}}1
             {∈}{{$\in$}}1
             {≈}{{$\approx$}}1
             {₁}{{$_1$}}1
             {λ}{{$\lambda$}}1
             {ₗ}{{$_\ell$}}1
             {∇}{{$\nabla$}}1
             {→}{{$\rightarrow$}}1
             {±}{{$\pm$}}1
             {⌈}{{$\lceil$}}1
             {⌉}{{$\rceil$}}1
             {·}{{$\cdot$}}1
             {・}{{$\cdot$}}1
             {≪}{{$\ll$}}1
             {∏}{{$\prod$}}1
             {θ}{{$\theta$}}1
             {β}{{$\beta$}}1
             {ŵ}{{$\hat{w}$}}1
             {↔}{{$\leftrightarrow$}}1
             {⊗}{{$\otimes$}}1
             {…}{{$\dots$}}1
             {κ}{{$\kappa$}}1
             {×}{{$\times$}}1
             {≠}{{$\neq$}}1
             {μ}{{$\mu$}}1
             {↦}{{$\mapsto$}}1
             {←}{{$\gets$}}1
             {ϵ}{{$\epsilon$}}1
             {δ}{{$\delta$}}1
}

\begin{document}

\preprint{APS/123-QED}

\title{An LLM System for Autonomous Variational Quantum Circuit Design} 

\author{Kenya Sakka\textsuperscript{1}}
\author{Wataru Mizukami\textsuperscript{1}}
\author{Kosuke Mitarai\textsuperscript{1, 2}}

\affiliation{\textsuperscript{1}Center for Quantum Information and Quantum Biology,
  The University of Osaka, 1-2 Machikaneyama, Toyonaka 560-0043, Japan}
\affiliation{\textsuperscript{2}Graduate School of Engineering Science, The University of Osaka
1-3 Machikaneyama, Toyonaka, Osaka 560-8531, Japan}

\vspace{5pt}
\date{\today}

\begin{abstract}
The design of high performing quantum circuits remains largely dependent on human expertise.
We introduce an autonomous agentic framework that employs large language models (LLMs) to conduct iterative quantum circuit designs under explicit design constraints.
Our system integrates seven components: Exploration, Generation, Discussion, Validation, Storage, Evaluation, and Review.
These components form a closed-loop workflow that combines web-based knowledge acquisition, literature-grounded critique, executable code generation, and experimental feedback.
We evaluate the framework on two tasks: quantum feature map construction for quantum machine learning and ansatz generation for variational quantum eigensolver applications in quantum chemistry.
In image classification benchmarks, the best generated feature map outperforms representative quantum feature maps and, when scaled to larger qubit counts, surpasses the classical radial basis function kernel.
In molecular ground state estimation across seven molecules, the generated ansatz attains competitive accuracy with widely used chemically inspired and hardware-efficient constructions while satisfying the imposed scaling constraints.
These results establish LLM driven agentic system as a viable paradigm for automated quantum circuit design and illustrate how AI systems can participate in iterative scientific optimization workflows across scientific domains.
\end{abstract}

\maketitle

\section{Introduction}
\label{sec:introduction}

\begin{figure*}[t]
    \centering
    \includegraphics[width=1.0\linewidth]{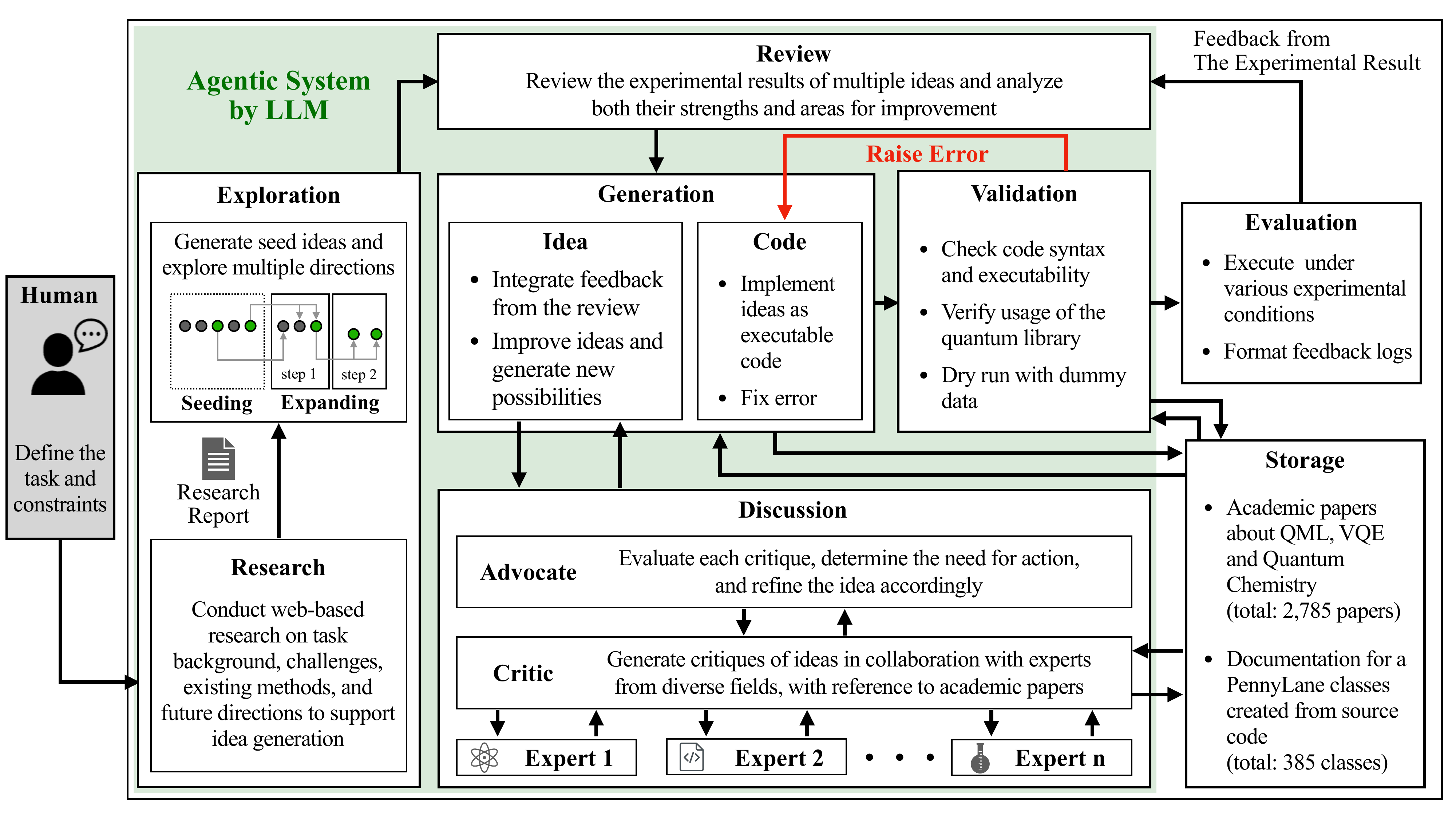}
    \caption{Overview of the proposed agentic system for automatic quantum circuit generation. The system autonomously conducts experiments and refines circuit designs through iterative collaboration among its components. The process consists of seven components: ``Exploration``, ``Generation'', ``Discussion'', ``Validation'', ``Evaluation'', ``Storage'', and ``Review''. The gray-shaded areas indicate the stages that require human involvement, such as selecting task type, and optionally defining constraints, whereas the other parts operate fully automatically through computer-based processes. The green-shaded regions denote modules powered by LLMs.}
    \label{fig:architecture}
\end{figure*}

The design of quantum circuits plays a central role in determining the practical performance of quantum algorithms.
In quantum machine learning (QML), the structure of a quantum feature map defines the geometry of the induced Hilbert space and directly influences generalization performance~\cite{havlivcek2019supervised,schuld2019quantum,mitarai2018quantum}.
In quantum chemistry, ans\"{a}tze determine the expressibility, trainability, and resource efficiency of variational quantum eigensolvers (VQE) used for molecular ground-state energy estimation~\cite{peruzzo2014variational}.
Despite advances in quantum hardware and algorithmic techniques, the quantum circuit design remains largely dependent on human intuition, domain-specific heuristics, and incremental modifications of established design paradigms~\cite{biamonte2017quantum, cerezo2021variational, bharti2022noisy}.

This reliance on manual design presents inherent challenges. 
In particular, despite numerous empirical successes, theoretical understanding of the relationship between quantum circuit structure and performance remains limited, making principled circuit design difficult both within specific tasks and across different application domains. 
Specialization is often unavoidable, yet it requires substantial human expertise and iterative experimentation, and the design burden increases with system size and circuit complexity.
Automated quantum architecture search methods have been proposed to expand exploration beyond manual heuristics, but such approaches typically operate within predefined templates~\cite{zhang2022differentiable, martyniuk2024quantum, nakaji2024gqe}.
More recently, large language models (LLMs) have demonstrated the ability to generate quantum programs and assist components of scientific workflows~\cite{basit2025pennylang, dupuis2024qiskit, zou2025agente, perezsanchez2026elagentequntur}.
However, most approaches automate isolated components, such as literature review or code generation, rather than embedding them within a closed-loop research cycle that systematically explores, evaluates, and refines circuit designs.

In our previous work~\cite{sakka2025automating}, we showed that LLMs can autonomously design quantum feature maps through iterative evaluation and refinement within an agentic framework.
That study demonstrated the feasibility of prompt-based circuit generation guided by quantitative feedback and achieved performance exceeding representative quantum baselines on benchmark datasets.
However, it was restricted to a single design task and relied primarily on the internal knowledge of the LLM during initial idea generation, without systematically integrating external domain knowledge.
The refinement process was driven mainly by performance-based feedback and lacked structured multi-perspective critique grounded in prior research.
Consequently, candidate quality varied, and computational effort was sometimes directed toward less promising directions, limiting the framework to a specific application setting.

In this work, we present an extended agentic framework that advances autonomous circuit design beyond task-specific pipelines.
The proposed system embeds LLMs within a structured research loop comprising seven components: ``Exploration``, ``Generation'', ``Discussion'', ``Validation'', ``Evaluation'', ``Storage'', and ``Review''. 
The ``Exploration'' component conducts web-based investigation before idea generation, surveying existing directions and expanding diverse seed designs.
The ``Discussion'' component assigns domain-specific expert roles and evaluates candidate ideas alongside retrieved literature, enabling structured multi-perspective critique.
The framework generates quantum circuits whose implementation logic is independent of a fixed number of qubits, ensuring scalability across system sizes.
Together, these mechanisms enable autonomous generation of circuit candidates while incorporating task performance, scalability, and resource efficiency into the design loop.

We evaluate the proposed framework on two conceptually distinct tasks: quantum feature map construction for QML and ansatz generation for molecular ground state energy estimation with the VQE.
In the QML setting, a central objective is to achieve performance that competes with or exceeds classical machine learning models.
Starting from diverse seed designs, the system iteratively refines candidate feature maps through quantitative feedback, yielding steady performance improvements over successive trials.
The best-performing feature map outperforms representative baseline quantum feature maps as well as the classical radial basis function (RBF) kernel on MNIST, Fashion-MNIST, and CIFAR-10.
Furthermore, the system produces executable feature map code whose logic remains valid across a range of qubit numbers, confirming that the learned design principles are scalable rather than tied to a fixed circuit size.
For molecular ground-state estimation, the best-performing ansatz outperforms several widely used hardware-efficient ans\"{a}tze, while remaining significantly more compact than chemically motivated constructions and achieving competitive accuracy with favorable parameter scaling.

These results demonstrate that structured agentic refinement can produce quantum circuits that balance performance and resource constraints across distinct scientific objectives.
Such automated circuit design frameworks are expected to serve as practical support tools for domain experts seeking to apply quantum computing within their own scientific fields.
By assisting researchers in navigating complex design spaces without requiring deep expertise in quantum circuit construction, these systems may help lower the barrier to interdisciplinary adoption of quantum technologies and facilitate broader exploration of quantum enhanced methods across diverse application domains.

\section{Preliminary}
\label{sec:preliminary}
This section outlines the fundamental concepts necessary for the subsequent discussion.

We begin by introducing quantum feature maps for QML and the variational quantum eigensolver for molecular ground state estimation, together with an overview of representative circuit design principles developed in previous studies.
Finally, we review recent efforts toward automating scientific discovery using artificial intelligence, and discuss emerging proposals in the context of quantum computing.

\subsection{Quantum Feature Map and Kernel Method}

In quantum machine learning, a quantum feature map is a data embedding circuit that maps a classical input $\mathbf{x}$ to a quantum state through an input dependent unitary $U(\mathbf{x})$~\cite{havlivcek2019supervised, schuld2019quantum, mitarai2018quantum, cerezo2022challenges}. Starting from the initial state $|0\rangle^{\otimes n}$, the resulting quantum feature state can be written as
\begin{equation}
\rho(\mathbf{x}) = U(\mathbf{x})|0\rangle^{\otimes n}\langle0|^{\otimes n} U(\mathbf{x})^\dagger.
\end{equation}
The design of $U(\mathbf{x})$ determines how classical information is embedded into Hilbert space and therefore strongly affects the expressivity and generalization behavior of the resulting model.

In this work, we focus on the quantum kernel approach, where the similarity between two inputs is measured by the overlap between their corresponding quantum feature states. A common choice is the Hilbert--Schmidt inner product
\begin{equation}
k(\mathbf{x}, \mathbf{x}') = \operatorname{Tr}[\rho(\mathbf{x}) \rho(\mathbf{x}')].
\label{eq:kernel}
\end{equation}
This kernel can be used with standard kernel based learning algorithms such as support vector machines or kernel ridge regression~\cite{havlivcek2019supervised, schuld2019quantum, cerezo2022challenges}. Because the predictive performance depends directly on the structure of the feature map circuit, the design of the quantum feature map is a central problem in QML and serves as one of the main targets of our automated circuit generation framework.

\subsection{Variational Quantum Eigensolver}

The VQE~\cite{peruzzo2014variational, mcclean2016theory, cerezo2021variational} is a hybrid quantum–classical algorithm that estimates the ground state energy of a given Hamiltonian $H$. 
It prepares a parameterized quantum state $|\psi(\boldsymbol{\theta})\rangle = U(\boldsymbol{\theta}) |0\rangle^{\otimes n}$ using an ansatz circuit $U(\boldsymbol{\theta})$ and iteratively optimizes the parameters $\boldsymbol{\theta}$ so that the expectation value $\langle \psi(\boldsymbol{\theta}) | H | \psi(\boldsymbol{\theta}) \rangle$ becomes minimal.
Once the optimization converges, the obtained energy corresponds to the ground-state energy, and the optimized state $|\psi(\boldsymbol{\theta})\rangle$ approximates the ground state of the Hamiltonian.

A key factor determining the performance of VQE is the design of the ansatz circuit. 
chemically inspired ans\"{a}tze, such as the unitary coupled cluster singles and doubles (UCCSD)~\cite{Lee2018JCTC, romero2018strategies}, can capture the electronic structure efficiently but may require deep circuits beyond the capability of current quantum hardware.
Hardware-efficient ans\"{a}tze~\cite{kandala2017hardware}, on the other hand, are shallow and better suited to current devices but often suffer from expressibility–trainability trade-offs~\cite{mcclean2018barren, cerezo2021cost}. 
Various adaptive strategies, such as ADAPT-VQE~\cite{grimsley2019adaptive} and its variants~\cite{tang2021qubit}, have been developed to balance these aspects by constructing the ansatz iteratively based on the energy gradient.

\subsection{Automated Science}
The use of AI for scientific discovery has expanded rapidly in recent years~\cite{wang2023scientific}, with LLMs in particular enabling new approaches to automating and accelerating scientific research~\cite{lu2024aiscientist, novikov2025alphaevolve, gottweis2025towards, mitchener2025kosmos, ghareeb2025robin, tie2025survey}.
In particular, LLMs have been used to automate and support key components of the research process, including literature analysis, hypothesis generation, experimental planning, software development, and the coordination of multi step scientific workflows.
Such approaches have been actively explored across a wide range of scientific domains, ranging from the automation of computational and simulation based research workflows~\cite{lu2024discopop, zhang2025large, swanson2025virtual} to the automation of experimental research involving interaction with physical systems and laboratory hardware~\cite{boiko2023autonomous, yoshikawa2023llm_chemistry_robotics, bran2024chemcrow}, with the aim of accelerating discovery and reducing manual intervention.

Motivated by these developments, similar ideas have begun to emerge in the context of quantum computing, where the design and optimization of quantum algorithms present a natural target for automation.
Even prior to advent of LLMs, quantum architecture search methods have been proposed to automatically design quantum circuits tailored to specific tasks~\cite{zhang2022differentiable, martyniuk2024quantum}.
More recently, approaches have been introduced in which quantum circuits are represented as sequences of tokens, analogous to natural language, enabling transformer based models~\cite{vaswani2017attention} to generate quantum circuits directly~\cite{nakaji2024gqe, minami2025generative}.
However, these methods typically rely on predefined circuit templates and therefore require task specific tuning to achieve satisfactory performance.

In the context of LLM-based approaches, several studies have explored the use of LLMs for the automatic generation of quantum programs, ranging from low-level quantum assembly languages such as QASM~\cite{fu2025qagent} to higher-level quantum programming languages~\cite{basit2025pennylang, dupuis2024qiskit}. 
LLM-based coding agents such as AlphaEvolve have further demonstrated the ability to generate and optimize quantum circuit code for approximating Hamiltonian evolution across a broad range of time scales and Hamiltonian parameters~\cite{novikov2025alphaevolve, Zhang2025QuantumComputationMolecularGeometry}. 
These studies highlight the promise of large-scale computational code search for quantum simulation, but are primarily designed for optimization within a fixed target task.
In particular, frameworks such as AlphaEvolve achieve strong performance in well-defined settings, where the search is driven by iterative, diff-based code refinement of candidate programs under task-specific evaluation pipelines, often requiring large-scale simulation-based evaluation of numerous candidates.

In contrast, our framework targets general-purpose quantum circuit design, rather than task-specific constructions such as feature maps or VQE ans\"{a}tze, and adopts a language-centric search process in which candidate ideas are first formulated and refined in natural language.
This enables more interpretable, concept-level reasoning prior to code-level realization.
Furthermore, our approach improves computational efficiency by introducing a ``Discussion'' component that performs lightweight, reasoning-based critique and refinement of candidate ideas, allowing promising designs to be filtered before costly simulation-based validation.

Concurrently, in our previous work~\cite{sakka2025automating}, we moved beyond using LLMs as tools for individual stages of the research process and proposed a framework that automates the entire research cycle for quantum feature map design, encompassing idea generation, experimentation, and iterative refinement.

Despite these advances, existing automated science frameworks in quantum computing remain largely limited to specific problem settings and controlled simulation environments. Extending such frameworks to more general circuit design and to realistic noisy settings involving quantum hardware remains an open challenge.


\section{LLM-based system for automatic generation of quantum circuit}\label{sec:methods}
Our agentic system for automatic generation of quantum circuit consists of seven components: ``Exploration'', ``Generation'', ``Discussion'', ``Storage'', ``Validation'', ``Evaluation'', and ``Review'', as shown in Fig.~\ref{fig:architecture}.
The `Exploration` component executes only before the first trial.
A single trial comprises processing by the remaining six components. By iteratively repeating this process while incorporating feedback from the results, the system progressively improves performance.
For reproducibility, the complete prompts, configuration files,
and implementation details are publicly available in the repository~\cite{astronaut_repository}.
In this section, we first describe the architecture of the present system in a self-contained manner and then summarize the main differences from our previous framework~\cite{sakka2025automating}.

\subsection{Exploration} \label{sec:methodsExploration}
In the ``Exploration'' component, the system first performs a comprehensive web-based search to generate a research report related to the given task. Based on this report, it then generates multiple seed ideas and evaluates them to filter and select a subset of promising candidates, which serve as the foundation for future idea refinement and improvement.

\subsubsection{Research for task background}
Investigating long-term research trends and related domain knowledge associated with the user specified task can improve the directionality of the initial quantum circuit design ideas. To take advantage of this, we employ a research oriented agent~\citep{openai2023deep} particularly effective for comprehensive exploration. This agent generates a research report that can inform and support the generation of seed ideas.

The agent system includes both a web search tool and a code interpreter, enabling it to retrieve relevant information from the web and perform basic analyses through code execution during the report generation process.
When constructing the research report, we instruct the LLM to first describe the background, current state, challenges, and insights \& recommendations of the task specified by the user. Subsequently, we instruct it to summarize existing quantum circuit designs in the form of a markdown table, including the core structure or design principle, advantages, limitations, and suitability for iterative optimization for each circuit.

\subsubsection{Idea Seeding \& Expanding}
Based on the research report, the system generates and selects seed ideas. At this stage, the goal is to explore a broad range of directions by generating multiple diverse candidate ideas and evaluating them to identify those with high potential for quantum circuit design. To prioritize responsiveness and efficiency, this phase does not involve external knowledge-based refinement or a result review process. Instead, ideas are generated directly from the research report followed immediately by experimental evaluation.

Based on the evaluation results, the number of seed ideas is gradually reduced while assessing the performance and characteristics of ideas from various directions. Through this process, the system selects an initial idea that will be further refined by the agentic system, and passes it to the ``Generation'' component. In this study, the number of seed ideas was systematically reduced in three stages: from 10, to 5, and finally to 2.

\subsection{Generation} \label{sec:methodsGeneration}
In the ``Generation'' component, the system uses LLMs for idea generation and code implementation.
This process begins with the generation of multiple candidate ideas. Based on the results from the ``Exploration'' component, review comments from the ``Review'' component, and the external knowledge, the system then engages in a discussion phase to refine the ideas (details are provided in Section~\ref{sec:methodsDiscussion}).
Finally, these refined ideas are implemented as Python programs.

\subsubsection{Idea generation}
The first process of the ``Generation'' component is generation of candidate ideas.
Here, the system instructs an LLM to generate ideas for quantum circuits. In this study, the term quantum circuit refers to quantum feature maps for QML and ans\"{a}tze for VQE.
For the first trial, a prompt is designed to instruct the LLM to generate ideas based on the evaluation results of all ideas produced by the ``Exploration'' component.
This process serves as the foundation for subsequent improvements.
In the subsequent trials, the system prompts the LLM to refine the design of the circuit based on the review comments provided from the ``Review'' component of previous trial.
Throughout the trials, we prompt the LLM to generate multiple ideas simultaneously to encourage diversity, while gradually narrowing the exploration scope through iterative prompt instructions.
Task-specific design constraints are also supplied through prompts so that the generated ideas remain compatible with the evaluation setting adopted for each task. The concrete constraints used in the QML and VQE experiments are described in Section~\ref{sec:experiments}.
The LLM structures each output into four components: an overview, a detailed explanation, corresponding mathematical expressions in TeX format, and a set of key sentences for search relevant academic papers.

\subsubsection{Implementation}
After completion of idea generation and refinement by the ``Discussion'' component, the system uses an LLM to generate a Python program that implements the corresponding quantum circuits.
First, all necessary classes for implementation were extracted by the LLM from the list of quantum libraries stored in the ``Storage'' component (Detailed in \ref{sec:methodsStorage}).
The LLM receives the idea name, detailed explanation and mathematical expression as inputs.
This process enables the generation of executable code that is compatible with the specific version of the selected library.
In addition, the LLM is explicitly instructed, through the implementation prompt, to generate code that is independent of the number of qubits, ensuring flexibility and scalability across different circuit sizes. 
Each element in the class list includes the defined class name in program, arguments (name, type, requirement and description) and brief description.
Implementation templates presented in Listing~\ref{lst:BaseFeatureMapCode} and Listing~\ref{lst:BaseAnsatzCode} are provided to guide the generation process. 
Finally, the LLM generates an executable quantum circuit program based on the prompts that include an idea explanation, an extracted method list and a program template.

\subsection{Discussion} \label{sec:methodsDiscussion}
In the ``Discussion'' component, the system improves the draft ideas generated by the ``Generation'' component by incorporating critical perspectives. This component is designed to address the difficulty that an LLM alone faces in objectively identifying fundamental issues in ideas. It aims to incorporate a process inspired by human research practices, where ideas are critically evaluated from multiple perspectives through collaboration with several experts. 
First, an LLM assigned the role of critic initiates a question and answer process by designating LLMs with various expert roles across different domains. Through these interactions, the system identifies scientific challenges and limitations of the ideas.
The LLM of advocate role then responds by deciding whether to accept each criticism and proposes refined versions of the ideas accordingly. By repeating this cycle of critical discussion, the system aims to enhance the draft ideas from a variety of perspectives.

\subsubsection{Critic} \label{sec:methodsCritic}
The objective of the critic role is to improve the quality and direction of a generated idea by offering critical commentary on its technical and logical issues in collaboration with various domain experts. The process begins by providing the LLM with the idea generated by the ``Generation'' component, reference papers retrieved from the ``Storage'' component.
Based on this information, the LLM constructs a list of questions directed to the expert role.
This list includes multiple expert profiles and corresponding questions, and responses are collected from each expert (see the Section \ref{sec:methodsExpert} for details).
Through several rounds of expert interaction, the critic role compiles a final list of comments on the idea along with their justifications, which is then passes to the advocate role.
To guide this process, the critic role follows a set of principles emphasizing scientific rigor, constructive feedback, and objectivity.
The critic is expected to identify blind spots, question assumptions, and evaluate practical limitations such as scalability or feasibility.
The strategy involves analyzing the core concept, assessing the methodology, and situating the idea within the broader literature, while offering specific, actionable suggestions for improvement.
In this study, up to two rounds of expert questioning were allowed per critic round, and both the number of expert questions and the number of final comments were limited to three.

\subsubsection{Expert} \label{sec:methodsExpert}
The objective of the expert role is to engage in question and answer interactions from the perspective of a domain expert, based on academic paper, in order to support the critic role in constructing a question list grounded in diverse scientific knowledge.
The LLM serving as the expert is provided with a prompt that includes the expert's assigned role (e.g., ``quantum algorithm benchmarking expert specializing in ansatz expressibility and symmetry effects in VQE optimization'', ``quantum chemistry theorist specializing in electronic structure and symmetry-breaking phenomena'', etc.), the content of the question, and relevant reference materials.
By passing pairs of expert roles and corresponding questions generated by the critic role as prompts, the system enables flexible switching among various expert perspectives during the interaction.
Reference materials are retrieved by using the question content as a search query and performing vector based search on a VectorDB containing academic paper information (see the Section \ref{sec:methodsStorage} for details).
In this study, the maximum number of academic papers retrieved per search query was set to three.

\subsubsection{Advocate} \label{sec:methodsAdvocate}
The objective of the advocate role is to assess whether the critiques provided by the critic role based on scientific reasoning necessitate revisions, and to refine the idea accordingly.
The LLM assigned to the advocate role receives the list of critiques generated by the critic role LLM.
For each item in the list, it determines whether a response is required and provides justification for that decision.
Based on the critiques deemed necessary to address, the advocate then refines the original idea. To guide this process, the advocate follows a set of principles that emphasize critical evaluation of feedback, selective incorporation of scientifically valid suggestions, and preservation of the core concept.
Revisions are made iteratively, balancing responsiveness to critique with the defense of well founded ideas.
The improved idea is passed back to the critic role, initiating the next round of discussion.
In this study, the maximum number of discussion rounds per idea set to three.

\subsection{Storage} \label{sec:methodsStorage}
The ``Storage'' component provides external knowledge to compensate for the limited internal knowledge of LLMs and to ensure compatibility with rapid evolving quantum software libraries.
As in our previous framework, the storage component stores relevant academic paper and technical documentation of the program libraries used for implementation.
Academic papers are stored in a vector database (VectorDB)~\citep{johnson2017billion, wang2021milvus}, which enables context-aware retrieval using natural language queries through dense vector representations~\citep{karpukhin2020dense}. This allows the system to retrieve relevant prior work efficiently during idea generation and refinement.
In the present framework, the documentation for program libraries is relatively small in size and is therefore stored locally in JSON format.
In the present framework, the documentation for program libraries is handled differently in order to improve validation efficiency. Since this documentation is relatively small in size and frequently accessed during code validation, it is stored locally in JSON format rather than in the VectorDB. Furthermore, because the access pattern is limited to either retrieving the full documentation or querying documentation for a specific class, a simple key-value structure is sufficient.

\subsection{Validation} \label{sec:methodsValidation}
The ``Validation'' component iteratively fixes the code until it becomes executable by static and dynamic analysis with external knowledge.
It consists of three static validation steps followed by one dynamic validation step.

As in our previous framework, the first two static validation steps perform basic Python-level checks, ensuring that the generated code can be compiled and conforms to Python syntax rules.
If these checks pass, the system extracts PennyLane class names from the generated code and retrieves their corresponding documentation from the ``Storage'' component.
Based on the associated classes and their documentation, the system performs syntactic analysis to verify the correctness of class usage, including class initialization, argument names, and argument types.
After passing all three static validation steps, the system proceeds to dynamic validation, where the code is executed using small dummy data to ensure it runs without errors.

If an error occurs at any stage of the validation process, the generated source code and the associated error messages are fed back to the LLM, which attempts to correct and regenerate the code. As in the previous framework, this correction loop is repeated until all validation steps pass or a predefined maximum number of retries is reached, which is set to three in this study.
In the present framework, the LLM is additionally allowed to access a web search tool during this correction process. This extension enables the model to leverage broader reference information beyond the technical documentation stored in the ``Storage'' component, particularly for resolving Python-level implementation errors that are not specific to quantum libraries.

\subsection{Evaluation} \label{sec:methodsEvaluation}
The ``Evaluation’’ component assesses the performance of quantum circuits using validated code. After the evaluation, the system feeds back the results to ``Review'' component which guide idea generation in the next trial. 
The specific evaluation metrics must be tailored to the research objective; for instance, a quantum feature map might be evaluated using accuracy, whereas a VQE ansatz would be evaluated based on the average difference between the VQE energy and the exact ground state energy.
Furthermore, the evaluation can extend beyond performance-based metrics to include resource considerations, such as gate counts or optimization difficulties, to assess hardware friendliness.

\subsection{Review} \label{sec:methodsReview}
The ``Review'' component analyzes the outcomes of the most recent trial based on feedback provided by the ``Evaluation'' component and summarizes key factors contributing to success and areas for improvement.
As in our previous framework, this analysis is performed by an LLM that considers both the current and past trials and provides guidance for subsequent idea refinement.
The prompt includes multiple candidate ideas sorted by performance, together with their explanations, mathematical formulations, and implementation code generated in the ``Generation'' component, as well as task-specific feedback information.
For the quantum feature map generation task, this feedback consists of execution time and classification performance metrics.
For the ansatz generation task, it includes scalability orders of gate and parameter counts, the average number of optimization iterations and computation time obtained by averaging over all bond lengths for each molecule, the cost function value, and the estimation error for each bond length.
The results of this review are used as supplementary input to the ``Generation'' component in the next trial.

Three remarks are in order.
First, we include computational cost related metrics in the feedback information to discourage the generation of unnecessarily complex quantum circuits.
As in our previous framework, training time is provided to prevent reliance on computationally expensive constructions.
In the ansatz generation task, the feedback further includes scalability orders of gate and parameter counts, and the prompt explicitly encourages hardware-efficient ans\"{a}tze.
While such designs are not inherently required for achieving high performance, they are essential for maintaining feasible optimization and execution times, particularly in quantum chemistry applications.

Second, hardware information is explicitly included in the prompt to ensure that the review process remains aligned with the intended scope of quantum feature map and ansatz construction.
As in the previous framework, this prevents the system from drifting toward broader quantum computing challenges that fall outside the focus of this study.
By explicitly encoding device-level constraints in the prompt, the same hardware assumptions are consistently enforced across both the ``Review'' and ``Generation'' components.

Lastly, for the ansatz generation task, we introduce domain-specific review criteria that go beyond performance-based evaluation.
Specifically, it verifies whether the ansatz allows for meaningful gradient computation and proper parameter optimization (Tier 1) and whether it adheres to fundamental physical and chemical principles while maintaining computational efficiency (Tier 2). 
These review criteria encourage the circuit to preserve particle number, total spin, and spatial symmetries; that it remains size-consistent and smoothly varying across molecular geometries; and that it can be initialized from chemically meaningful reference states.
By explicitly embedding these review criteria in the prompt, the system is guided to design ans\"{a}tze that are not only hardware-efficient but also physically interpretable and practically optimizable.

\subsection{Differences from the Previous Agentic System} \label{sec:diffSystem}
The proposed agentic system builds upon our previously introduced framework~\cite{sakka2025automating}, while incorporating several extensions that broaden its applicability and improve robustness and efficiency. In particular, the target task is extended from quantum feature map generation to the automatic construction of ansatz for VQE in quantum chemistry.

To reduce sensitivity to initial ideas and improve robustness, we introduce a new ``Exploration'' component that performs preliminary investigation of related tasks and structured exploration of candidate design directions prior to the first trial. This component provides a more diverse and informed initialization for subsequent optimization.
We further introduce a ``Discussion'' component that enables multi perspective evaluation of candidate ideas. In this component, multiple expert roles representing distinct domain specific viewpoints collaboratively assess and critique generated ideas through structured discussion, allowing diverse evaluation criteria to be incorporated during refinement.
At the same time, several components from the previous framework are simplified or removed based on empirical observations. The ``Scoring'' component, which evaluated generated ideas using LLM based scores for originality, feasibility, and versatility, is removed in the present framework. This decision is motivated by its high API cost and the limited impact of the resulting scores on downstream performance.
In addition, the mechanism in the ``Review'' component that explicitly determined the direction of idea refinement based on experimental results is removed. As shown by the ablation study in our previous work, this mechanism did not yield a clear performance improvement, and its removal simplifies the refinement process without degrading performance.
To reduce API cost and accelerate validation, the code validation strategy is modified. Instead of LLM based evaluation of library usage, the present framework employs static validation using a curated set of pre constructed technical documents.
Finally, to support ansatz generation for VQE, the scope of the ``Storage'' component is extended to include academic literature on quantum chemistry and VQE, as well as expanded PennyLane documentation covering chemistry modules and variational algorithms.

These changes preserve the overall agentic structure of the original framework while significantly extending its generality and applicability.

\section{Experiments}
\label{sec:experiments}
We evaluate the effectiveness of our LLM-based agentic system using two major tasks. 
The first task evaluates the generation of quantum feature maps through standard image classification benchmarks, which is the same task considered in our previous study. 
The second task evaluates the generation of quantum ans\"{a}tze through a series of experiments on standard molecule datasets.
Additional experimental details are summarized in Appendix~\ref{sec:addExpSetup}.

\subsection{Large language model}
We use ``o3-deep-research'' and ``GPT-5'' series of OpenAI's LLMs in the agentic system, switching between them depending on the task.
Specifically, the system uses the ``o3-deep-research-2025-06-26'' model to generate research reports for seed idea generation.
This model allows access to both web search and code interpreter tools to facilitate research tasks.
The ``gpt-5-2025-08-07'' model, which excels in reasoning tasks, is used for review, idea generation and code generation.
The ``gpt-5-mini-2025-08-07'' model, a highly versatile general purpose model, is employed for discussion and validation tasks. Furthermore, during the validation task, the LLM used for code fixing is allowed to access the web search tool.
The lightweight ``gpt-5-nano-2025-08-07'' model is used for academic paper summary tasks.
The \textit{temperature} parameter, which controls the randomness of output, is not supported in the ``o3-deep-research'' and ``GPT-5'' series models, and thus the output exhibits a fixed level of randomness.
Complete hyperparameter settings are provided in Appendix~\ref{sec:hyperparameters}.

\subsection{External knowledge}
We use the arXiv API to retrieve PDF files of academic papers.
We restrict the search criteria to the quant-ph category, which corresponds to quantum physics, and set the target period from January 1, 2013, to April 30, 2025.
We specify ``Quantum Machine Learning'', ``Variational Quantum Eigensolver'', ``VQE'' and ``Quantum Chemistry'' as the search keyword and find 2,785 papers.
To store them in the database, we extract text from the retrieved PDFs and segment it into chunks of 1,024 tokens.
We then convert these text segments into 1536-dimensional vector representations using OpenAI's ``text-embedding-3-small'' model.

We use the source code of PennyLane version 0.39.0 as a reference for software documentation.
Since PennyLane is open-source software, we can access both its source code and documentation publicly.
However, the documentation is not always up-to-date and lacks sufficient detail for the purpose of this study.
To address this, we create reference information for the LLM directly from the source code.
First, we begin by statically extracting all class definitions from the top-level and quantum chemistry modules of PennyLane, along with their class names, docstrings, and source code.
Next, the extracted information for each class is passed to the LLM, which transforms it into a structured format consisting of a class description, the call name used in the program, and for each argument its name, type, whether it is required, and a description.
The corresponding source code of each class is also provided to the LLM, allowing the generation of accurate reference documentation even when docstrings are missing or outdated.
As a result, documentation for a total of 385 classes was generated and saved locally in JSON format.
The example of JSON file described in Appendix~\ref{sec:storedProgramDocument}.

\subsection{Evaluation settings}
\subsubsection{Quantum feature map for QML}\label{sec:evalQFMSettings}
As the first evaluation task, we consider quantum feature map generation for image classification, following the same experimental setup as in our previous work.
We evaluate the generated feature maps on the MNIST~\citep{lecun1998mnist} dataset using a support vector machine (SVM) as the downstream classifier.
To assess generalizability, the final evaluation is further conducted on the Fashion-MNIST~\citep{xiao2017fashion} and CIFAR-10~\citep{krizhevsky2009learning} datasets.
As in our previous work~\cite{sakka2025automating}, we restrict the generated quantum feature maps to purely quantum data embeddings by prohibiting nonlinear classical transformations and trainable parameters in the feature map itself. This is because, if such nonlinear preprocessing were allowed, the system could in principle rely on a highly expressive classical embedding, such as a neural network, and use the quantum circuit only as a small additional component, which would fall outside the intended scope of quantum feature map design considered in this study.
Kernel values are computed from exact statevector simulations using PennyLane's~\citep{Bergholm2018} \texttt{lightning.qubit} device.
Since GPU acceleration offers limited advantage for the relatively small statevector sizes considered in the QML experiments, all kernel computations are performed on CPUs.
The resulting kernel matrices are then used to train the SVM classifier.
Detailed descriptions of the datasets, preprocessing, downstream models, and baseline methods are provided in Appendix~\ref{sec:imageDatasets} and Appendix~\ref{sec:featureMapBaseline}.

Additionally, we assess scalability with respect to system size.
Our agentic system is instructed through a prompt to generate logic and code that does not depend on a fixed number of qubits.
We also verify that the generated quantum feature map follows a scalable logic that does not depend on the number of qubits, one of the key device-level constraints.
Specifically, we vary the number of qubits from two to fourteen in increments of two and confirm that the resulting program remains executable across different system size. 
Using the different qubit configurations, we then evaluate how the accuracy changes by testing the model on the test sets of all three benchmark datasets.

\subsubsection{Ansatz for VQE}
\label{sec:evalVQESettings}
As the second evaluation task, we consider automatic ansatz generation for VQE in quantum chemistry.
We use seven molecular systems for ansatz evaluation and refinement: $\mathrm{H_{2}}$, $\mathrm{H_{4}}$, $\mathrm{H_{6}}$, \textrm{LiH}, $\mathrm{BeH_{2}}$, $\mathrm{H_{2}O}$, and $\mathrm{N_{2}}$.
For all systems, the molecular Hamiltonians are constructed with the minimal basis set STO-3G~\cite{Hehre1969}, using the Hartree–Fock method implemented in the PySCF package~\cite{Sun2020} for mean-field electronic structure calculations.
Fermionic operators are subsequently mapped to qubit operators through the Jordan--Wigner transformation~\cite{Jordan1928}.
To keep the evaluation loop computationally feasible, we restrict the generated ans\"{a}tze to single parametrized circuit constructions and disallow approaches that require repeated circuit growth or repeated optimization subroutines, such as ADAPT-VQE~\citep{grimsley2019adaptive}.
We also require the total number of single-qubit rotation gates, CNOT gates, and trainable parameters to scale as $O(N^2)$ or lower with respect to the number of qubits $N$, so that the search remains focused on hardware-efficient circuit families.
As a result, many chemically inspired ans\"{a}tze with less favorable scaling, including UCCSD~\cite{romero2018strategies}, are outside the target design space of the present experiment.
To control computational cost while ensuring systematic coverage of molecular geometries, the bond length of each molecule is restricted to ten values.
Specifically, we select the bond length corresponding to the minimum electronic energy of each molecule and nine additional values equally spaced between predefined minimum and maximum distances.
These bond length values were consistently applied across all trials and evaluations within our proposed agentic system.
The details of the molecular settings are described in Appendix~\ref{sec:moleculeSettings}.

The generated circuits are evaluated based on the accuracy of the VQE energy, scaling of the gate count and trainable parameters with respect to qubits, and the number of iteration needed for VQE optimization. 
For the accuracy, we evaluate it by the average difference between the VQE energy and the exact ground state energy:
\begin{equation}
\frac{1}{n \cdot m} \sum_{i=1}^{m} \sum_{j=1}^{n} \left| E_{\mathrm{Exact}}^{(i,j)} - E_{\mathrm{VQE}}^{(i,j)} \right|,
\end{equation}
where $m$ is the number of molecules evaluated, $n$ is the number of atomic configurations per molecule, $E_{\mathrm{Exact}}^{(i,j)}$ denotes the exact energy for the $i$-th molecule at the $j$-th bond length, $E_{\mathrm{VQE}}^{(i,j)}$ denotes the corresponding VQE energy.
In this study, the value of $m$ is seven and the value of $n$ is ten.
This metric serves as the primary optimization objective for the agentic system and is used to rank candidate ans\"{a}tze during the automated search.
Although the aggregation of energy errors across different molecules provides a simple scalar objective, the absolute energy scales vary among molecular systems, and therefore the averaged error does not fully capture instance-specific behavior.
To complement this aggregated metric, the agent is additionally provided with molecule- and bond-length-specific energy errors during the review process, enabling more detailed analysis of performance on individual molecular configurations.
We perform all VQE energy calculations using PennyLane's \texttt{lightning.gpu} device, where energy expectation values are computed directly from exact statevector simulations without finite-shot statistical errors.
GPU acceleration is used to efficiently evaluate the larger statevector computations arising during VQE optimization.

The exact ground state energies are calculated using the PySCF package, with the computational method chosen according to the size and complexity of each molecular system.
Full configuration interaction (FCI) is employed for small molecules such as $\mathrm{H_{2}}$, $\mathrm{H_{4}}$, and $\mathrm{H_{6}}$, where exact diagonalization is computationally feasible.
For larger systems, complete active space configuration interaction (CASCI) is used, since performing FCI would be intractable.
In the CASCI calculations, core orbitals are frozen to reduce computational cost: one core orbital is frozen for LiH (Li 1s), $\mathrm{BeH_{2}}$ (Be 1s), and $\mathrm{H_{2}O}$ (O 1s), while two core orbitals are frozen for $\mathrm{N_{2}}$ (N 1s on each atom).
The details of molecular settings are described in Appendix~\ref{sec:moleculeSettings}.

The optimization in the VQE procedure is performed using the SciPy library~\citep{2020SciPy-NMeth}.
Specifically, the Sequential Least Squares Programming (SLSQP) algorithm~\citep{Kraft1988SQP} is employed, with the maximum number of iterations set to 2000 and the convergence tolerance specified as $1.0 \times 10^{-6}$.
For each molecule, perform the optimization in order of increasing bond length.
The optimized result from one bond length is used as the initial value for the next longer bond length.
For the shortest bond length, the initial values are randomly assigned within the range $[0.0, 0.1]$.
The same random seed was used for all experiments.
Although VQE performance can depend on the initial parameter initialization, we did not perform multiple optimization runs with different random seeds or select the best-performing seed.
All reported results therefore correspond to a single optimization run for each evaluation.

Finally, after the optimization of generated ansatz are finished, we compare the performance against several representative ans\"{a}tze that have been widely adopted in the literature.
Specifically, we consider three chemistry-inspired ans\"{a}tze (UCCSD~\cite{romero2018strategies}, All Single Double~\cite{arrazola2022universal}, and kUpCCSD~\cite{Lee2018JCTC}) and three hardware-efficient or symmetry-preserving designs (GateFabric~\cite{Anselmetti2021}, Particle U1~\cite{Barkoutsos2018PRA}, and Particle U2~\cite{Barkoutsos2018PRA}).
These ans\"{a}tze span a broad range of structural complexity and expressivity, providing a balanced benchmark for assessing both accuracy and circuit efficiency.
The summaries of these ans\"{a}tze are described in Appendix~\ref{sec:ansatzBaseline}.

\section{Results}
\label{sec:results}
In this section, we first confirm that our agentic system can iteratively improve performance and generate high performance quantum circuits in two tasks: feature map generation and ansatz generation. We then present a detailed analysis of the results for each individual task. Finally, we report the results of ablation study conducted to evaluate the effectiveness of the newly introduced ``Exploration'' and ``Discussion'' component.

\subsection{Performance evaluation of the agentic system}
\label{sec:resultsAgent}
\begin{figure}[ht]
    \centering
    \includegraphics[width=1.0\linewidth]{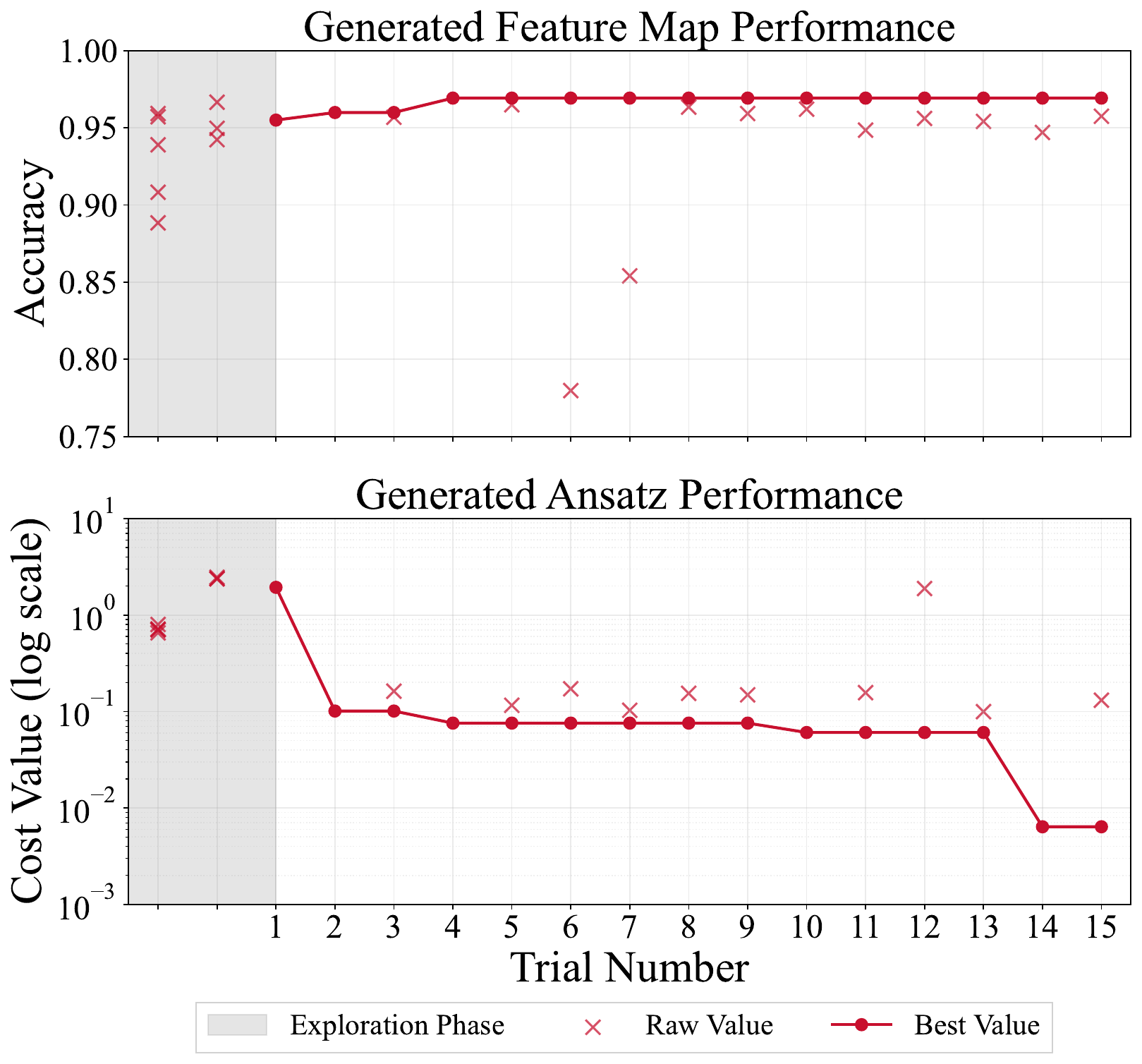}
    \caption{The performance trajectories of the quantum feature map and ansatz over 15 trials. The gray-shaded region represents the exploration phase, during which diverse candidates were generated to explore various directions inspired by the research report. The red crosses denote the raw performance values in each trial, while the red circles indicate the best values obtained up to that trial. The upper panel shows the SVM classification results obtained using the generated quantum feature map, and the lower panel shows the VQE results using the generated ansatz.}
    \label{fig:performance_trajectory}
\end{figure}

The results of iterative refinement over 15 trials using our agentic system are shown in Fig.~\ref{fig:performance_trajectory}.
The upper panel shows the trajectories of accuracy on the MNIST test datasets obtained with the generated quantum feature map.
The lower panel shows the trajectories of the cost function values computed using the  generated ansatz for estimating the molecule ground state described in Section \ref{sec:evalVQESettings}.
A higher accuracy value (closer to 1.0) indicates better performance, while a lower cost function value corresponds to higher performance.
Both tasks demonstrate consistent performance improvement across successive trials. 
One of the target benchmarks for evaluating the accuracy of VQE is the chemical accuracy threshold ($1.6 \times 10^{-3}$ Hartree).
In our best performing ansatz, the final cost value reached $6.4 \times 10^{-3}$ Hartree.
Although this value falls slightly short of the chemical accuracy threshold, it is worth noting that the molecular set used to compute the cost includes challenging systems for which chemically motivated ans\"{a}tze such as UCCSD do not consistently achieve chemical accuracy in our benchmarks (see Fig.~\ref{fig:pec_difference}).

Furthermore, to evaluate the robustness of the proposed agentic system with respect to initialization and stochasticity, we conducted four additional independent experiments in addition to the  best-performing result shown in Fig.~\ref{fig:performance_trajectory}.
The corresponding performance trajectories are presented in Appendix~\ref{sec:robustnessModel}.
All experiments exhibited performance improvements through the iterative trials, confirming the robustness of our agentic system. Further analysis of agentic behavior is collected in Appendix~\ref{sec:detailedAnalysisAgent}.

\subsection{Evaluation results of the generated quantum feature map}
\label{sec:resultsQFM}
The iterative refinement process identified a feature map that consistently outperformed the others, ultimately reaching the best performance at trial 4. 
The evaluation results and comparisons with baseline methods are summarized in Table \ref{tab:compare_results}.
All baseline quantum feature maps were computed on a noiseless simulator with 10 qubits.
The proposed method was evaluated on noiseless simulators with 10 and 14 qubits.
In a comparison under the same system size of 10 qubits, the generated quantum feature map outperformed all baseline quantum feature maps and our previous system across all datasets.
Furthermore, when extending the system size to 14 qubits, it surpassed the classical RBF kernel on the full test datasets.

\begin{table*}
\caption{Generated quantum feature map performance on different datasets.}
\label{tab:compare_results}
\setlength{\tabcolsep}{4pt}
\begin{ruledtabular}
\begin{tabular}{llrrr}
Type & Method & \multicolumn{1}{c}{MNIST} & \multicolumn{1}{c}{Fashion-MNIST} & \multicolumn{1}{c}{CIFAR-10} \\
\hline
Classical & Linear kernel
  & 0.9385 $\pm$ 0.0002 & 0.8437 $\pm$ 0.0009 & 0.4087 $\pm$ 0.0011 \\
 & Polynomial kernel
  & 0.9667 $\pm$ 0.0058 & 0.8702 $\pm$ 0.0030 & 0.5375 $\pm$ 0.0014 \\
 & Sigmoid kernel
  & 0.9343 $\pm$ 0.0002 & 0.8189 $\pm$ 0.0120 & 0.4079 $\pm$ 0.0006 \\
 & RBF kernel
  & 0.9765 $\pm$ 0.0005 & 0.8864 $\pm$ 0.0014 & 0.5669 $\pm$ 0.0085 \\
Quantum & ZZ feature map \cite{havlivcek2019supervised}
  & 0.9255 $\pm$ 0.0009 & 0.8252 $\pm$ 0.0023 & 0.3907 $\pm$ 0.0016 \\
 & NPQC feature map \cite{haug2021quantum}
  & 0.9644 $\pm$ 0.0028 & 0.8749 $\pm$ 0.0026 & 0.4903 $\pm$ 0.0188 \\
 & YZCX feature map \cite{haug2021quantum}
  & 0.9727 $\pm$ 0.0006 & 0.8778 $\pm$ 0.0049 & 0.4753 $\pm$ 0.0341 \\
 & Our previous system \cite{sakka2025automating}
  & 0.9731 $\pm$ 0.0008 & 0.8835 $\pm$ 0.0021 & 0.5290 $\pm$ 0.0030 \\
 & Generated (ours, 10 qubits)
  & \textbf{0.9772} $\pm$ 0.0002 & 0.8880 $\pm$ 0.0014 & 0.5450 $\pm$ 0.0057 \\
 & Generated (ours, 14 qubits)
  & 0.9767 $\pm$ 0.0002 & \textbf{0.8888} $\pm$ 0.0007 & \textbf{0.5734} $\pm$ 0.0006 \\
\end{tabular}
\end{ruledtabular}
\end{table*}

\begin{figure}[ht]
    \centering
    \includegraphics[width=1.0\linewidth]{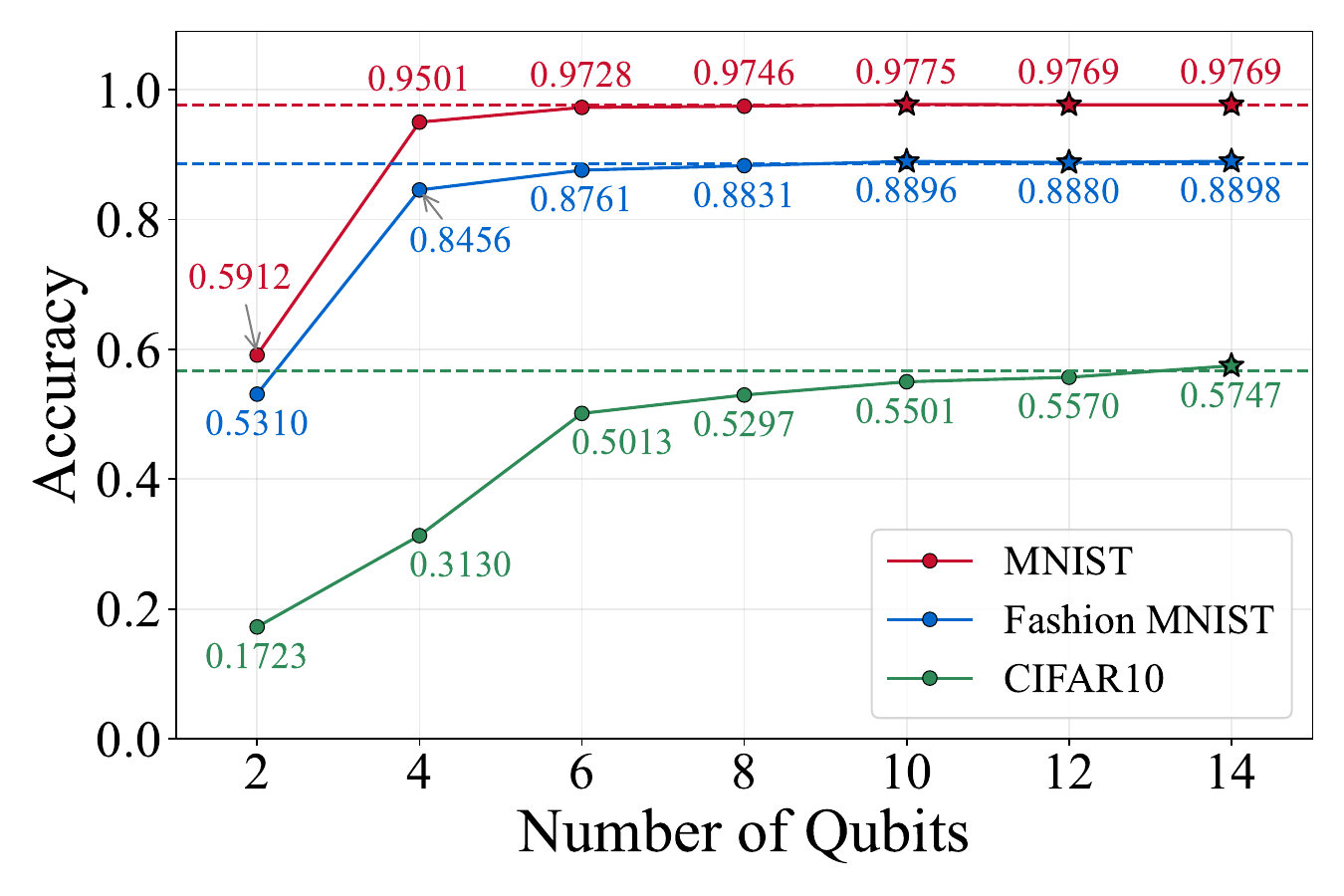}
    \caption{Evaluation of the scalability of the best-performing quantum feature map generated by the agentic system. The x-axis represents the number of qubits. The y-axis indicates the classification accuracy on the test datasets. The dotted line represents the mean accuracy of the RBF kernel, and the star markers indicate cases where the quantum feature map outperform the RBF kernel.}
    \label{tab:qubit_scalability}
\end{figure}

We next investigate the scalability of the generated feature map with respect to the number of qubits.
Fig.~\ref{tab:qubit_scalability} illustrates how the accuracy on the test datasets change as the number of qubits is increased from two to fourteen in increments of two.
Within the tested range, the generated code was confirmed to be executable without errors.
The results show that the accuracy of the generated feature maps improves as the number of qubits increases, and notably, when the qubit count reaches 6, the accuracy already surpasses that of baseline quantum feature maps using 10 qubits, as reported in Table \ref{tab:compare_results}.
This scalability is likely attributed to the ability of the generated quantum feature map to dynamically adjust its circuit structural configuration according to the number of qubits.

The circuit diagram and implementation of the best-performing quantum feature map are provided in Appendix~\ref{sec:qkBestSolutions}. The generated circuit uses a repeated data re-uploading structure~\cite{perez2020data} with nontrivial entangling and rotation patterns. However, despite its strong empirical performance, we found it difficult to extract a clear and generalizable design insight from the resulting structure alone, and therefore leave a more detailed interpretation as future work. At the same time, this also highlights a potential advantage of automated circuit search: it may uncover effective circuit structures that are not easily reached through conventional human-designed heuristics.

For completeness, the ``Exploration'' component initially produced five seed quantum feature maps reflecting different design directions.
These seed circuits differ in their strategies for data preprocessing, encoding, and entangling patterns, and their circuit diagrams and algorithmic summaries are provided in Appendix~\ref{sec:appendixQunatumFM}, especially Appendix~\ref{sec:featureMapSeed}.

\subsection{Evaluation results of the generated VQE ansatz}
\label{sec:resultsVQE}

\begin{figure}[t]
    \centering
    \includegraphics[width=\linewidth]{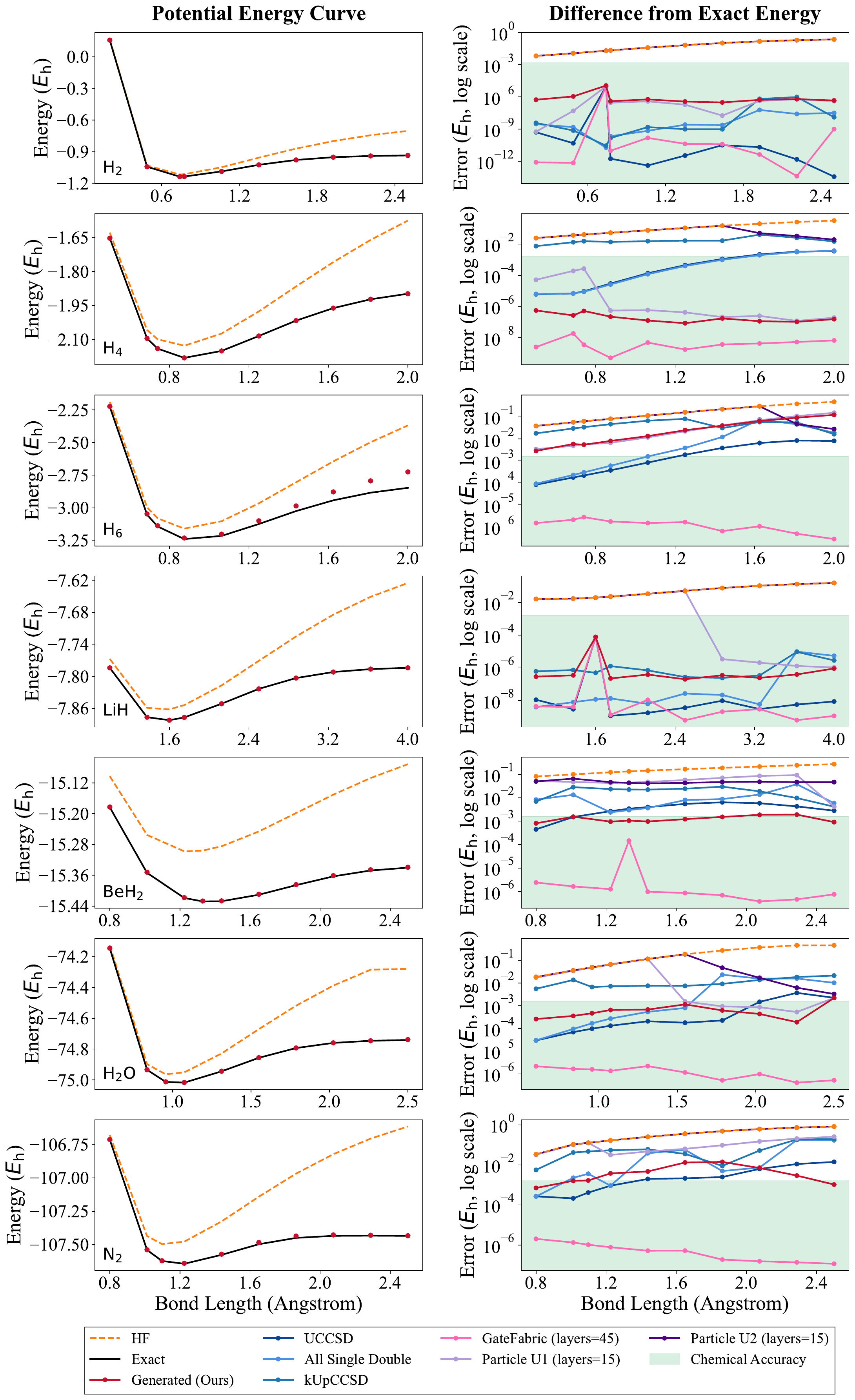}
    \caption{Ground-state potential energy curves (PECs) estimated by VQE (left) and energy errors relative to exact energy (FCI or CASCI) (right). The right panel plots the errors for several well-known ans\"{a}tze as well as for the best-performing ansatz generated by our system. From top to bottom, the plots correspond to H$_{2}$, H$_{4}$, H$_{6}$, LiH, BeH$_{2}$, H$_{2}$O and N$_{2}$ molecules. The green shaded area indicates the threshold for chemical accuracy.}
    \label{fig:pec_difference}
\end{figure}

We first evaluated the accuracy of the ground-state energy estimation performed by the VQE using the ansatz generated by our agentic system, whose iterative refinement process reached its minimum cost value at trial 14.
For comparison with representative hardware-efficient baselines, we set the depth of GateFabric to 45 and that of Particle U1 and Particle U2 to 15, so that each baseline used more trainable parameters than our generated ansatz. Since increasing the number of parameters generally improves the achievable energy accuracy in noiseless VQE, this choice provides a conservative comparison in which the baseline circuits are not disadvantaged by underparameterization.
The results are presented on the left side of Fig.~\ref{fig:pec_difference}.
In addition, the right side of Fig.~\ref{fig:pec_difference} shows the energy difference from the exact energies, enabling direct comparison across methods.
The detailed experimental conditions are described in Section \ref{sec:evalVQESettings}.

As observed in the left side of Fig.~\ref{fig:pec_difference}, our ansatz achieves high accuracy for all molecules except for the bond stretched region of H$_{6}$.
Hydrogen chain systems are well known to exhibit strong static correlation as the interatomic distance increases. 
This trend is consistent with that observed for other well established ans\"{a}tze, highlighting the intrinsic difficulty of capturing multi-reference correlation within a compact variational circuit.
As shown on the right side of Fig.~\ref{fig:pec_difference}, the generated ansatz maintains chemical accuracy ($\le1.6\times10^{-3}$ Hartree) across almost all molecules, except for certain stretched configurations of H$_{6}$ and N$_{2}$.

When compared with other hardware-efficient ans\"{a}tze that were configured with a similar number of parameters, the generated ansatz performed worse than GateFabric but outperformed Particle U1 and Particle U2 under most conditions. 
When compared with the chemically inspired ans\"{a}tze, the generated ansatz showed lower accuracy for molecules such as hydrogen chains, for which both approaches achieved high estimation accuracy. However, it exhibited competitive performance for the other molecules. In addition, for larger bond lengths of BeH$_2$, H$_2$O and N$_2$, the generated ansatz outperformed UCCSD and achieved chemical accuracy.  
Overall, our ansatz accurately reproduces the exact ground-state energy.

\begin{figure}[ht]
    \centering
    \includegraphics[width=1.0\linewidth]{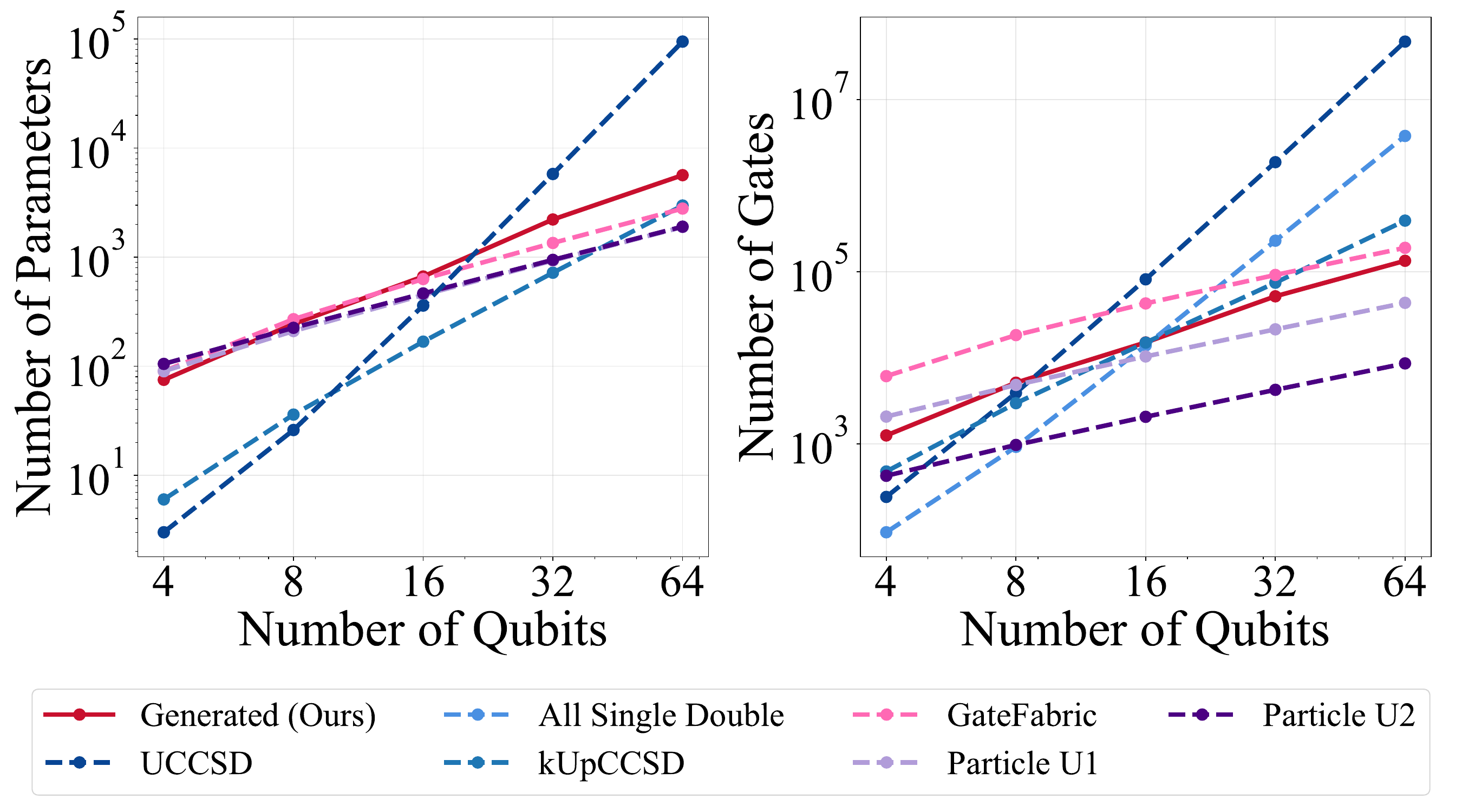}
    \caption{Scalability analysis of the well-known and our best ans\"{a}tze. The scaling of the number of trainable parameters (left) and the gate counts (right) with respect to qubit count. The gate counts were evaluated after decomposing the circuits into the gate set \{\(R_X, R_Y, R_Z\), CNOT\}}
    \label{fig:vqe_qubits_depths_params}
\end{figure}

We next analyze the scalability of the optimized ansatz in terms of trainable parameter count and gate count as the system size increases.
Fig.~\ref{fig:vqe_qubits_depths_params} presents this scaling behavior.
The number of qubits in the ansatz was varied as 4, 8, 16, 32, and 64, and comparisons were made with representative ans\"{a}tze commonly used in quantum chemistry calculations.
GateFabric, Particle U1 and Particle U2, which are known as hardware-efficient ans\"{a}tze, allow the circuit depth to be adjusted through the parameter depth, which controls the number of structural repetitions.

The left panel shows the the number of trainable parameters as a function of the number of qubits. The generated ansatz required more trainable parameters than chemically inspired ans\"{a}tze when implemented on small scale device, such as those with four qubits.
However, since the number of parameters scales as $O(L \cdot N)$ with respect to the depth of layer $L$ and the number of qubits $N$.
Additionally, the depth of the generated ansatz is fixed 10 when the number of qubits greater than 42.
The generated ansatz becomes more parameter efficient than other ans\"{a}tze, as the device size increases. 
The right panel shows the number of required gates as a function of the number of qubits.
The total gate count was obtained by decomposing all gates in the ansatz into single rotation gate (\(R_X, R_Y, R_Z\)) and two qubit gate (CNOT) and counting their occurrences.
The gate count increases approximately as $O(L \cdot N)$.
This trend is consistent with that observed for the trainable parameters, suggesting that the generated ansatz becomes increasingly efficient compared with UCCSD as the number of qubits grows and the system scales to larger devices.
Based on these results, we confirmed that our agentic system successfully generated hardware-efficient ansatz in accordance with the given prompt instructions.
However, we found that some of the Tier 2 review criteria introduced in the ``Review'' component, particularly particle-number, spin, and spatial-symmetry preservation, were not consistently satisfied by the generated circuits.
Unlike gate-count and parameter-scaling constraints, which were explicitly evaluated and fed back to the agent during the review process, these physical properties were specified only as textual review instructions and were not quantitatively verified at each iteration.
This observation suggests that, when such physical constraints are critical, incorporating explicit symmetry evaluations and feeding the resulting metrics back into the ``Review'' component may provide a more effective mechanism for enforcing them during autonomous circuit optimization.

The circuit diagram and implementation details of the best performing VQE ansatz are provided in Appendix~\ref{sec:vqeBestSolutions}.
The generated ansatz combines multiple entangling patterns while maintaining favorable scaling of parameters and gates, and additional analyses of its components are reported in Appendix~\ref{sec:paramAnsatz}.
At the same time, as in the QML setting, we found it difficult to draw a strong and general conclusion from the detailed circuit structure itself, so we treat these architectural observations as descriptive rather than as definitive design principles.
This limitation may also reflect a strength of the agentic search process, namely its ability to discover effective circuit organizations that do not follow familiar human-designed templates.

For completeness, the ``Exploration'' component initially produced five seed ans\"{a}tze spanning different circuit construction principles.
These seed ans\"{a}tze adopt distinct strategies in entangling topology, parameter sharing, symmetry preservation, and structural organization, while all satisfying the imposed \(O(N^2)\) resource constraint on trainable parameters and gate count.
Circuit diagrams and algorithmic summaries of the five seed ans\"{a}tze are provided in Appendix~\ref{sec:appendixVQE}, especially Appendix~\ref{sec:ansatzSeed}.

\subsection{Ablation study}
\label{sec:ablationStudy}

\begin{figure*}[t]
    \centering
    \includegraphics[width=0.8\linewidth]{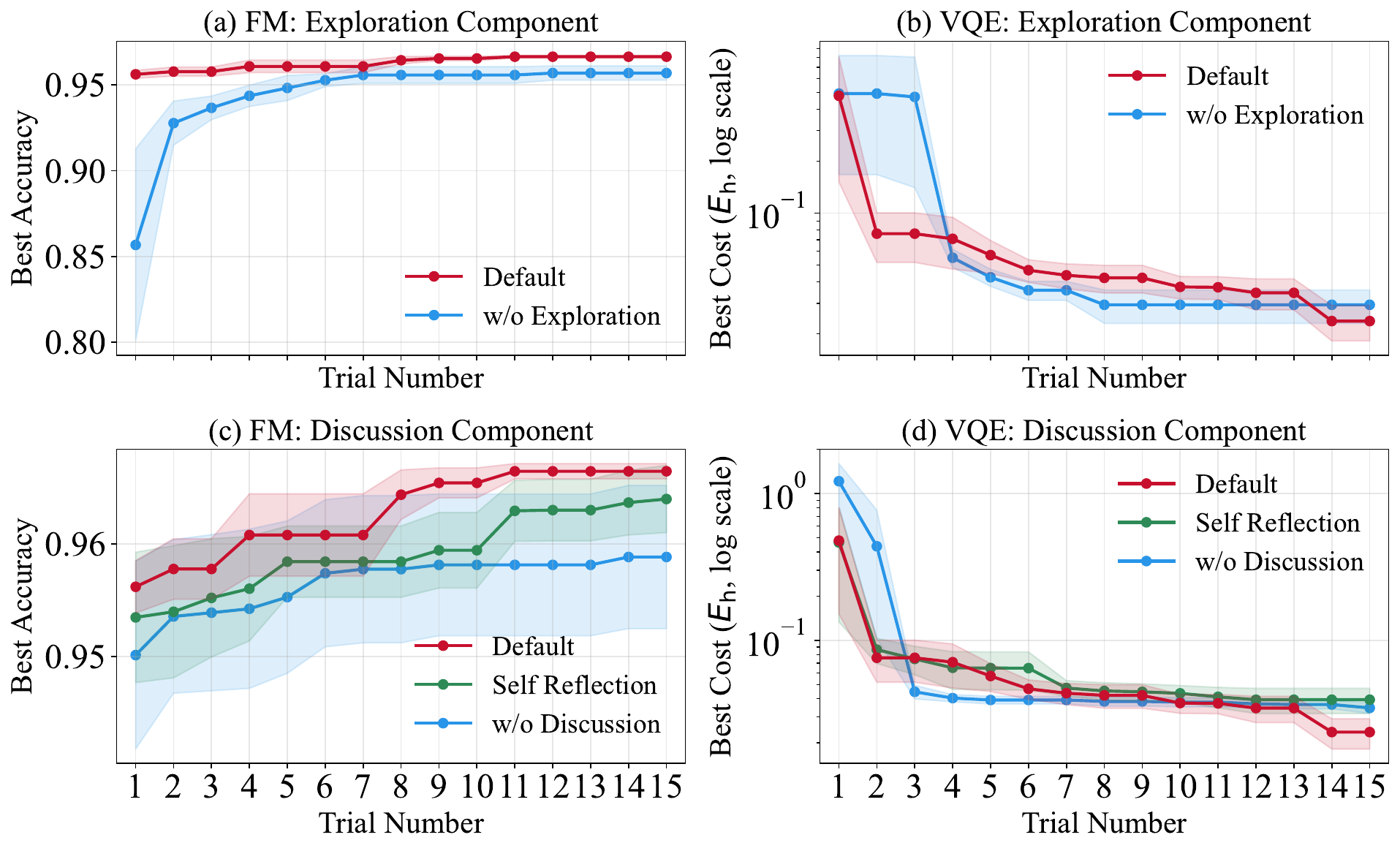}
    \caption{Results of the ablation study for the quantum feature map and VQE ansatz generation tasks. Solid lines show the mean of the best performance across five independent runs for each trial. Shaded bands denote the standard error of the mean. The two graphs in the upper panel compare the system with the version without the ``Exploration'' component. The two graphs in the lower panel compare different refinement strategies: the proposed ``Discussion'' component based on multi-role LLM interactions, a self-reflection in which a single LLM simply refines its own outputs, and a setting without any refinement step after initial generation (``w/o Discussion'').}
    \label{fig:ablationSutudy}
\end{figure*}

In this study, we performed ablation studies on the ``Exploration'' and ``Discussion'' component. 
To assess their effectiveness, we compared variants with these components removed or modified, and the results are presented in Fig.~\ref{fig:ablationSutudy}.
For each condition, we ran five independent experiments and report the mean and standard error across trials.

The ``Exploration'' component was introduced to sophisticated initial ideas based on a comprehensive survey of prior research.
As shown in the upper panel of Fig.~\ref{fig:ablationSutudy}, the ``Exploration'' component improves the quality and stability of initial candidates.
In the quantum feature map task, the Trial 1 accuracy increases from $0.8567 \pm 0.0626$ without Exploration to $0.9562 \pm 0.0026$ with the default setting, indicating higher performance together with a substantial reduction in variance.
In the ansatz generation task, the initial energy is also lower with Exploration, changing from $-1.5444 \pm 0.5720$ to $-1.7320 \pm 0.6596$, indicating improved initial solutions.
Despite these improvements, the relationship between initial and final performance remains weak when Exploration is used.
In the feature map task, the correlation is $0.0034$, indicating essentially no predictive relationship. 
In contrast, without Exploration, the correlation is $-0.5723$, suggesting an unstable search process in which strong initial candidates do not reliably lead to better final outcomes.
A similar trend is observed in the ansatz generation task: the correlation is $0.53$ with Exploration but becomes $-0.16$ without it, indicating reduced consistency in the optimization process.
These observations indicate that the benefit of the Exploration component does not primarily arise from improved initialization, but rather from stabilizing the search trajectory and improving the overall optimization dynamics.
Details of the research reports generated through the ``Exploration'' phase were summarized in Appendix~\ref{sec:researchReport}.

Finally, to verify the effectiveness of the ``Discussion'' component, we compared the default system with two ablated conditions.
In the first condition, the multi-role discussion was replaced by a self-reflection mechanism~\cite{madaan2023self}, where a single LLM refines its own outputs based solely on generic improvement instructions, as adopted in our previous system.
In the second condition, the refinement step was entirely removed, and ideas were evaluated immediately after initial generation.
The results are shown in the lower panel of Fig.~\ref{fig:ablationSutudy}.
Although a certain degree of variance was observed across trails in both tasks, the final performance consistently improved when the ``Discussion'' component was enabled.
This result suggests that literature-grounded multi-perspective critique provides more reliable guidance for iterative refinement than reflection strategies that rely only on the model's internal reasoning.

Since these results are based on only five runs, they should be interpreted as preliminary, and further validation with a larger number of experiments is required.

\section{Conclusion}
\label{sec:conclusion}
In this study, we presented an enhanced version of our previously developed agentic system for autonomous quantum circuit design. 
The framework employs a LLM to iteratively generate, evaluate, and refine quantum feature maps for QML and ans\"{a}tze for VQE together with executable code 
To improve idea generation and refinement, we introduced an ``Exploration'' component that systematically surveys design strategies through web based research, and a ``Discussion'' component that incorporates literature informed expert level feedback into the iterative optimization process.

The best quantum feature map generated by the framework outperformed representative baseline quantum feature maps as well as classical methods, including the widely used RBF kernel, across all benchmark datasets. 
In the VQE ansatz design task, the system autonomously and iteratively improved circuit performance while satisfying the structural constraints of hardware-efficient ans\"{a}tze. 
For the benchmark molecules, the generated ansatz outperformed several representative hardware-efficient ans\"{a}tze and achieved competitive performance relative to chemically inspired ans\"{a}tze.
These results demonstrate that the proposed approach can generate accurate and hardware-efficient quantum circuits while explicitly controlling parameter and gate scaling under resource constraints.

While these results are encouraging, several challenges remain before such systems can be deployed on practical quantum hardware.
Extending the approach to device specific circuit generation that accounts for connectivity and noise characteristics will be essential for practical deployment. 
In particular, ensuring robustness under noisy feedback signals remains a key challenge for stable iterative refinement on real devices.

Beyond VQE, the generated ans\"{a}tze may also serve as compact state-preparation circuits for a broader class of quantum algorithms.
In particular, quantum phase estimation and recent hybrid quantum chemistry methods often benefit from high-quality approximate wavefunctions that can be prepared with shallow and hardware-efficient circuits~\cite{kanno2023quantum,robledo2025chemistry,motta2023bridging}.
Investigating whether agentically generated ans\"{a}tze can provide improved state preparation for such workflows represents a promising direction for future work.

This work focused on well defined benchmark settings, including QML and ground state energy estimation, where quantitative evaluation is established. 
A natural next challenge within AI for Science is the identification of classically intractable yet quantum accessible problems in which autonomous circuit design may yield practical quantum advantage. 
Progress in this direction will require more capable agentic optimization strategies and benchmark tasks that meaningfully reflect the capabilities of near term and future quantum hardware. 
At the same time, as LLMs can generate a vast number of circuit candidates, improving the interpretability of design intentions and developing theoretical analyses of successful circuits will be essential.
Such advances may help move automated quantum circuit generation beyond empirical optimization toward deeper theoretical understanding.

\section{Code and data availability}
\label{sec:sourceCode}
The source code of the agentic system and the experimental results reported in this study are publicly available at \url{https://github.com/Qyusu/astronaut}.

\begin{acknowledgments}
This work is supported by MEXT Quantum Leap Flagship Program (MEXT Q-LEAP) Grant Nos. JPMXS0120319794 and JPMXS0118067394, JST NEXUS Grant No. JPMJNX26C6, JST Moonshot R\&D Grant No. JPMJMS256J, and JST COI-NEXT Grant No. JPMJPF2014.
K.M. is supported by JSPS KAKENHI Grant No. 23H03819, 24K16980, and JST CREST Grant Number JPMJCR24I4.
\end{acknowledgments}

\bibliography{references}

\clearpage
\appendix
\onecolumngrid

\part*{Appendix}

\renewcommand{\arraystretch}{1.4}
\begin{tabularx}{\textwidth}{@{}X r@{}}

\ref{sec:addExpSetup}. Additional experimental setup \dotfill & \pageref{sec:addExpSetup} \\
\quad \ref{sec:hyperparameters}. Hyperparameters \dotfill & \pageref{sec:hyperparameters} \\
\quad \ref{sec:storedProgramDocument}. Stored program document \dotfill & \pageref{sec:storedProgramDocument} \\
\quad \ref{sec:detailedQMLSettings}. Detailed settings for quantum feature map experiment \dotfill & \pageref{sec:detailedQMLSettings} \\
\quad \ref{sec:detailedVQESettings}. Detailed settings for VQE experiment \dotfill & \pageref{sec:detailedVQESettings} \\

\ref{sec:detailedAnalysisAgent}. Detailed analysis of agentic system \dotfill & \pageref{sec:detailedAnalysisAgent} \\
\quad \ref{sec:robustnessModel}. Robustness of our agentic system \dotfill & \pageref{sec:robustnessModel} \\
\quad \ref{sec:researchReport}. Classification of generated research reports \dotfill & \pageref{sec:researchReport} \\
\quad \ref{sec:expertRole}. Patterns in expert roles \dotfill & \pageref{sec:expertRole} \\

\ref{sec:appendixQunatumFM}. Generation of quantum feature maps for QML \dotfill & \pageref{sec:appendixQunatumFM} \\
\quad \ref{sec:featureMapSeed}. Seed quantum feature maps \dotfill & \pageref{sec:featureMapSeed} \\
\quad \ref{sec:qkBestSolutions}. Best-performing quantum feature map \dotfill & \pageref{sec:qkBestSolutions} \\
\quad \ref{sec:comparsionKernel}. Comparison of kernel matrices \dotfill & \pageref{sec:comparsionKernel} \\

\ref{sec:appendixVQE}. Generation of ans\"{a}tze for VQE \dotfill & \pageref{sec:appendixVQE} \\
\quad \ref{sec:ansatzSeed}. Seed ans\"{a}tze \dotfill & \pageref{sec:ansatzSeed} \\
\quad \ref{sec:vqeBestSolutions}. Best-performing ansatz \dotfill & \pageref{sec:vqeBestSolutions} \\
\quad \ref{sec:paramAnsatz}. Effect of modules in the best generated ansatz \dotfill & \pageref{sec:paramAnsatz} \\

\end{tabularx}

\newpage

\section{Additional experimental setup}
\label{sec:addExpSetup}

\subsection{Hyperparameters}
\label{sec:hyperparameters}
In our developed agent system, all nine types of hyperparameters are set at runtime. Table \ref{tab:hyperparams} summarizes their overview and the values used in this study.
\begin{table}[ht]
\caption{Hyperparameters for our agent system.}
\label{tab:hyperparams}
\setlength{\tabcolsep}{3pt}
\renewcommand{\arraystretch}{1.05}
\small
\begin{threeparttable}
\begin{ruledtabular}
\begin{tabular}{llll}
Component & Name & Description & Value \\
\hline
All
& max\_trial\_num
& Maximum number of experimental trials per run.
& 15 \\
Exploration
& research\_mode
& Research mode controlling exploration depth (none, light, or deep).
& deep \\
& expanding\_steps
& Step size for idea expansion within the exploration component.
& [5, 3] \\
Generation
& max\_idea\_num
& Maximum number of ideas generated simultaneously per trial.
& 2 \\
& max\_reflection\_round
& Maximum number of reflection rounds per trial.
& 2 \\
& max\_paper\_per\_query
& Maximum number of papers retrieved per search query.
& 3 \\
Evaluation-FM
& n\_qubits\tnote{a}
& Number of qubits used in evaluation runs.
& 10 \\
Evaluation-VQE
& n\_qubits\tnote{a}
& Number of qubits determined by the target molecule.
& \{4, 8, 10, 12\} \\
Review
& max\_suggestion\_num
& Maximum number of improvement suggestions during idea review.
& 3 \\
\end{tabular}
\end{ruledtabular}
\begin{tablenotes}\footnotesize
\item[a] The circuit templates for both the quantum feature map for QML and ansatz for VQE are independent of the number of qubits.
The qubit counts reported here correspond to the configurations evaluated during the quantum circuit search process performed by the proposed agentic framework.
\end{tablenotes}
\end{threeparttable}
\end{table}

\subsection{Stored program document}
\label{sec:storedProgramDocument}
The libraries related to quantum computation are currently under active development, and major updates that break backward compatibility are not uncommon. Consequently, LLMs generate code, it is expected that supplementing their internal knowledge with accurate external information will enable more stable and executable code generation. In this study, to support LLM based code generation and perform static validation of the generated code, we constructed reference documentation from the source code of PennyLane~\cite{Bergholm2018} version 0.39.0 and stored it in local storage.

The generated documentation targets all 385 classes defined in the ``qml.'' module, which provides general purpose quantum operations, and the ``qml.qchem.'' module, which supports quantum chemistry computations. The documentation is automatically generated by a LLM from the source code. For each class, the model extracts and generates information about the arguments (name, type, required or optional, and description) as well as explanations regarding the usage and functionality. Each result is defined in JSON format, with the class name as the key and the detailed description as the value. As example of the generated documentation for the CNOT gate is shown in Listing~\ref{lst:pennylaneDocExample}.

\newpage

\lstdefinelanguage{json}{
    basicstyle=\ttfamily\footnotesize,
    numbers=none,
    stepnumber=1,
    showstringspaces=false,
    breaklines=true,
    breakatwhitespace=false,
    breakindent=2em,
    columns=fullflexible,
    literate=
      {⟨}{{$\langle$}}1
      {⟩}{{$\rangle$}}1
}

\begin{center}
\captionof{lstlisting}{Example of local stored program documents}
\label{lst:pennylaneDocExample}

\begin{tcolorbox}[colframe=black, colback=white, boxrule=0.5pt,
                  left=2mm, right=2mm, top=1mm, bottom=1mm]
\begin{lstlisting}[language=json]
{
  "qml.CNOT": {
    "args": [
      {
        "name": "wires",
        "type": "Sequence[int]",
        "required": true,
        "description": "A sequence of two integers specifying the wires (qubits) that the CNOT gate acts on. The first wire in the sequence is the control qubit, and the second is the target qubit. For example, [0, 1] applies the CNOT with control on wire 0 and target on wire 1."
      },
      {
        "name": "id",
        "type": "Any or None",
        "required": false,
        "description": "An optional identifier for the operation instance. This can be used to uniquely tag or reference the operation within a quantum circuit. If not provided, defaults to None."
      }
    ],
    "description": "The qml.CNOT method constructs a controlled-NOT (CNOT) quantum gate operation acting on two specified wires (qubits). The CNOT gate flips the state of the target qubit (second wire) if and only if the control qubit (first wire) is in the |1⟩ state. The method requires exactly two wires, with the first being the control and the second the target. The operation does not take any trainable parameters. The CNOT gate is fundamental in quantum computing for creating entanglement and implementing universal quantum logic. The method can be used within a quantum circuit to apply the CNOT operation between two qubits, and optionally, an identifier can be supplied for tracking or referencing the operation."
  }
}
\end{lstlisting}
\end{tcolorbox}
\end{center}

\subsection{Detailed settings for quantum feature map experiment}
\label{sec:detailedQMLSettings}
This subsection provides detailed experimental settings for the quantum feature map generation task, which are largely consistent with those used in our previous work.

\subsubsection{Benchmark datasets for image classification}
\label{sec:imageDatasets}
In this study, we used three types of image classification datasets to evaluate the generated quantum feature maps.
The details of datasets are described below.

\begin{itemize}
    \item \textbf{MNIST}~\citep{lecun1998mnist}: The MNIST dataset consists of 70,000 handwritten digit images, each of which is a 28$\times$28 grayscale image labeled with one of the digits from 0 to 9. 
    MNIST is officially divided into 60,000 training images and 10,000 test images. 
    Our agentic system uses a sampled subset of the training data and do not use any of the test set. 
    When it finishes the iterative improvements of the feature maps, we evaluate the feature maps generated from the system on the entire test dataset.
    
    \item \textbf{Fashion-MNIST}~\citep{xiao2017fashion}: The Fashion-MNIST dataset comprises 70,000 images related to fashion, each of which is a 28$\times$28 grayscale image categorized into one of ten classes, such as clothing and footwear. This dataset was used solely for the purpose of evaluating the generalization ability of the quantum feature map generated by our agentic system.
    
    \item \textbf{CIFAR-10}~\citep{krizhevsky2009learning}: The CIFAR-10 dataset consists of 60,000 color images, where each image is a 32$\times$32 RGB image. The dataset includes ten classification labels, such as automobiles, birds, and other object categories. This dataset was used solely for the purpose of evaluating the generalization ability of the quantum feature map generated by our agentic system.
\end{itemize}

\subsubsection{Dataset split and evaluation protocol}
For the quantum feature map generation task, the MNIST dataset is used during the iterative generation process of the agentic system.
To enhance computational efficiency during the iterative quantum circuit generation process in our agentic system, we use only the MNIST training dataset.
This dataset is then divided into 36,000 images for training, 12,000 for validation, and 12,000 for testing.
After the feature map generation is completed, the best-performing feature maps are evaluated on the full MNIST test set.
To assess generalization performance, the same feature maps are additionally evaluated on the Fashion-MNIST and CIFAR-10 datasets.
These evaluation protocols are consistent across all experiments.

\subsubsection{Input data preprocessing}
To make the image inputs compatible with a qubit limited quantum feature maps, we apply principal component analysis (PCA) to reduce the dimensionality of the image data.
This is a standard approach when constructing quantum feature map for relatively high-dimensional inputs such as images~\citep{huang2021power}.
In this work, we use the top 80 principal components as input features and normalize the data to the range $[0,1]$.
This preprocessing procedure is fixed throughout all experiments. The agentic system does not observe the original image data at any stage and is not allowed to modify the preprocessing pipeline.

\subsubsection{Kernel evaluation and downstream classifier}
The effectiveness of the quantum feature maps using an support vector machine (SVM) as the downstream model in the ``Evaluation'' component. 
Kernel values are computed using the Hilbert--Schmidt inner product and assembled into a kernel matrix.
All kernel evaluations are performed using a noiseless simulator provided by PennyLane's \texttt{lightning.qubit} device.

Using the computed kernel matrix, we then train an SVM classifier using \texttt{scikit-learn}~\citep{scikit-learn}.
We choose the SVM due to its efficiency and reproducibility in kernel-based learning tasks. The SVM has a hyperparameter \texttt{C} that controls the strength of regularization. To determine the optimal value of \texttt{C}, we perform a hyperparameter search in the range from 0.0 to 100.0 using Optuna~\citep{akiba2019optuna}. This search is conducted using 10,000 samples randomly selected from the training dataset.

\subsubsection{Benchmark quantum feature map designs}
\label{sec:featureMapBaseline}
We used the three representative quantum feature maps as baseline methods to compare the performance of the quantum feature map generated by our agentic system.
For an input vector $\boldsymbol{x}=(x_1,\ldots,x_d)\in\mathbb{R}^d$ and $n$ qubits, each baseline feature map is defined as a unitary embedding $U_{\phi}^{(L)}(\boldsymbol{x})$.
Here, $L$ denotes the number of repetitions (layers) of the fundamental feature map block in the circuit, which controls the circuit depth and expressive power.
The three baselines are defined as follows:

\begin{enumerate}
  \item \textbf{ZZ Feature Map}~\citep{havlivcek2019supervised}:
  The ZZ feature map is defined as
    \begin{equation}
    U_{\mathrm{ZZ}}^{(L)}(\boldsymbol{x})
    =
    \prod_{\ell=1}^{L}
    \left\{
    \left[
    \prod_{i=1}^{n-1}
    \exp\!\left(
    -i\, (\pi - x_i)(\pi - x_{i+1})\, Z_i Z_{i+1}
    \right)
    \right]
    \left[
    \prod_{i=1}^{n}
    \left(R_Z(x_i) H\right)_i
    \right]
    \right\}
    \end{equation}
    In each feature map block, Hadamard gates are applied to all qubits, followed by $Z$ rotations $R_Z(x_i)$. The block then applies nearest neighbour Ising type entanglers,
    $\mathrm{ZZ}(\theta_{i,i+1})=\exp[-i\theta_{i,i+1} Z_i Z_{i+1}/2]$, 
    to every adjacent pair, with $\theta_{i,i+1}=2(\pi-x_i)(\pi-x_{i+1})$. 
    This standard construction combines linear $Z$ phase encoding with quadratic ZZ phases, and repeating the block $L$ times increases expressivity.

  \item \textbf{Natural Parameterized Quantum Circuit (NPQC) Feature Map} ~\cite{haug2021quantum}: 
    The NPQC feature map constructed on an even number of qubits $n$, and defined as
        \begin{equation}
        \begin{aligned}
        U_{\mathrm{NPQC}}^{(L)}(\boldsymbol{x})
        &=
        \left[
        \prod_{\ell=1}^{L}
        \left\{
        \prod_{i=0,2,\ldots,n-2}
        \Big(
        R_Y(\tfrac{\pi}{2})_{i}\,
        \mathrm{CZ}_{\,i,\,t(i,\ell)}\,
        R_Y(c\,x_{k(i,\ell)}+\tfrac{\pi}{2})_{i}\,
        R_Z(c\,x_{k(i,\ell)+1}+\tfrac{\pi}{2})_{i}
        \Big)
        \right\}
        \right]
        \\
        &\qquad
        \left[
        \prod_{q=0}^{n-1}
        R_Z(c\,x_{2q+2}+\tfrac{\pi}{2})_{q}\,
        R_Y(c\,x_{2q+1}+\tfrac{\pi}{2})_{q}
        \right].
        \end{aligned}
        \end{equation}
    Here, $c$ is a global scaling factor controlling the kernel length scale. 
    The data indices $k(i,\ell)$ cycle sequentially through the components of $\boldsymbol{x}$ modulo $d$, so that the input is reuploaded across the circuit. 
    Entanglement is applied only from even indexed qubits $i$ to targets $t(i,\ell)$ specified by the NPQC pattern. In repetition $\ell$, the target is shifted by $2s(\ell)+1$ sites with wrap around, where $s(\ell)$ is the largest integer such that $2^{s(\ell)}$ divides $\ell$ (equivalently, the number of trailing zeros in the binary representation of $\ell$). 
    This hierarchical pairing yields the NPQC family with an analytically tractable identity quantum Fisher information at a reference point, which leads to an approximately isotropic radial basis function like kernel, as discussed in~\cite{haug2021quantum}.

  \item \textbf{YZ-CX Parameterized Quantum Circuit (YZCX) Feature Map}~\citep{haug2021quantum}:
    The YZCX feature map is defined as
        \begin{equation}
        \begin{aligned}
        U_{\mathrm{YZCX}}^{(L)}(\boldsymbol{x})
        &=
        \prod_{\ell=0}^{L-1}
        \Bigg[
        \prod_{i=0}^{n-1}
        \Big(
        R_Y(c\,x_{k(i,\ell)})_i\,
        R_Y(\alpha_{i,\ell})_i\,
        R_Z(c\,x_{k(i,\ell)+1})_i\,
        R_Z(\beta_{i,\ell})_i
        \Big)
        \\
        &\qquad\qquad\times
        \prod_{\substack{i=0,\ldots,n-2\\ i \equiv \ell\ (\mathrm{mod}\ 2)}}
        \mathrm{CNOT}_{i,i+1}
        \Bigg]
        \end{aligned}
        \end{equation}
    Here, $c$ is a global scaling factor controlling the effective kernel length scale, and the indices $k(i,\ell)$ cycle sequentially through the components of $\boldsymbol{x}$ modulo $d$, so that the input is reuploaded throughout the circuit. 
    The angles $\alpha_{i,\ell},\beta_{i,\ell}\in[0,2\pi)$ are fixed random parameters drawn once from a uniform distribution (with a prescribed seed) and kept constant during training. 
  
\end{enumerate}

\subsubsection{Template for quantum feature map implementation}
The implementation template provided to the LLM together with the ideas generated by the ``Generation'' component is shown in Listing~\ref{lst:BaseFeatureMapCode}.
The template defines an empty class that inherits from the feature map base class provided by QXMT, an experiment management tool for quantum machine learning (\url{https://github.com/Qyusu/qxmt}).
The base class implements the interfaces and method required for evaluation in the ``Evaluation'' component.
The LLM is then instructed to implement the quantum feature map using PennyLane, one of the quantum computing libraries, including all necessary external library imports.

\begin{lstlisting}[style=codeFormat, caption={The template for implementing generated feature map idea}, label=lst:BaseFeatureMapCode]
import numpy as np
from qxmt.constants import PENNYLANE_PLATFORM
from qxmt.feature_maps import BaseFeatureMap

# above are the default imports. DO NOT REMOVE THEM.
# new imports can be added below this line if needed.


class SeedFeatureMap(BaseFeatureMap):
    """Seed feature map class.

    Args:
        BaseFeatureMap (_type_): base feature map class

    Example:
    """

    def __init__(self, n_qubits: int) -> None:
        """Initialize the Seed feature map class.

        Args:
            n_qubits (int): number of qubits
        """
        super().__init__(PENNYLANE_PLATFORM, n_qubits)
        # hyperparameters
        self.n_qubits: int = n_qubits

    def feature_map(self, x: np.ndarray) -> None:
        """Create quantum circuit of feature map.
        The input data is a sample of MNIST image data. It is decomposed into 80 features by PCA.

        Args:
            x (np.ndarray): input data shape is (80,).
        """
        # define your quantum feature map here
        pass
\end{lstlisting}

\subsection{Detailed settings for VQE experiment}
\label{sec:detailedVQESettings}
This subsection provides supplementary details for the VQE experiments introduced in this work.
While the main experimental settings are described in the main text, we summarize here the molecular configurations and benchmark ansatz constructions used for evaluation.

\subsubsection{Molecular settings}
\label{sec:moleculeSettings}
In this study, we calculated the ground-state energies of hydrogen chain and small molecular systems (H$_2$, H$_4$, H$_6$, LiH, BeH$_2$, H$_2$O, and N$_2$).  
The computational settings for each molecule are summarized below.

\begin{itemize}
  \item \textbf{H$_2$:} 
  Atomic composition: H--H; charge=0; multiplicity=1;  
  active electrons=2; active orbitals=2; qubits=4;  
  bond length range=0.2--2.5~\AA; number of samples=10;  
  equilibrium bond length=0.741~\AA;  
  exact energy reference: FCI.

  \item \textbf{H$_4$:} 
  Atomic composition: linear H$_4$ chain; charge=0; multiplicity=1;  
  active electrons=4; active orbitals=4; qubits=8;  
  bond length range=0.5--2.0~\AA; number of samples=10;  
  equilibrium bond length=0.741~\AA;  
  exact energy reference: FCI.

  \item \textbf{H$_6$:}  
  Atomic composition: linear H$_6$ chain; charge=0; multiplicity=1;  
  active electrons=6; active orbitals=6; qubits=12;  
  bond length range=0.5--2.0~\AA; number of samples=10;  
  equilibrium bond length=0.741~\AA;  
  exact energy reference: FCI.

  \item \textbf{LiH:} 
  Atomic composition: Li--H; charge=0; multiplicity=1;  
  active electrons=2; active orbitals=5; qubits=10;  
  bond length range=1.0--4.0~\AA; number of samples=10;  
  equilibrium bond length=1.595~\AA;  
  exact energy reference: CASCI.

  \item \textbf{BeH$_2$:}  
  Atomic composition: linear Be--H--H; charge=0; multiplicity=1;  
  active electrons=4; active orbitals=6; qubits=12;  
  bond length range=0.8--2.5~\AA; number of samples=10;  
  equilibrium bond length=1.326~\AA;  
  exact energy reference: CASCI.

  \item \textbf{H$_2$O:}  
  Atomic composition: bent H--O--H (104.5$^\circ$); charge=0; multiplicity=1;  
  active electrons=8; active orbitals=6; qubits=12;  
  bond length range=0.6--2.5~\AA; number of samples=10;  
  equilibrium bond length=0.957~\AA;  
  exact energy reference: CASCI.

  \item \textbf{N$_2$:}  
  Atomic composition: N$\equiv$N; charge=0; multiplicity=1;  
  active electrons=6; active orbitals=6; qubits=12;  
  bond length range=0.8--2.5~\AA; number of samples=10;  
  equilibrium bond length=1.098~\AA;  
  exact energy reference: CASCI.
\end{itemize}

\subsubsection{Benchmark ansatz designs}
\label{sec:ansatzBaseline}
We used the following six representative ans\"{a}tze as baseline methods to evaluate the advantages of the ansatz generated by our agentic system.
These baseline methods were implemented using the built in ansatz templates provided by PennyLane~\cite{Bergholm2018}.
We used PennyLane version 0.39.0 for all experiments.
The $N$ denotes the number of qubits used to represent the active space after the fermion to qubit mapping, so $N$ coincides with the number of active spin orbitals up to possible reductions by qubit tapering or frozen orbitals.
The number of electrons is denoted by $\eta$.
Occupied spin orbitals in the Hartree Fock reference are labeled by $i,j,\dots$, virtual spin orbitals by $a,b,\dots$, and general spin orbitals by $p,q,r,s,\dots$.
The Hartree--Fock reference state is $\ket{HF}$.
The number of repeated layers is denoted by $k$ or $L$, depending on the ansatz.
For GateFabric, Particle U1, and Particle U2, $\mathcal{E}$ denotes the set of nearest-neighbor qubit pairs.
Fermionic creation and annihilation operators operators are $a^{\dagger}$ and $a$, and $h.c.$ denotes Hermitian conjugation.
For all ans\"{a}tze considered below, the variational state is prepared as $\ket{\psi(\boldsymbol{\theta})} = U(\boldsymbol{\theta})\ket{HF}$, with the specific from of $U(\boldsymbol{\theta})$ defined for each ansatz.

\begin{enumerate}
  \item \textbf{Unitary Coupled Cluster Singles and Doubles (UCCSD) ansatz}~\cite{romero2018strategies}:
  The UCCSD ansatz is the chemically motivated VQE baseline obtained by exponentiating anti Hermitian single and double excitation operators. UCCSD is the chemically motivated VQE ansatz obtained by exponentiating anti-Hermitian single and double excitation operators acting on the Hartree--Fock reference state. It inherits the physical intuition of classical CCSD while providing a variationally optimizable wavefunction suitable for quantum computers. Here $T_1$ and $T_2$ denote the single and double excitation operators, respectively, whose amplitudes are treated as variational parameters.
    \begin{equation}
    U_{\mathrm{UCCSD}}(\boldsymbol{\theta})
    =
    \exp{(T_{1}(\boldsymbol{\theta})+T_{2}(\boldsymbol{\theta})-h.c.)}
    \end{equation}
  The number of parameters equals the number of single plus double amplitudes, scaling as $O(N_{occ}N_{virt} + N_{occ}^{2}N_{virt}^{2})$, which is $O(N^{4})$ in the worst case. With a first order Trotterization and standard fermion to qubit mappings, the two qubit gate count and circuit depth typically scale as $O(N^{5})$.

  \item \textbf{All Single Double ansatz}~\cite{arrazola2022universal}:
  The all single double ansatz generalizes UCCSD by including all symmetry-preserving single and double excitations within the active space, without restricting to occupied to virtual transitions relative to the reference. This reduces the dependence on the occupied–virtual partition of the reference state and increases expressivity, which can be advantageous for systems with strong static correlation, at the cost of a larger search space.
    \begin{equation}
    U_{\mathrm{ASD}}(\boldsymbol{\theta})
    =
    \exp{\Big(
    \sum_{pq}\theta_{p}^{q}a_{q}^{\dagger}a_{p}
    +\frac{1}{4}\sum_{pqrs}\theta_{pq}^{rs}a_{r}^{\dagger}a_{s}^{\dagger}a_{q}a_{p}
    -h.c.
    \Big)}
    \end{equation}
  The generalized singles contribute $O(N^{2})$ parameters and generalized doubles contribute $O(N^{4})$, giving a total of $O(N^{4})$ parameters with a larger prefactor than UCCSD. Implementation costs are dominated by the number of excitation terms, leading to a typical gate count scaling around $O(N^{5})$.

  \item \textbf{k-UpCCSD ansatz}~\cite{Lee2018JCTC}:
  The k-UpCCSD ansatz is a sparse unitary coupled cluster construction based on the observation that a significant fraction of electron correlation can be captured by generalized single excitations together with pair-preserving double excitations. A single UpCCSD layer applies these excitations once, and the ansatz increases expressivity systematically by repeating the layer $k$ times. Here $T_{1}^{\ell}$ is the single excitation operator in layer $\ell$, and $T_{pair}^{\ell}$ is the pair-preserving double excitations.
    \begin{equation}
    U_{\mathrm{kUpCCSD}}(\boldsymbol{\theta})
    =
    \prod_{\ell=0}^{k-1}\exp{(T_{1}^{\ell}(\boldsymbol{\theta})+T_{pair}^{\ell}(\boldsymbol{\theta})-h.c.)}
    \end{equation}
  Each layer contains $O(N^{2})$ parameters from generalized singles plus pair doubles, so the total parameter count scales as $O(kN^{2})$. The circuit depth and two qubit gate count increase linearly with $k$, and remain substantially lower than UCCSD, with overall scaling on the order of $O(kN^{2})$.
  
  \item \textbf{GateFabric ansatz}~\cite{Anselmetti2021}:
  GateFabric is a hardware efficient, symmetry preserving ansatz built from local two qubit blocks arranged in a brick-wall nearest-neighbor pattern and stacked in layers. Each block is particle-number conserving, ensuring that the variational search remains within the physically relevant electron-number sector. The tiled structure makes the circuit shallow and well aligned with local connectivity, while layer repetition controls expressivity.
    \begin{equation}
    U_{\mathrm{GF}}(\boldsymbol{\theta})
    =
    \prod_{\ell=0}^{L-1}\prod_{(i,j)\in\mathcal{E}}G_{ij}(\theta_{ij}^{\ell})
    \end{equation}
  where $G_{ij}$ is a particle conserving local block with a constant number of angles. There are $O(N)$ blocks per layer, each contributing a constant number of parameters, so the parameter count scales as $O(LN)$. The gate count and depth scale similarly as $O(LN)$.
  
  \item \textbf{Particle U1 ansatz}~\cite{Barkoutsos2018PRA}:
  The Particle U1 ansatz uses particle conserving exchange type two qubit gates, ensuring exact U(1) particle-number symmetry throughout the optimization. By restricting the evolution to the physically relevant particle-number sector, it often improves optimization stability while maintaining a hardware-efficient layered structure.
    \begin{equation}
    U_{\mathrm{pU1}}(\boldsymbol{\theta})
    =
    \prod_{\ell=0}^{L-1}\prod_{(i,j)\in\mathcal{E}}U_{ij}^{ex}(\theta_{ij}^{\ell}),
    \end{equation}
  with $U_{ij}^{ex}(\theta)=\exp{[ i\theta(a_{i}^{\dagger}a_{j}+a_{j}^{\dagger}a_{i}) ]}$ mixing $\ket{01}$ and $\ket{10}$ only. Each layer contains $O(N)$ nearest-neighbor exchange gates, each with one parameter, giving $O(LN)$ parameters in total. The gate count and depth are also $O(LN)$.
  
  \item \textbf{Particle U2 ansatz}~\cite{Barkoutsos2018PRA}:
  Particle U2 extends Particle U1 by replacing each exchange gate with a two-parameter particle-number-conserving block, increasing expressivity while maintaining the same U(1) symmetry sector.
    \begin{equation}
    U_{\mathrm{pU2}}(\boldsymbol{\theta})
    =
    \prod_{\ell=0}^{L-1}\prod_{(i,j)\in\mathcal{E}}U_{ij}^{ex}(\theta_{ij}^{\ell},\phi_{ij}^{\ell}),
    \end{equation}
  where $U_{ij}^{ex}$ denotes a two-parameter particle-number-conserving exchange block acting on the nearest-neighbor pair $(i,j)$. Because each nearest-neighbor pair contributes a constant number of variational parameters, the parameter count scales as $O(LN)$, with a larger constant prefactor than Particle U1. The gate count scales similarly as $O(LN)$.
\end{enumerate}

\subsubsection{Template for VQE ansatz implementation}
The implementation template provided to the LLM together with the ideas generated by the ``Generation'' component is shown in Listing~\ref{lst:BaseAnsatzCode}.
The template defines an empty class that inherits from the ansatz base class implemented in QXMT. This base class provides the interfaces and methods required for evaluation in the ``Evaluation'' component, allowing generated ans\"{a}tze to be seamlessly integrated into the experimental pipeline.
The LLM is then instructed to implement the ansatz using PennyLane, one of the quantum computing libraries, including all necessary external library imports.

\begin{lstlisting}[style=codeFormat, caption={The template for implementing generated ansatz idea}, label=lst:BaseAnsatzCode]
from typing import cast

import pennylane as qml
from qxmt.ansatze import BaseVQEAnsatz
from qxmt.devices import BaseDevice
from qxmt.hamiltonians.pennylane.molecular import MolecularHamiltonian

# above are the default imports. DO NOT REMOVE THEM.
# new imports can be added below this line if needed.


class BaseAnsatz(BaseVQEAnsatz):
    """Base ansatz class.

    Args:
        BaseVQEAnsatz (_type_): base ansatz class

    Example:
    """

    def __init__(
        self, device: BaseDevice, hamiltonian: MolecularHamiltonian
    ) -> None:
        super().__init__(device, hamiltonian)
        self.hamiltonian = cast(MolecularHamiltonian, self.hamiltonian)
        self.n_qubits = self.hamiltonian.get_n_qubits()
        self.n_params = 1  # number of parameters to be optimized. This value calculated in init method.

    def circuit(self, params: qml.numpy.ndarray) -> None:
        """Define your ansatz here.

        Args:
            params (qml.numpy.ndarray): parameters to be optimized. The shape is (n_params,).
                Note:
                    The 'params' argument is directly updated by the optimization library.
                    Do not reassign or modify 'params' itself.
                    When using the parameters, access them by their index (e.g., params[0], params[1], ...).
        """
        # define your ansatz here
        pass
\end{lstlisting}

\section{Detailed analysis of agentic system}
\label{sec:detailedAnalysisAgent}
This section presents additional analyses and experimental results that complement the main findings of this work.
We first examine the robustness of the proposed agentic system with respect to stochasticity arising from the use of LLMs by analyzing multiple independent runs.
We then provide qualitative analyses of the intermediate outputs by the ``Exploration'' and ``Discussion'' components, including the characteristics of generated research reports and the distribution of expert roles involved in multi-perspective evaluation.

Furthermore, we summarize the seed ideas that led to the best-performing quantum feature map and ansatz for VQE, providing insight into the initial hypotheses explored by the system.
Additional investigations are conducted on the properties of the generated feature maps through kernel matrix analysis.
Finally, we analyze the impact of optional circuit modules included in the best-performing ansatz, focusing on the relationship between parameter count, gate count, and estimation accuracy.

\subsection{Robustness of our agentic system}
\label{sec:robustnessModel}
\begin{figure}[ht]
    \centering
    \includegraphics[width=1.0\linewidth]{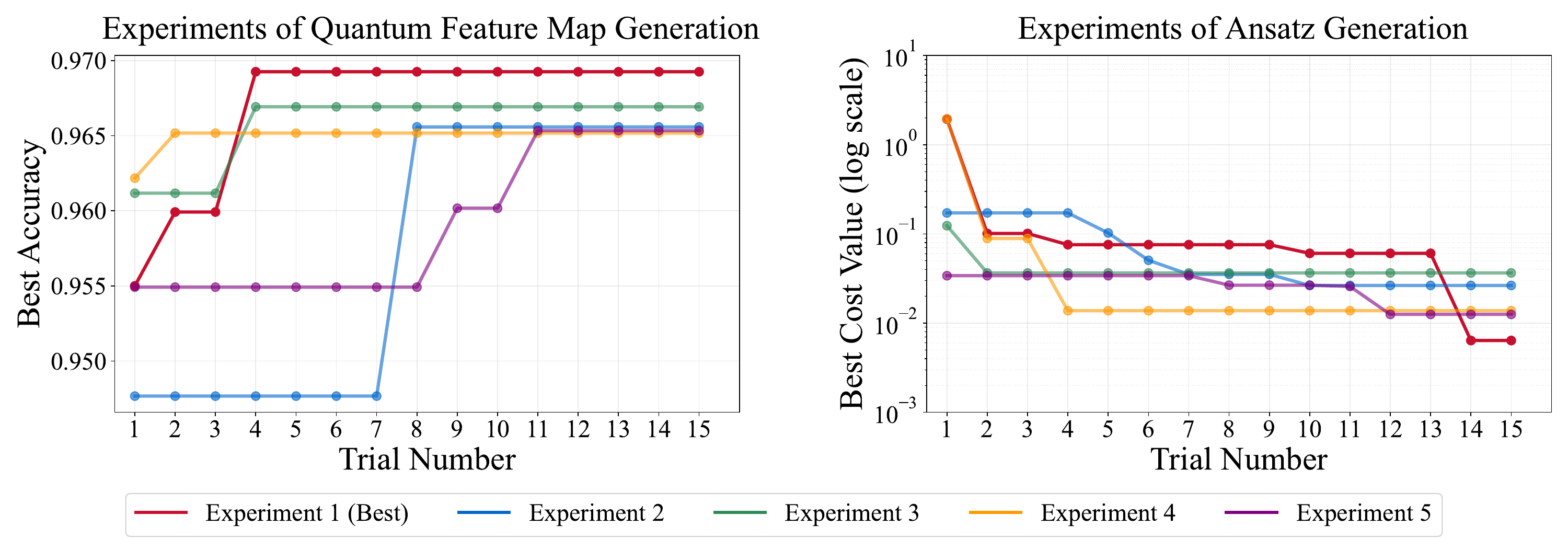}
    \caption{Results from five independent experiments. Each color represents one experiment, and the red line denotes the best solution. The x-axis indicates the trial number, and the y-axis shows the performance metric. (Left) Quantum feature map generation task, where the metric is the accuracy of the sampled test dataset (see Section \ref{sec:evalQFMSettings}). (Right) Ansatz generation task, where the metric is the cost value calculated for the target molecules (see Section \ref{sec:evalVQESettings})}
    \label{tab:all_trajectory}
\end{figure}

Since the outputs of LLMs exhibit a certain degree of randomness, we evaluated whether the developed agentic system can consistently improve performance across different tasks.
To this end, we conducted five independent experiments, including the one that generated the best solution reported in the main text and in Sections \ref{sec:qkBestSolutions} and \ref{sec:vqeBestSolutions}.
All experiments were performed under the same conditions described previously.
In what follows, we present the results for two representative tasks.
For clarity, the exploration phase, which aims to search for diverse idea directions, is omitted from the figures.

\vspace{1em}

First, in the quantum feature map generation task for image classification, the results of five independent experiments are shown in the left side of Fig.~\ref{tab:all_trajectory}.
Although the first trial already achieved an accuracy exceeding 90\%, all experiments demonstrated steady improvements across successive trials.
These results confirm that, for this task, the agentic system is capable of consistently enhancing accuracy from diverse initial ideas.

\vspace{1em}

Next, in the ansatz generation task for ground state estimation by VQE, five independent experiments are also shown in the right side of Fig.~\ref{tab:all_trajectory}.
In the first trial, the cost function values differed by approximately two orders of magnitude.
Nevertheless, all experiments exhibited multiple instances of improvement across successive trials, ultimately reducing the cost function to below $5 \times 10^{-2}$. Notably, the experiment shown in red, which produced the best final solution, had the second highest cost function value in its first trial among all five experiments.
This suggests that the relationship between the quality of the initial idea and the final outcome is not straightforward.
This behavior suggests that a detailed examination of the refinement process is required, but it is plausibly attributed to the newly introduced ``Discussion'' component, which enables critical and even negative assessment of generated ideas during iterative refinement.
At the same time, the initial idea plays a critical role in guiding the overall search trajectory, suggesting that further large-scale experiments are required to better understand this relationship.
Overall, these results demonstrate that the agentic system can consistently improve performance in the VQE task as well, even from diverse starting points.

\subsection{Classification of generated research reports}
\label{sec:researchReport}
\begin{figure}[ht]
    \centering
    \includegraphics[width=1.0\linewidth]{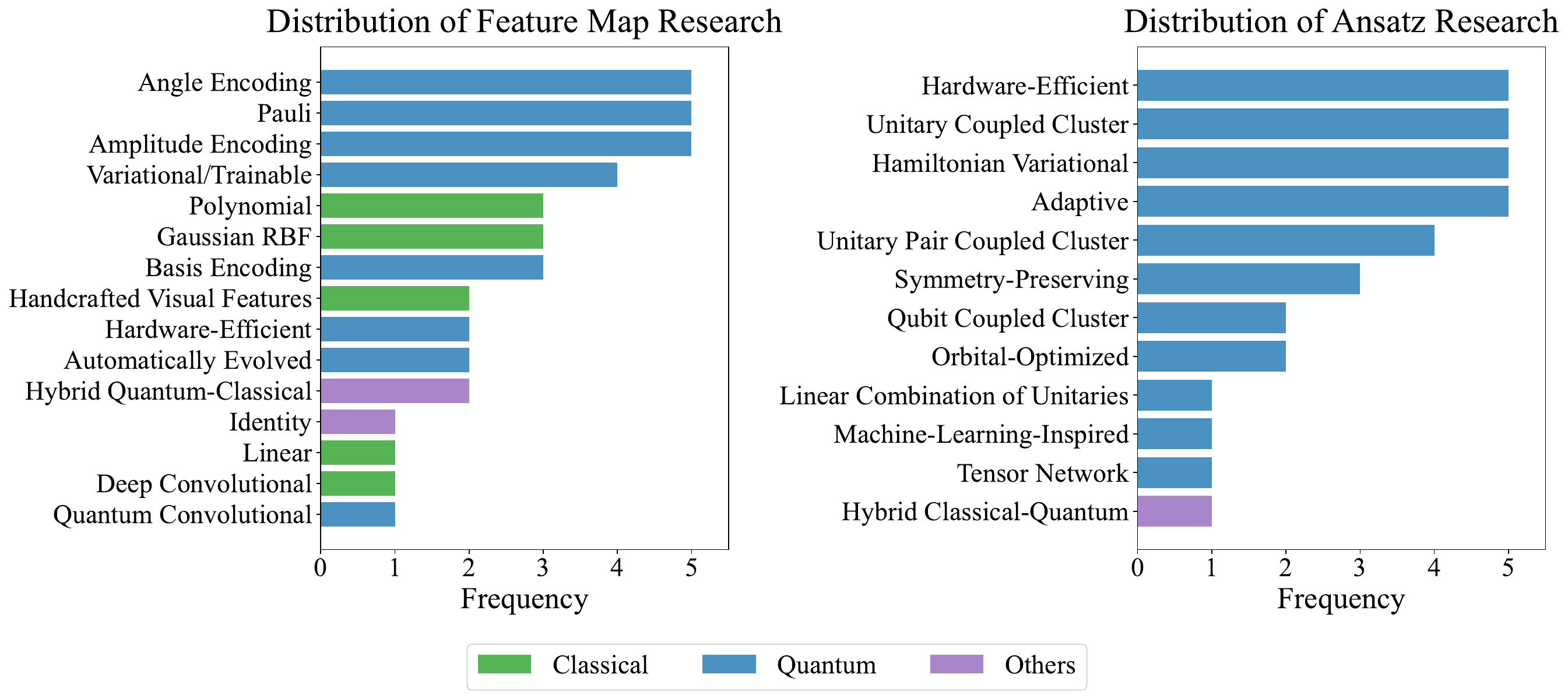}
    \caption{Distribution of previously reported methods investigated in the ``Exploration'' component. Each graph shows the frequency of these methods, categorized by type (Classical, Quantum, Others), in the tasks of quantum feature map generation (left) and ansatz generation for VQE (right).}
    \label{fig:report_distribution}
\end{figure}

We analyzed the research reports generated for the five independent experiments described in Appendix \ref{sec:robustnessModel}, conducted for both the quantum feature map generation task and the ansatz generation task for VQE.
First, we confirmed that all reports followed the structure specified in prompt: General Background (Background, Current State, Challenges, Insights\&Recommendations), Method Comparison (Name, Core Structure, Pros\&Cons, Suitability), and Exploratory Investigation.

\vspace{1em}

Next, we closely examined how each report summarized and compared existing methods as instructed in the prompt.
Fig.~\ref{fig:report_distribution} shows the distribution of the frequency with existing methods were mentioned in the reports for each task.
The vertical axis lists the method names, horizontal axis represents their frequency. Because five independent experiments were conducted for each task, the horizontal axis ranges from 0 to 5.
The analysis of prior research was not limited to quantum approaches; instead, it covered a broad range of related methods, with the color of each bar indicating the application domain.
From the graphs, we observed that prominent methods appeared in nearly all reports for both tasks.
In addition, methods that appeared only once or twice were often specialized approaches that were adapted to the construction of the feature maps or ans\"{a}tze.
Established methods are expected to be identified regardless of the depth of the search, and their repeated appearance across multiple reports supports the validity of these results.

\vspace{1em}

In the ``Exploratory Investigation'' section of the reports, the first part typically summarized the historical development of each task, followed by a review of current research trends and potential future directions.
The summaries of the early historical aspects of the tasks showed little variation among reports, suggesting that these portions can be commonly reused.
Furthermore, the sections on research trends and future work do not appear to require high-frequency updates in each execution.

\vspace{1em}

Overall, the ``Exploration'' component was able to conduct a comprehensive literature review in accordance with the prompt instructions.
As indicated by the ablation study results in Section~\ref{sec:ablationStudy}, incorporating these research reports contributed positively to task performance improvement. 
However, further optimization is required to improve both computational and economic efficiency.
The deep-research model used to generate the reports provides detailed analyses but requires long execution times and incurs higher API costs than other conversational models (approximately \$4 and 30 minutes per execution in this task).
On the other hand, we confirmed that the core background information, representative methods, and historical context of each task remained consistent across executions. 
Therefore, a more efficient design would be to provide a default research report as a base resource, allowing users to selectively perform additional investigations only when necessary.

\subsection{Patterns in expert roles}
\label{sec:expertRole}
\begin{figure}[ht]
    \centering
    \includegraphics[width=1.0\linewidth]{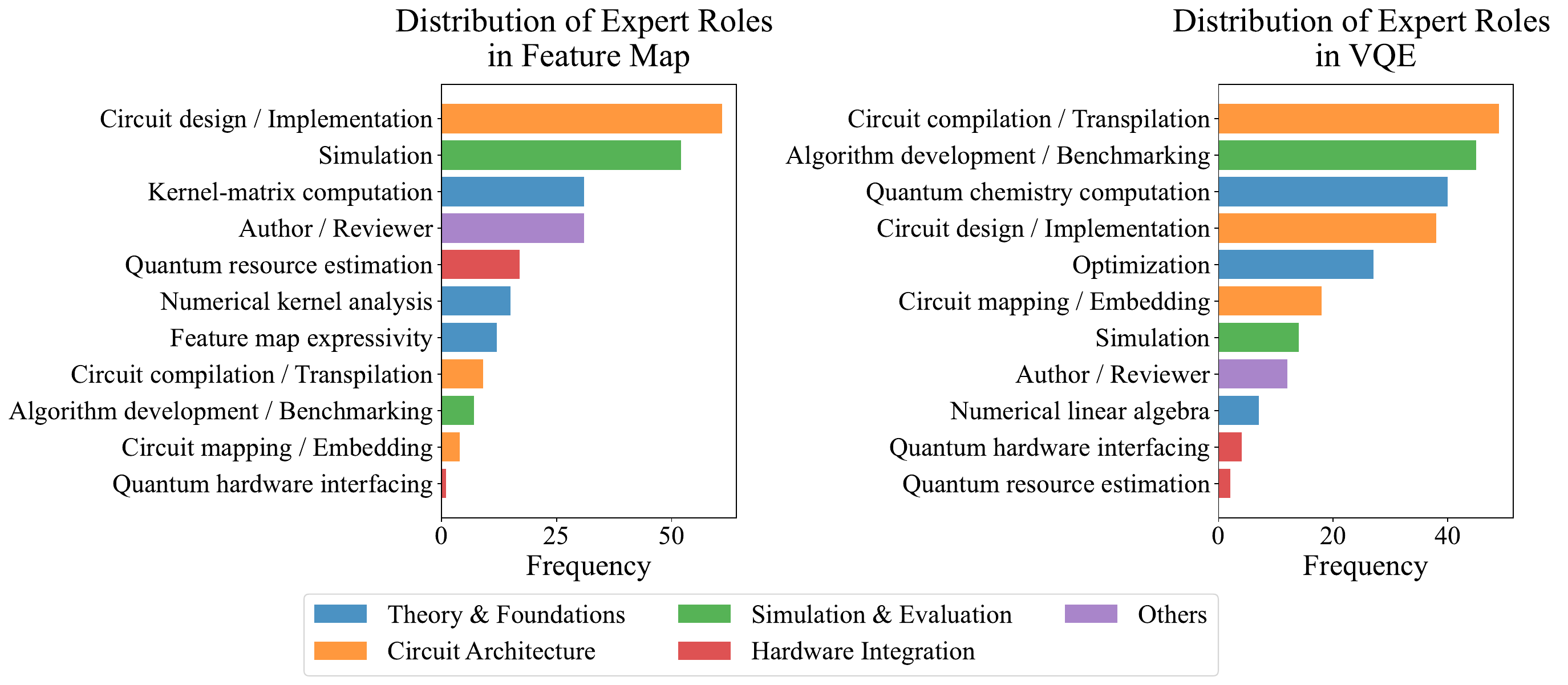}
    \caption{Distribution of expert roles that generated in ``Discussion'' component. Each graph shows the frequency of generated roles, categorized by their assigned roles, in the tasks of quantum feature map generation (left) and ansatz generation for VQE (right).}
    \label{fig:expert_distribution}
\end{figure}

We examined the distribution of expert roles that were assigned as the targets of questions in the ``Critic'' step of ``Discussion'' component (see Section \ref{sec:methodsExpert} of the main text for details of expert role).
Specifically, for both the quantum feature map generation task and the ansatz generation task for VQE, we extracted the raw text of all assigned expert roles from 15 trials of the experiments that produced the best-performing circuits.
As a result, 125 expert assignments were identified in the feature map generation task and 112 in the ansatz generation task.

\vspace{1em}

Each expert role was manually categorized, and the resulting distributions are shown in Fig.~\ref{fig:expert_distribution}.
Because multiple categories could be assigned to a single expert role, the total counts in the graphs do not necessarily match the total number of extracted expert role texts.
For each task, we manually defined eleven role categories based on the content of the experts.
In addition, we established five higher-level categories (Theory \& Foundations, Simulation \& Evaluation, Circuit Architecture, Hardware Integration, and Others), which are represented by colors in the graphs.
The assignment of categories to expert roles was performed manually.

\vspace{1em}

We analyzed the results shown on the left side of Fig.~\ref{fig:expert_distribution}, which correspond to the quantum feature map generation task.
Expert assignments related to quantum circuit implementation and simulation were dominant, while experts assigned to authors of specific methods or those estimating quantum resource requirements such as gate counts were also frequently observed.
In contrast, the right side of Fig.~\ref{fig:expert_distribution}, which corresponds to the ansatz generation task, showed a higher frequency of expert assignments related to circuit architecture.
In addition, in the ansatz generation task, the number of cases in which authors of specific methods were assigned decreased, whereas experts with broader knowledge of general benchmarking were assigned more frequently.

\vspace{1em}

However, along with expert assignments, there were also cases in which additional instructions were issued to perform simplified simulations, resource estimations, or program generation.
As a future work, further prompt tuning is required to ensure that only appropriate expert assignments are made accurately.

\section{Generation of quantum feature maps for QML}
\label{sec:appendixQunatumFM}

\subsection{Seed quantum feature maps}
\label{sec:featureMapSeed}
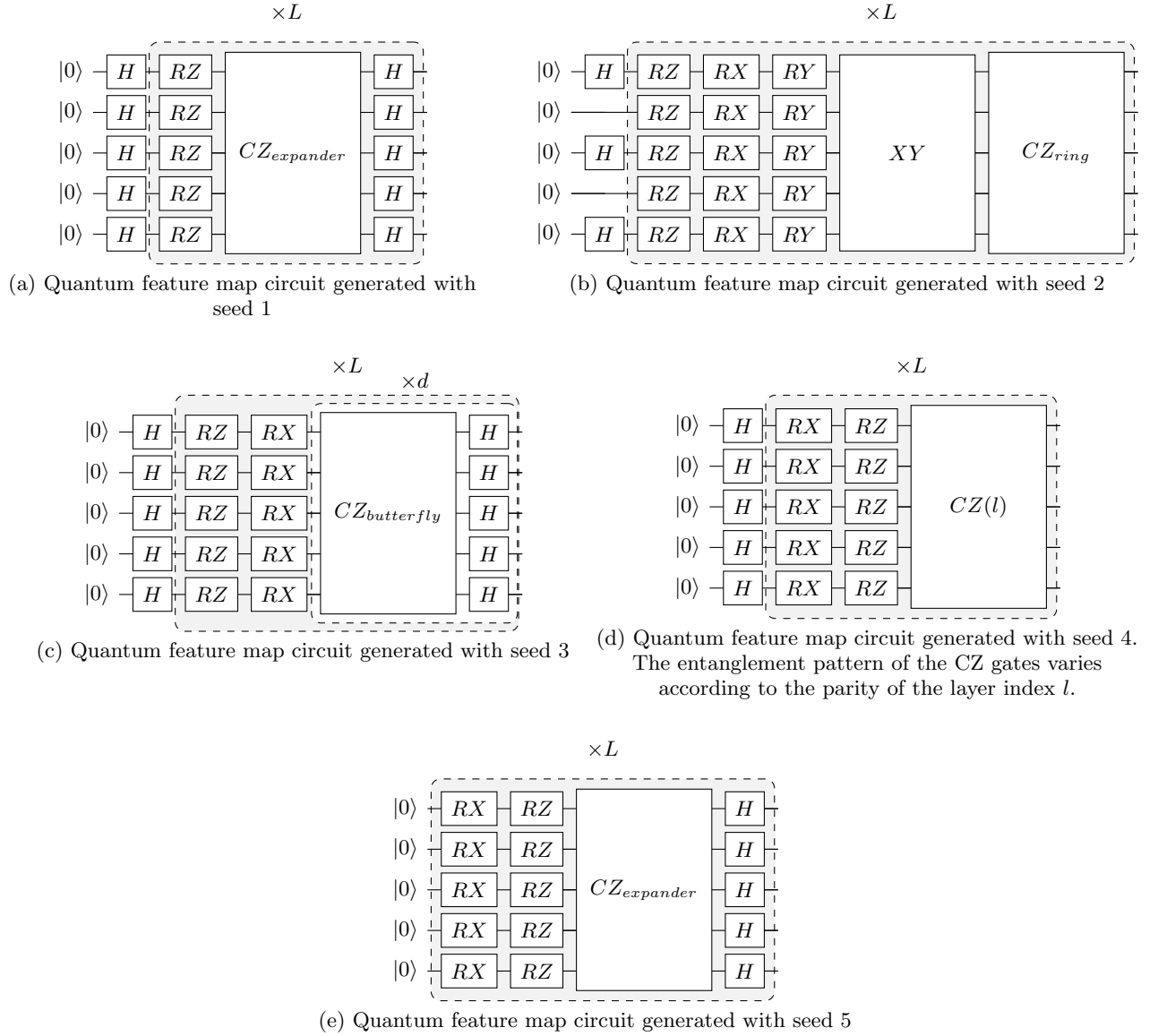
\begin{figure}[htbp]
\centering

\begin{minipage}[t]{0.4\textwidth}
    \centering
    \vspace{0pt}
    \begin{quantikz}[row sep={0.6cm,between origins}, column sep=0.2cm, thin lines]
        \lstick{$\ket{0}$} & \gate{H} & \gate{RZ} \gategroup[wires=5,steps=3,
            style={dashed,rounded corners,fill=gray!10,
                   inner xsep=1pt,inner ysep=1pt},
            background,
            label style={label position=above, yshift=0.2cm}
        ]{$\times L$} & \gate[wires=5][2.0cm]{CZ_{expander}} & \gate{H} & \qw \\
        \lstick{$\ket{0}$} & \gate{H} & \gate{RZ} & \qw & \gate{H} & \qw \\
        \lstick{$\ket{0}$} & \gate{H} & \gate{RZ} & \qw & \gate{H} & \qw \\
        \lstick{$\ket{0}$} & \gate{H} & \gate{RZ} & \qw & \gate{H} & \qw \\
        \lstick{$\ket{0}$} & \gate{H} & \gate{RZ} & \qw & \gate{H} & \qw 
    \end{quantikz}
    \par\smallskip
    {\small (a) Quantum feature map circuit generated with seed 1}
\end{minipage}%
\hspace{0.015\textwidth}%
\begin{minipage}[t]{0.55\textwidth}
    \centering
    \vspace{0pt}
    \begin{quantikz}[row sep={0.6cm,between origins}, column sep=0.2cm, thin lines]
        \lstick{$\ket{0}$} & \gate{H} & \gate{RZ} \gategroup[wires=5,steps=5,
            style={dashed,rounded corners,fill=gray!10,
                   inner xsep=1pt,inner ysep=1pt},
            background,
            label style={label position=above, yshift=0.2cm}
        ]{$\times L$}  & \gate{RX} & \gate{RY} & \gate[wires=5][2.0cm]{XY} & \gate[wires=5][2.0cm]{CZ_{ring}} & \qw \\
        \lstick{$\ket{0}$} & \qw & \gate{RZ} & \gate{RX} & \gate{RY} & \qw & \qw & \qw \\
        \lstick{$\ket{0}$} & \gate{H} & \gate{RZ} & \gate{RX} & \gate{RY} & \qw & \qw & \qw \\
        \lstick{$\ket{0}$} & \qw & \gate{RZ} & \gate{RX} & \gate{RY} & \qw & \qw & \qw \\
        \lstick{$\ket{0}$} & \gate{H} & \gate{RZ} & \gate{RX} & \gate{RY} & \qw & \qw & \qw
    \end{quantikz}
    \par\smallskip
    {\small (b) Quantum feature map circuit generated with seed 2}
\end{minipage}

\vspace{1.5em}

\begin{minipage}[t]{0.45\textwidth}
    \centering
    \vspace{0pt}
    \begin{quantikz}[row sep={0.6cm,between origins}, column sep=0.2cm, thin lines]
        \lstick{$\ket{0}$} & \gate{H} & \gate{RZ} \gategroup[wires=5,steps=4,
            style={dashed,rounded corners,fill=gray!10,
                   inner xsep=1pt,inner ysep=4pt},
            background,
            label style={label position=above, yshift=0.2cm}
        ]{$\times L$} & \gate{RX} & \gate[wires=5][2.0cm]{CZ_{butterfly}}\gategroup[wires=5,steps=2,
            style={dashed,rounded corners,fill=white,
                   inner xsep=0.5pt,inner ysep=0.5pt},
            background,
            label style={label position=above, yshift=0.1cm}
        ]{$\times d$} & \gate{H} & \qw \\
        \lstick{$\ket{0}$} & \gate{H} & \gate{RZ} & \gate{RX} & \qw & \gate{H} & \qw \\
        \lstick{$\ket{0}$} & \gate{H} & \gate{RZ} & \gate{RX} & \qw & \gate{H} & \qw \\
        \lstick{$\ket{0}$} & \gate{H} & \gate{RZ} & \gate{RX} & \qw & \gate{H} & \qw \\
        \lstick{$\ket{0}$} & \gate{H} & \gate{RZ} & \gate{RX} & \qw & \gate{H} & \qw
    \end{quantikz}
    \par\smallskip
    {\small (c) Quantum feature map circuit generated with seed 3}
\end{minipage}%
\hspace{0.015\textwidth}%
\begin{minipage}[t]{0.45\textwidth}
    \centering
    \vspace{0pt}
    \begin{quantikz}[row sep={0.6cm,between origins}, column sep=0.2cm, thin lines]
        \lstick{$\ket{0}$} & \gate{H} & \gate{RX} \gategroup[wires=5,steps=3,
            style={dashed,rounded corners,fill=gray!10,
                   inner xsep=1pt,inner ysep=1pt},
            background,
            label style={label position=above, yshift=0.2cm}
        ]{$\times L$} & \gate{RZ} & \gate[wires=5][2.0cm]{CZ(l)} & \qw \\
        \lstick{$\ket{0}$} & \gate{H} & \gate{RX} & \gate{RZ} & \qw & \qw \\
        \lstick{$\ket{0}$} & \gate{H} & \gate{RX} & \gate{RZ} & \qw & \qw \\
        \lstick{$\ket{0}$} & \gate{H} & \gate{RX} & \gate{RZ} & \qw & \qw \\
        \lstick{$\ket{0}$} & \gate{H} & \gate{RX} & \gate{RZ} & \qw & \qw 
    \end{quantikz}
    \par\smallskip
    {\small (d) Quantum feature map circuit generated with seed 4. The entanglement pattern of the CZ gates varies according to the parity of the layer index $l$.}
\end{minipage}

\vspace{1.5em}

\begin{minipage}[t]{0.95\textwidth}
    \centering
    \vspace{0pt}
    \begin{quantikz}[row sep={0.6cm,between origins}, column sep=0.2cm, thin lines]
        \lstick{$\ket{0}$} & \gate{RX} \gategroup[wires=5,steps=4,
            style={dashed,rounded corners,fill=gray!10,
                   inner xsep=1pt,inner ysep=1pt},
            background,
            label style={label position=above, yshift=0.2cm}
        ]{$\times L$} & \gate{RZ} & \gate[wires=5][2.0cm]{CZ_{expander}} & \gate{H} & \qw \\
        \lstick{$\ket{0}$} & \gate{RX} & \gate{RZ} & \qw & \gate{H} & \qw \\
        \lstick{$\ket{0}$} & \gate{RX} & \gate{RZ} & \qw & \gate{H} & \qw \\
        \lstick{$\ket{0}$} & \gate{RX} & \gate{RZ} & \qw & \gate{H} & \qw \\
        \lstick{$\ket{0}$} & \gate{RX} & \gate{RZ} & \qw & \gate{H} & \qw 
    \end{quantikz}
    \par\smallskip
    {\small (e) Quantum feature map circuit generated with seed 5}
\end{minipage}

\caption{Circuit diagrams of the five seed quantum feature maps.}
\label{fig:seedFeatureMap}
\end{figure}

For the experiment in which the best-performing result was obtained, we describe the details of the five seed quantum circuits generated in the ``Exploration'' component.
Fig.~\ref{fig:seedFeatureMap} shows the circuit diagrams of the five seed quantum feature maps using a five-qubits system as an illustrative example.
Note that the structure of these feature maps does not depend on the number of qubits.

The source code of the five generated seed feature maps is available in the artifact repository (\url{https://github.com/Qyusu/astronaut-artifact}) under
\texttt{version\_2\_0/generated\_code/best\_quantum\_kernel\_results}.

\vspace{1em}

The seed 1 feature map, shown in Fig.~\ref{fig:seedFeatureMap} (a).
It begins by applying Hadamard gates to all qubits, followed by $L$ repeated blocks consisting of data dependent \(R_Z\) gates and entangling operations.
In each layer, the PCA features are reuploaded, and linear \(R_Z\) encodings scaled by layer specific coefficients are applied.
This is followed by a sparse \(CZ\) network determined by a set of precomputed step sizes, after which Hadamard gates are again applied to all qubits.
The step sized used in each layer are selected so as to avoid repeatedly coupling only specific pairs of qubits.


\vspace{1em}

The seed 2 feature map, shown in Fig.~\ref{fig:seedFeatureMap} (b).
It begins by applying Hadamard gates only to qubits with even indices.
In each layer, linear data encoding is performed via single qubit rotations about three axes, \(R_Z\), \(R_X\), and \(R_Y\), followed by entangling operations implemented with iSWAP family two qubit gates.
The entangling block is realized using a two round brickwork pattern.
In the first round, nearest-neighbor even odd pairs, for example 0–1 and 2–3, are coupled in parallel.
In the second round, the pairing is shifted by one site, coupling pairs such as 1–2 and 3–4, as well as the boundary pair.
This is followed by the application of \(CZ\) gates between qubits separated by distance two.
At the end of each layer, a permutation operation is inserted that cyclically shifts the qubit indices at the program level by a fixed amount.
As a result, different pairs of qubits become adjacent in the subsequent layer.


\vspace{1em}

The seed 3 feature map, shown in Fig.~\ref{fig:seedFeatureMap} (c).
It begins by applying a Haar wavelet transformation like linear preprocessing step to the input vector~\cite{Haar1910, Viola2001}.
This operation decomposes the original features into multi scale average and difference components, yielding a hierarchical representation prior to quantum encoding.
The circuit then applies Hadamard gates to all qubits.
In each layer, the transformed features are encoded via linear angle rotations using \(R_Z\) and \(R_X\) gates, with layer dependent coefficients that decrease with scale.
Entanglement is introduced through a hierarchical radix 2 pairing pattern: at each substage, disjoint pairs of qubits within blocks of a given size are coupled by \(CZ\) gates, and the block size doubles at the next substage, followed by Hadamard gates.
Finally, between consecutive layers, a fixed cyclic permutation of the qubit indices is applied at the program level.


\vspace{1em}

The seed 4 feature map, shown in Fig.~\ref{fig:seedFeatureMap} (d).
It begins by applying a fixed linear variance equalization to the input vector, $\tilde{x} = (x - \mu) / \sigma$, using predefined constant values $\mu$ and $\sigma$ for all 80 features.
The circuit then applies Hadamard gates to all qubits.
In each layer, the normalized features are encoded via linear angle rotations using \(R_X\) followed by \(R_Z\) gates.
Entanglement alternates between two patterns depending on the parity of the layer index: even indexed layers implement a ring pattern that applies \(CZ\) gates between nearest-neighbors on a circular chain, whereas odd indexed layers implement a star pattern that applies \(CZ\) gates between a single hub qubit and all other qubits.
Finally, between consecutive layers, a fixed cyclic permutation of the qubit indices is applied at the program level.


\vspace{1em}

The seed 5 feature map, shown in Fig.~\ref{fig:seedFeatureMap} (e).
It is built from $L$ sequential upload blocks that alternate data encoding, permutation, and sparse entanglement.
In each layer, the assigned input features are uploaded using linear angle rotations implemented by \(R_X\) followed by \(R_Z\) gates on each qubit.
After the upload, a fixed cyclic permutation of the qubit indices is applied at the program level.
This relabeling step changes which qubits become neighbors in the subsequent entangling stage.
Entanglement is then introduced by a sparse \(CZ\) network defined through a small set of precomputed step sizes on a ring, resulting in a low number of two qubit couplings per layer.
Finally, Hadamard gates are applied to all qubits.


\subsection{Best-performing quantum feature map}
\label{sec:qkBestSolutions}
The trajectory of the accuracy obtained by applying our agentic system to the quantum feature map task is shown in top of Fig.~\ref{fig:performance_trajectory}.
At trial 4, the system achieved best performance on the test dataset.
The generated source code and logs are shown in Listing ~\ref{lst:BestFeatureMapCode}, ~\ref{lst:InitFeatureMapIdea} and ~\ref{lst:ReflectedFeatureMapIdea}.
We note that this quantum feature map was generated in a fully automated manner by an LLM.
In the following, we first describe its structural characteristics.
However, understanding why the resulting circuit achieves strong empirical performance on the present task requires a separate theoretical analysis.

\begin{figure}[ht]
\centering

\begin{quantikz}[row sep = {0.6cm,between origins}, column sep=0.8cm, thin lines]
    \lstick{$\ket{0}$} & \gate[wires=10][2.0cm]{U_{passA}} \gategroup[wires=10,steps=1,
            style={dashed,rounded corners,fill=gray!10,
                   inner xsep=1pt,inner ysep=2pt},
            background,
            label style={label position=above, yshift=0.2cm}
        ]{$\times L_p$} & \gate[wires=10][2.0cm]{U_{passB}} \gategroup[wires=10,steps=1,
            style={dashed,rounded corners,fill=gray!10,
                   inner xsep=1pt,inner ysep=2pt},
            background,
            label style={label position=above, yshift=0.2cm}
        ]{$\times L_p$} & \qw \\
    \lstick{$\ket{0}$} & \qw & \qw & \qw \\
    \lstick{$\ket{0}$} & \qw & \qw & \qw \\
    \lstick{$\ket{0}$} & \qw & \qw & \qw \\
    \lstick{$\ket{0}$} & \qw & \qw & \qw \\
    \lstick{$\ket{0}$} & \qw & \qw & \qw \\
    \lstick{$\ket{0}$} & \qw & \qw & \qw \\
    \lstick{$\ket{0}$} & \qw & \qw & \qw \\
    \lstick{$\ket{0}$} & \qw & \qw & \qw \\
    \lstick{$\ket{0}$} & \qw & \qw & \qw
\end{quantikz}
\caption{Overview of the best feature-map quantum circuit.}
\label{fig:fmCircuitOverall}
\end{figure}
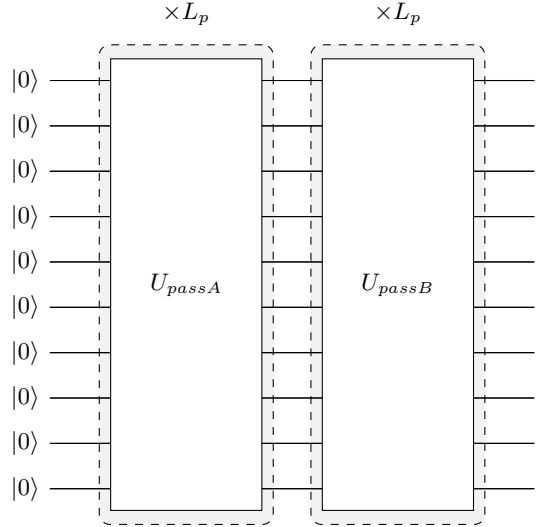

\begin{figure}[ht]
\centering

\adjustbox{scale=0.9}{
\begin{minipage}{0.34\textwidth}
\centering
\begin{quantikz}[row sep = {0.8cm,between origins}, column sep=0.35cm, transform shape, thin lines]
    \centering
    \lstick{$\ket{0}$} & \gate{H} & \swap{5} & \qw & \qw & \qw & \qw & \gate{RY} & \gate{RZ} & \qw \\
    \lstick{$\ket{0}$} & \gate{H} & \qw & \swap{5} & \qw & \qw & \qw & \gate{RY} & \gate{RZ} & \qw \\
    \lstick{$\ket{0}$} & \gate{H} & \qw & \qw & \swap{5} & \qw & \qw & \gate{RY} & \gate{RZ} & \qw \\
    \lstick{$\ket{0}$} & \gate{H} & \qw & \qw & \qw & \swap{5} & \qw & \gate{RY} & \gate{RZ} & \qw \\
    \lstick{$\ket{0}$} & \gate{H} & \qw & \qw & \qw & \qw & \swap{5} & \gate{RY} & \gate{RZ} & \qw \\
    \lstick{$\ket{0}$} & \gate{H} & \targX{} & \qw & \qw & \qw & \qw & \gate{RY} & \gate{RZ} & \qw \\
    \lstick{$\ket{0}$} & \gate{H} & \qw & \targX{} & \qw & \qw & \qw & \gate{RY} & \gate{RZ} & \qw \\
    \lstick{$\ket{0}$} & \gate{H} & \qw & \qw & \targX{} & \qw & \qw & \gate{RY} & \gate{RZ} & \qw \\
    \lstick{$\ket{0}$} & \gate{H} & \qw & \qw & \qw & \targX{} & \qw & \gate{RY} & \gate{RZ} & \qw \\
    \lstick{$\ket{0}$} & \gate{H} & \qw & \qw & \qw & \qw & \targX{} & \gate{RY} & \gate{RZ} & \qw
\end{quantikz}
\par\smallskip
{\small (a) Pass B ($l \bmod 2 = 1$)}
\end{minipage}
\hfill
\begin{minipage}{0.65\textwidth}
\centering
\begin{quantikz}[row sep = {0.8cm,between origins}, column sep=0.2cm, transform shape, thin lines]
    \centering
    \lstick{$\ket{0}$} & \gate{H} & \ctrl{5} & \qw & \qw & \qw & \qw & \gate{RZ} & \gate{RY} & \qw \\
    \lstick{$\ket{0}$} & \gate{H} & \qw & \ctrl{5} & \qw & \qw & \qw & \gate{RZ} & \gate{RY} & \qw \\
    \lstick{$\ket{0}$} & \gate{H} & \qw & \qw & \ctrl{5} & \qw & \qw & \gate{RZ} & \gate{RY} & \qw \\
    \lstick{$\ket{0}$} & \gate{H} & \qw & \qw & \qw & \ctrl{5} & \qw & \gate{RZ} & \gate{RY} & \qw \\
    \lstick{$\ket{0}$} & \gate{H} & \qw & \qw & \qw & \qw & \ctrl{5} & \gate{RZ} & \gate{RY} & \qw \\
    \lstick{$\ket{0}$} & \gate{H} & \gate{ZZ_{\pi/2}} & \qw & \qw & \qw & \qw & \gate{RZ} & \gate{RY} & \qw \\
    \lstick{$\ket{0}$} & \gate{H} & \qw & \gate{ZZ_{\pi/2}} & \qw & \qw & \qw & \gate{RZ} & \gate{RY} & \qw \\
    \lstick{$\ket{0}$} & \gate{H} & \qw & \qw & \gate{ZZ_{\pi/2}} & \qw & \qw & \gate{RZ} & \gate{RY} & \qw \\
    \lstick{$\ket{0}$} & \gate{H} & \qw & \qw & \qw & \gate{ZZ_{\pi/2}} & \qw & \gate{RZ} & \gate{RY} & \qw \\
    \lstick{$\ket{0}$} & \gate{H} & \qw & \qw & \qw & \qw & \gate{ZZ_{\pi/2}} & \gate{RZ} & \gate{RY} & \qw
\end{quantikz}
\par\smallskip
{\small (b) Pass B ($l \bmod 2 = 0$)}
\end{minipage}
}
\caption{Detailed circuit of Pass B. (a) The index of layer is odd. (b) The index of layer is even.}
\label{fig:fmCircuitPassB}
\end{figure}
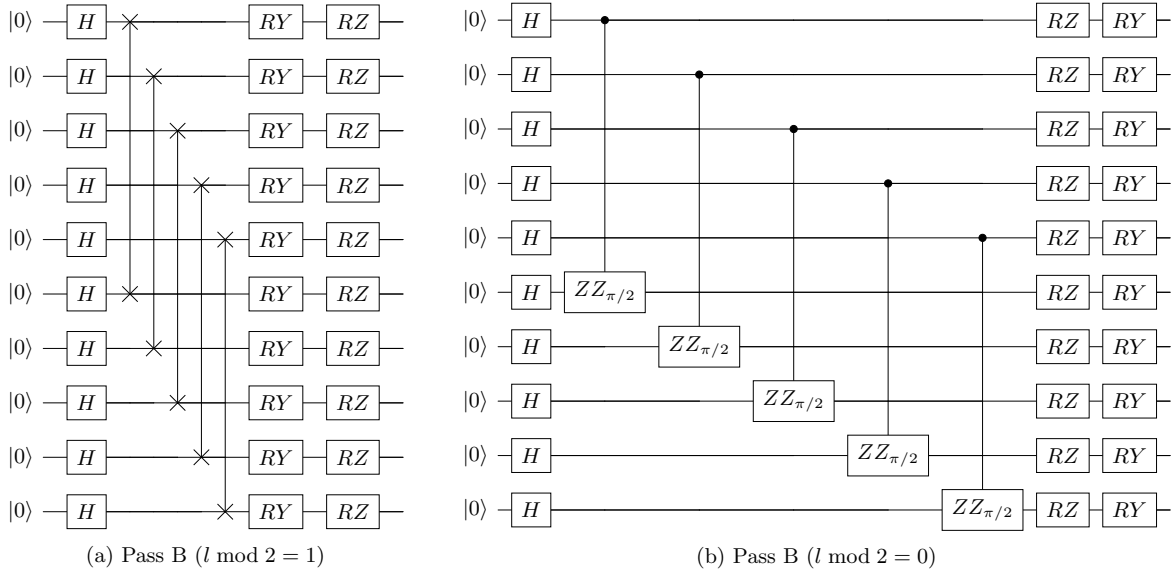

The corresponding quantum circuit diagram is shown in Fig.~\ref{fig:fmCircuitOverall}, ~\ref{fig:fmCircuitPassA} and ~\ref{fig:fmCircuitPassB}. 
The Fig.~\ref{fig:fmCircuitOverall} illustrates the overall structure of the quantum circuit.
In this example, the diagram is drawn assuming a 10 qubit system.
However, the quantum feature map generated by the system does not depend on the number of qubits.
The circuit consists of two main components, $U_{\mathrm{passA}}$ and $U_{\mathrm{passB}}$, each with a repetitive structure.
The number of repetitions, $L_{p}$, is determined by the input data dimension $F$ and the number of qubits $Q$ in the device, and is given by
\begin{equation}
    L_p = \lceil F / \max(1, Q) \rceil .
\end{equation}
By employing data re-uploading~\cite{perez2020data} in this manner, the input data can be embedded into the quantum circuit without any loss of information.
Both passes employ circulant interaction graphs defined by step sets \( \mathcal{S}_A \) and \( \mathcal{S}_B \). 
The step sets are chosen using a score based on the separation between the largest eigenvalue and the remaining eigenvalues of the associated adjacency matrix. 

\begin{figure}[ht]
\centering
\adjustbox{scale=0.9}{
    \begin{quantikz}[row sep = {0.8cm,between origins}, column sep=0.1cm, thin lines]
        \lstick{$\ket{0}$} & \gate{H} & \ctrl{5} & \qw & \qw & \qw & \qw & \gate[style={draw=green}]{H} & \ctrl[style={draw=green}]{5} & \gate[style={draw=green}]{H} & \qw & \qw & \qw & \qw & \qw & \qw & \qw & \qw & \qw & \qw & \gate{RX} & \gate{RZ} \qw \\
        \lstick{$\ket{0}$} & \gate{H} & \qw & \ctrl{5} & \qw & \qw & \qw & \qw & \qw & \qw & \qw & \ctrl[style={draw=blue}]{5} & \qw & \qw & \qw & \qw & \qw & \qw & \qw & \qw & \gate{RX} & \gate{RZ} & \qw \\
        \lstick{$\ket{0}$} & \gate{H} & \qw & \qw & \ctrl{5} & \qw & \qw & \qw & \qw & \qw & \qw & \qw & \qw & \ctrl[style={draw=red}]{5} & \qw & \qw & \qw & \qw & \qw & \qw & \gate{RX} & \gate{RZ} & \qw \\
        \lstick{$\ket{0}$} & \gate{H} & \qw & \qw & \qw & \ctrl{5} & \qw & \qw & \qw & \qw & \qw & \qw & \qw & \qw & \gate[style={draw=green}]{H} & \ctrl[style={draw=green}]{5} & \gate[style={draw=green}]{H} & \qw & \qw & \qw & \gate{RX} & \gate{RZ} & \qw \\
        \lstick{$\ket{0}$} & \gate{H} & \qw & \qw & \qw & \qw & \ctrl{5} & \qw & \qw & \qw & \qw & \qw & \qw & \qw & \qw & \qw & \qw & \qw & \ctrl[style={draw=blue}]{5} & \qw & \gate{RX} & \gate{RZ} & \qw \\
        \lstick{$\ket{0}$} & \gate{H} & \gate{Z} & \qw & \qw & \qw & \qw & \qw & \gate[style={draw=green}]{ZZ_{\theta_g}} & \qw & \qw & \qw & \qw & \qw & \qw & \qw & \qw & \qw & \qw & \qw & \gate{RX} & \gate{RZ} & \qw \\
        \lstick{$\ket{0}$} & \gate{H} & \qw & \gate{Z} & \qw & \qw & \qw & \qw & \qw & \qw & \gate[style={draw=blue}]{H} & \gate[style={draw=blue}]{ZZ_{\theta_b}} & \gate[style={draw=blue}]{H} & \qw & \qw & \qw & \qw & \qw & \qw & \qw & \gate{RX} & \gate{RZ} & \qw \\
        \lstick{$\ket{0}$} & \gate{H} & \qw & \qw & \gate{Z} & \qw & \qw & \qw & \qw & \qw & \qw & \qw & \qw & \gate[style={draw=red}]{ZZ_{\theta_r}} & \qw & \qw & \qw & \qw & \qw & \qw & \gate{RX} & \gate{RZ} \\
        \lstick{$\ket{0}$} & \gate{H} & \qw & \qw & \qw & \gate{Z} & \qw & \qw & \qw & \qw & \qw & \qw & \qw & \qw & \qw & \gate[style={draw=green}]{ZZ_{\theta_g}} & \qw & \qw & \qw & \qw & \gate{RX} & \gate{RZ} & \qw \\
        \lstick{$\ket{0}$} & \gate{H} & \qw & \qw & \qw & \qw & \gate{Z} & \qw & \qw & \qw & \qw & \qw & \qw & \qw & \qw & \qw & \qw & \gate[style={draw=blue}]{H} & \gate[style={draw=blue}]{ZZ_{\theta_b}} & \gate[style={draw=blue}]{H} & \gate{RX} & \gate{RZ}& \qw 
    \end{quantikz}
}
\caption{Detailed circuit of Pass A}
\label{fig:fmCircuitPassA}
\end{figure}

The Fig.~\ref{fig:fmCircuitPassA} shows the detailed structure of $U_{\mathrm{passA}}$.
Pass~A constitutes the first stage of the feature map and is primarily responsible for embedding pairwise feature correlations through structured two-qubit interactions.
Each layer in Pass~A follows the general sequence
\[
H^{\otimes Q} \rightarrow M_\ell^{(A)} \rightarrow \mathcal{Z}^{(3)}_\ell \rightarrow E_\ell^{(A)}(x),
\]
where \( \ell \) indexes the layer within Pass~A.
At the beginning of each layer, a Hadamard operation \( H^{\otimes Q} \) creates a uniform superposition.
The mixer layer \( M_\ell^{(A)} \) is implemented as a product of controlled phase operations applied over all edges,
\[
M_\ell^{(A)} = \prod_{(u,v)\in \mathcal{E}_A(\ell)} \mathrm{CZ}(u,v),
\]
introducing controlled phase entanglement across the qubits connected by the edges.
Here, the edge set \( \mathcal{E}_A(\ell) \) is defined as the circulant edge set generated from the step set \( \mathcal{S}_A \), cyclically shifted by the layer index \( \ell \).

A distinctive component of Pass~A is the \( \mathcal{Z}^{(3)}_\ell \) module.
Each edge \( (u,v) \) is assigned one of three classes \( c \in \{0,1,2\} \) using a deterministic mapping based on qubit indices and the current layer index.
For each class, the injected two-qubit interaction differs slightly: (i) \(c=0\) uses the standard \( \mathrm{IsingZZ}(\theta) \) coupling (red color in Fig.~\ref{fig:fmCircuitPassA}), (ii) \(c=1\) applies an \(H\)-conjugated coupling on the target qubit, thereby realizing a ZX-type interaction (blue color in Fig.~\ref{fig:fmCircuitPassA}), and (iii) \(c=2\) applies an \(H\)-conjugated coupling on the control qubit, thereby realizing a XZ-type interaction (green color in Fig.~\ref{fig:fmCircuitPassA}).
The rotation angle \( \theta \) applied to each pair is computed as
\[
\theta_{uv,\ell} = \gamma_A f_{uv}(x) + \sigma_{uv,\ell},
\]
where \( f_{uv}(x) \) combines two input features \( x_u \) and \( x_v \) according to a class dependent rule (sum or difference), and \(\sigma_{uv,\ell} = \pm \delta_0 (-1)^\ell\) introduces a small deterministic bias alternating across layers.
Here, \( \gamma_A \) and \( \delta_0 \) are prededined constant hyperparameters specified prior to circuit execution.

\vspace{1em}

After these entangling interactions, each qubit is subjected to two types of single rotation gates,
\[
E_\ell^{(A)}(x) =
\begin{cases}
R_X(\alpha_A x_j) R_Z(\beta_A r_{j,\ell} x_j), & \text{if } \ell \text{ is odd}, \\[4pt]
R_Z(\beta_A r_{j,\ell} x_j) R_X(\alpha_A x_j), & \text{if } \ell \text{ is even},
\end{cases}
\]
where \( r_{j,\ell} = 1 + \epsilon\, s_{j,\ell} \) is a deterministic scaling factor with \( s_{j,\ell} \in \{+1,0,-1\} \), and \( \epsilon \) is a fixed constant.

\vspace{1em}

The Fig.~\ref{fig:fmCircuitPassB} shows the detailed structure of $U_{\mathrm{passB}}$.
Pass~B constitutes the second stage of the feature map and complements Pass~A.
Each layer in Pass~B follows the sequence
\[
H^{\otimes Q} \rightarrow M_\ell^{(B)} \rightarrow E_\ell^{(B)}(x),
\]
In Pass~B, each layer also begins with a Hadamard operation \( H^{\otimes Q} \), just like in Pass~A.
The mixer layer \( M_\ell^{(B)} \) alternates its entangling operation depending on the layer index.
For odd numbered layers, the mixer applies \( \mathrm{iSWAP} \) gates over all edges:
\[
M_\ell^{(B)} = \prod_{(u,v)\in \mathcal{E}_B(\ell)} \mathrm{iSWAP}(u,v),
\]
which introduce complex-valued exchange interactions.
For even numbered layers, the mixer instead applies Ising-type couplings with fixed strength \( \pi/2 \),
\[
M_\ell^{(B)} = \prod_{(u,v)\in \mathcal{E}_B(\ell)} \mathrm{IsingZZ}\!\left(\tfrac{\pi}{2}\right)_{(u,v)},
\]
which restore a real-valued correlation structure complementary to the iSWAP layers.

\vspace{1em}

Following the mixer, each qubit is subjected to two types of single rotation gates analogous in structure to that of Pass~A, but operating on a different set of rotation axes:
\[
E_\ell^{(B)}(x) =
\begin{cases}
R_Y(\alpha_B x_j) R_Z(\beta_B r'_{j,\ell} x_j), & \text{if } \ell \text{ is odd}, \\[4pt]
R_Z(\beta_B r'_{j,\ell} x_j) R_Y(\alpha_B x_j), & \text{if } \ell \text{ is even},
\end{cases}
\]
where \( r'_{j,\ell} \) follows the same functional form as in Pass~A, with the class dependent sign pattern shifted relative to that of Pass~A.

\vspace{1em}

Further details of the implementation, including the configuration of hyperparameters, can be found in the source code provided in Listing~\ref{lst:BestFeatureMapCode}.

\begin{lstlisting}[style=codeFormat, caption={Best performing feature map code generated by our agentic system}, label=lst:BestFeatureMapCode]
import numpy as np
from qxmt.constants import PENNYLANE_PLATFORM
from qxmt.feature_maps import BaseFeatureMap

# above are the default imports. DO NOT REMOVE THEM.
# new imports can be added below this line if needed.
import math
from itertools import combinations
import pennylane as qml


class TriMix2Pass3ColorExpander(BaseFeatureMap):
    """TriMix-2Pass-3Color-Expander feature map.

    Structure:
        - Two passes (A, B) with re-uploading to cover all features.
        - Layer block: H^⊗Q -> mixer M_ell -> 3-color ZZ injection (Pass A only) -> axis-staggered single-qubit encodings.
        - Circulant expander edge sets chosen by a small FFT-based spectral heuristic (deterministic).

    Notes:
        - No measurements are performed in this circuit; it defines a feature embedding only.
        - Works for any number of qubits; defaults tuned for Q≈10.

    Args:
        n_qubits (int): number of qubits
        deg_A (int, optional): even degree for expander graph in Pass A (via generator steps). Defaults to 4.
        deg_B (int, optional): even degree for expander graph in Pass B (via generator steps). Defaults to 4.
        gamma_A (float, optional): two-qubit encoding strength in A's 3-color ZZ injections. Defaults to pi/4.
        alpha_A (float, optional): RX encoder coefficient in Pass A. Defaults to pi/2.
        beta_A (float, optional): RZ encoder coefficient in Pass A. Defaults to pi.
        alpha_B (float, optional): RY encoder coefficient in Pass B. Defaults to pi/sqrt(2).
        beta_B (float, optional): RZ encoder coefficient in Pass B. Defaults to pi/sqrt(3).
        epsilon (float, optional): small color-doping factor for RZ encoders. Defaults to 1/16.
        delta0 (float, optional): small ZZ-bias used in A's 3-color injection. Defaults to pi/64.
        use_fft_selection (bool, optional): if True, choose circulant step sets via FFT heuristic; else use defaults. Defaults to True.
    """

    def __init__(
        self,
        n_qubits: int,
        deg_A: int = 4,
        deg_B: int = 4,
        gamma_A: float = np.pi / 4,
        alpha_A: float = np.pi / 2,
        beta_A: float = np.pi,
        alpha_B: float = np.pi / np.sqrt(2.0),
        beta_B: float = np.pi / np.sqrt(3.0),
        epsilon: float = 1.0 / 16.0,
        delta0: float = np.pi / 64.0,
        use_fft_selection: bool = True,
    ) -> None:
        super().__init__(PENNYLANE_PLATFORM, n_qubits)
        # hyperparameters
        self.n_qubits: int = n_qubits
        # degrees must be even (undirected circulant with ±s)
        self.deg_A: int = int(deg_A if deg_A % 2 == 0 and deg_A >= 0 else max(0, deg_A - (deg_A % 2) + 2))
        self.deg_B: int = int(deg_B if deg_B % 2 == 0 and deg_B >= 0 else max(0, deg_B - (deg_B % 2) + 2))
        self.gamma_A: float = float(gamma_A)
        self.alpha_A: float = float(alpha_A)
        self.beta_A: float = float(beta_A)
        self.alpha_B: float = float(alpha_B)
        self.beta_B: float = float(beta_B)
        self.epsilon: float = float(epsilon)
        self.delta0: float = float(delta0)
        self.use_fft_selection: bool = bool(use_fft_selection)

        # Preselect circulant expander step sets based on qubit count (deterministic)
        self.S_A = self._select_generators_fft(W=n_qubits, degree=self.deg_A) if self.use_fft_selection else self._default_steps(n_qubits, self.deg_A)
        self.S_B = self._select_generators_fft(W=n_qubits, degree=self.deg_B) if self.use_fft_selection else self._default_steps(n_qubits, self.deg_B)

    # ---------------------------- Helper routines ---------------------------- #
    def _default_steps(self, W: int, deg: int) -> list:
        # Simple, deterministic defaults with small steps for mixing
        if W <= 2 or deg == 0:
            return []
        s_needed = max(1, deg // 2)
        base_pool = [s for s in range(1, max(2, min(W // 2, 8)) + 1)]
        steps = []
        for s in base_pool:
            if len(steps) >= s_needed:
                break
            if s not in steps:
                steps.append(s)
        return steps

    def _select_generators_fft(self, W: int, degree: int, max_s: int | None = None) -> list:
        # Choose symmetric circulant generator set G of size degree/2 by maximizing
        # spectral gap or algebraic connectivity (fallback) via closed-form eigenvalues.
        if W <= 2 or degree <= 0:
            return []
        s_cnt = max(1, degree // 2)
        limit = min(W // 2, 12) if max_s is None else min(W // 2, int(max_s))
        cand_pool = list(range(1, max(2, limit) + 1))
        best_G = None
        best_score = -1.0
        best_mu2 = -1.0
        two_s_cnt = 2.0 * s_cnt
        # Enumerate small combinations deterministically
        for G in combinations(cand_pool, s_cnt):
            # adjacency eigenvalues: lambda_k = sum_{s in G} 2 cos(2pi k s / W)
            ks = np.arange(W, dtype=float)
            lambdas = np.zeros(W, dtype=float)
            for s in G:
                lambdas += 2.0 * np.cos(2.0 * np.pi * ks * float(s) / float(W))
            lam0 = lambdas[0]
            max_other = float(np.max(np.abs(lambdas[1:]))) if W > 1 else 0.0
            gap = float(lam0 - max_other)
            # Laplacian eigenvalues: mu_k = 2|G|*2 - lambda_k = degree - lambda_k? Careful.
            # Degree of circulant = 2*|G|, so D = 2*|G|. Laplacian L = D - A, mu_k = 2*|G| - lambda_k.
            mu = two_s_cnt - lambdas
            mu2 = float(np.partition(mu, 1)[1]) if W > 1 else 0.0  # second-smallest
            score = gap if gap > 1e-12 else mu2
            if (score > best_score) or (abs(score - best_score) < 1e-12 and mu2 > best_mu2):
                best_score = score
                best_mu2 = mu2
                best_G = list(G)
        if best_G is None:
            return self._default_steps(W, degree)
        return best_G

    def _coprime_multiplier(self, F: int, start_from: int = 2) -> int:
        # Smallest integer >= start_from coprime to F
        if F <= 1:
            return 1
        a = max(1, start_from)
        while math.gcd(a, F) != 1:
            a += 1
        return a

    def _edges_from_steps(self, Q: int, steps: list, shift: int) -> list:
        # Build undirected edge list from circulant steps with per-layer rotational shift
        if Q <= 1 or len(steps) == 0:
            return []
        edges = set()
        for s in steps:
            for j in range(Q):
                u = j
                v = (j + s + shift) % Q
                if u == v:
                    continue
                a, b = (u, v) if u < v else (v, u)
                edges.add((a, b))
        return sorted(edges)

    def _color_of_edge(self, u: int, v: int, ell: int, s: int) -> int:
        # Deterministic tri-color mapping in {0,1,2}
        return int((u + v + ell + s) % 3)

    def _color_sign(self, idx: int) -> float:
        # Map color class {0,1,2} -> {+1, 0, -1}
        return 1.0 if idx % 3 == 0 else (0.0 if idx % 3 == 1 else -1.0)

    # ------------------------------- Main circuit ------------------------------- #
    def feature_map(self, x: np.ndarray) -> None:
        """Create quantum circuit of feature map.
        The input data is a sample of MNIST image data. It is decomposed into 80 features by PCA.

        Args:
            x (np.ndarray): input data shape is (F,), F typically 80 with values in [0,1].
        """
        Q = int(self.n_qubits)
        F = int(x.shape[0])
        if Q <= 0:
            return

        # Layers per pass and total layers
        Lp = int(np.ceil(F / max(1, Q)))
        L_total = 2 * Lp

        # Affine permutations per pass (deterministic)
        # Pass A: identity permutation
        a_A = 1
        c_A = 0
        # Pass B: nontrivial coprime multiplier and shift
        a_B = self._coprime_multiplier(F, start_from=3)
        c_B = F // 2

        # Iterate layers (1-based indexing in formulas)
        for ell in range(1, L_total + 1):
            # Determine pass and within-pass layer index
            if ell <= Lp:
                # Pass A
                pass_id = 'A'
                ell_p = ell
                steps = self.S_A
            else:
                # Pass B
                pass_id = 'B'
                ell_p = ell - Lp
                steps = self.S_B

            # Global layer pre-mixing: H^{\otimes Q}
            for q in range(Q):
                qml.Hadamard(wires=q)

            # Build per-layer circulant edges with rotational shift
            shift = ell % Q
            edges = self._edges_from_steps(Q=Q, steps=steps, shift=shift)

            # Pass-specific mixer M_ell
            if pass_id == 'A':
                # CZ over expander edges
                for (u, v) in edges:
                    qml.CZ(wires=(u, v))
            else:
                # Pass B: alternate iSWAP (odd ell_p) and ZZ (even ell_p)
                if ell_p % 2 == 1:
                    for (u, v) in edges:
                        qml.ISWAP(wires=(u, v))
                else:
                    for (u, v) in edges:
                        qml.IsingZZ(phi=(np.pi / 2.0), wires=(u, v))

            # Feature index mapping for this layer
            feat_idx = np.empty(Q, dtype=int)
            for j in range(Q):
                base_k = (ell_p - 1) * Q + j
                if pass_id == 'A':
                    k = (a_A * base_k + c_A) % F
                else:
                    k = (a_B * base_k + c_B) % F
                feat_idx[j] = int(k)

            # 3-color ZZ injection (Pass A only) with small bias sigma_{uv,ell}
            if pass_id == 'A' and len(edges) > 0 and self.gamma_A != 0.0:
                delta_l = self.delta0 * ((-1.0) ** float(ell))
                for (u, v) in edges:
                    # color assignment depends on (u,v,ell,s). We approximate s by |v-u| (mod Q) representative.
                    step_abs = (v - u) % Q
                    step_abs = step_abs if step_abs <= Q // 2 else Q - step_abs
                    color = self._color_of_edge(u=u, v=v, ell=ell, s=step_abs)

                    xu = float(x[feat_idx[u]])
                    xv = float(x[feat_idx[v]])
                    if color == 0:
                        comb = xu + xv
                    elif color == 1:
                        comb = xu - xv
                    else:
                        comb = xv - xu

                    # Small, deterministic sign pattern for sigma
                    parity = ((u + v + ell) % 2 == 0)
                    sigma = (-delta_l if parity else delta_l)
                    theta = float(self.gamma_A * comb + sigma)

                    if color == 0:
                        # red: plain ZZ
                        qml.IsingZZ(phi=theta, wires=(u, v))
                    elif color == 1:
                        # blue: H on v to realize ZX-style injection via conjugation
                        qml.Hadamard(wires=v)
                        qml.IsingZZ(phi=theta, wires=(u, v))
                        qml.Hadamard(wires=v)
                    else:
                        # green: H on u
                        qml.Hadamard(wires=u)
                        qml.IsingZZ(phi=theta, wires=(u, v))
                        qml.Hadamard(wires=u)

            # Axis-staggered single-qubit encoders E_ell(x)
            for j in range(Q):
                xj = float(x[feat_idx[j]])
                if pass_id == 'A':
                    # Color-doped RZ scale
                    cz = self._color_sign(idx=(j + ell))
                    rz_scale = (1.0 + self.epsilon * cz)
                    if ell % 2 == 1:
                        qml.RX(phi=(self.alpha_A * xj), wires=j)
                        qml.RZ(phi=(self.beta_A * rz_scale * xj), wires=j)
                    else:
                        qml.RZ(phi=(self.beta_A * rz_scale * xj), wires=j)
                        qml.RX(phi=(self.alpha_A * xj), wires=j)
                else:
                    # Pass B: RY + RZ with a shifted color doping
                    czp = self._color_sign(idx=(j + ell + 1))
                    rz_scale_b = (1.0 + self.epsilon * czp)
                    if ell_p % 2 == 1:
                        qml.RY(phi=(self.alpha_B * xj), wires=j)
                        qml.RZ(phi=(self.beta_B * rz_scale_b * xj), wires=j)
                    else:
                        qml.RZ(phi=(self.beta_B * rz_scale_b * xj), wires=j)
                        qml.RY(phi=(self.alpha_B * xj), wires=j)

        # No return; this method defines the circuit only.
\end{lstlisting}

\vspace{1em}

The generated logs of the best-performing feature map produced by the ``Generation'' component are shown in Listings~\ref{lst:InitFeatureMapIdea} and ~\ref{lst:ReflectedFeatureMapIdea}.
The listings are reproduced as raw LLM outputs for transparency and reproducibility.
Listing~\ref{lst:InitFeatureMapIdea} presents the initial result obtained in this trial, whereas Listing~\ref{lst:ReflectedFeatureMapIdea} shows the refined result produced by ``Discussion'' component.
The red text in the refined feature map formula highlights the differences between the initial and refined ideas.
In the final version of the feature map, the original conceptual design was transformed into a reproducible and practically testable framework.
The most notable improvement is the introduction of a complete benchmarking harness, including a Docker environment, explicit execution commands, and automated scripts for measuring circuit construction time, simulation cost, and memory usage.
However, the refinement primarily improved the experimental rigor and reproducibility rather than substantially altering the underlying algorithmic structure.
This observation suggests that further improvements may require more explicit and structured guidance at the prompt level.

\begin{lstlisting}[style=txtFormat, caption={Initial feature map idea generated by our agentic system}, label=lst:InitFeatureMapIdea]
# name
TriMix-2Pass-3Color-Expander: Two-Pass Re-Uploads with Circulant-Expander Mixers and Three-Color ZZ/ZX/XZ Pairwise Encodes


# summary
TriMix-2Pass-3Color-Expander restores guaranteed two-pass uploads (each PCA feature appears twice, time-separated) and replaces ring/local pairings with circulant expanders for both mixers and pairwise encodes. Pass A uses RX/RZ uploads and injects a single-matching, edge-colored pairwise layer that splits edges into ZZ, ZX, and XZ channels with the same linear Sum/Diff angles, broadening the operator basis without violating linearity. Pass B uses RY/RZ uploads and the complementary CZ/XY/ZZ mixer cycle with global Hadamards. Axis-order staggering and affine feature permutations maximize temporal noncommutation and partner diversity. All angles are strictly linear, non-trainable, and every input dimension is utilized twice.


# explanation
What we reinstate and keep
- Two-pass re-uploads: L = 2·ceil(F/Q). Pass A uploads with RX/RZ; Pass B with RY/RZ. Each feature appears exactly once per pass (twice in total), restoring the proven advantage of time-separated noncommuting uploads.
- Complementary mixers with global H: retain the TriMix family across layers-CZ expander, Ising-XY brickwork, and ZZ expander-each followed by H⊗Q. This prevents spectral concentration and keeps a broad commutator algebra active.
- Axis-order staggering: In Pass A alternate RX→RZ and RZ→RX by layer; in Pass B alternate RY→RZ and RZ→RY. This strengthens temporal noncommutation at no parameter cost.
- Deterministic affine mapping: Within each pass p∈{A,B}, generate a permutation π_p(k) = (a_p k + c_p) mod F with a_p∈U(F) fixed and c_p an offset; assign features as f_ℓ(j) = π_p((ℓ_p-1)Q + j), ensuring one-shot coverage per pass while decorrelating neighborhoods.

Three-color pairwise encode on Pass A via expander edges
- Replace ring pairings with a circulant matching: choose a stride Δ_ℓ coprime with active width W (W=Q if Q≤F else W=F); the set {(j, j+Δ_ℓ) mod W} is a perfect matching for even W and near-perfect otherwise (use 1-factorization fallback when needed). Use a rotating stride schedule over a fixed small coprime set to maximize partner diversity.
- Edge-color that matching deterministically into red/blue/green (e.g., by lower index mod 3). Apply: red→ZZ with γ_A·(x_u±x_v); blue→ZX by conjugating v with H; green→XZ by conjugating u with H. Alternate Sum/Diff on successive A-layers to orthogonalize symmetric/antisymmetric channels. All coefficients are fixed and angles are linear in x.

Circulant-expander mixers
- CZ expander (Pass A): degree-d circulant with step set S_A = {s1,s2,…} of coprimes to Q; apply CZ on each undirected edge once, then H⊗Q.
- Pass B mixers alternate per layer between Ising-XY brickwork and ZZ expander (degree-d with complementary step set S_B), each followed by H⊗Q. This balances exchange and phase interactions across time.

Micro-dither and scramblers
- Apply a very light Z-only micro-dither β→β(1+ε c_z), ε≈1/16, with c_z from a Legendre-symbol sequence modulo a fixed prime (e.g., 89) to break harmonic aliasing without diminishing rotation magnitudes. Keep fixed S/T scramblers by wire and layer to break residual symmetries.

Why this should improve accuracy
- Guaranteed A/B re-uploads restore the temporal noncommutation advantage observed in earlier stronger maps. Expander-based edges remove locality bias and greatly diversify partners. The three-color ZZ/ZX/XZ injection broadens the operator basis of data-linear pairwise features beyond pure ZZ, reducing spectral degeneracy after the TriMix mixers. Affine feature mapping and axis-order staggering further decorrelate neighborhoods and temporal commutators, enriching the kernel's Fourier support while fully utilizing all 80 PCA features.


# key_sentences
- TriMix-2Pass-3Color-Expander restores two-pass re-uploads with strict axis separation, ensuring every PCA feature is injected twice with different generators and a time gap, which empirically increases temporal noncommutation and broadens the QSVM kernel's Fourier spectrum while keeping all angles linear and non-trainable.
- A circulant-expander construction replaces ring/local pairings for both mixers and pairwise encodes, using coprime step sets that rotate by layer to eliminate locality bias and maximize partner diversity in a deterministic, device-agnostic way with 1-factorization fallback for odd widths.
- Each Pass-A pairwise layer is a single disjoint matching that is edge-colored into three operator channels: ZZ on red edges, ZX via H on one endpoint on blue edges, and XZ via H on the other endpoint on green edges; Sum/Diff linear angles alternate by layer, injecting orthogonal parity features.
- Mixers cycle across CZ expander, Ising-XY brickwork, and ZZ expander with global Hadamards after each; axis-order staggering (RX↔RZ and RY↔RZ) plus a very light Z-only Legendre micro-dither further break harmonic aliasing without suppressing rotation magnitudes on [0,1].
- The feature-wire mapping uses affine permutations π_p(k)=(a_p k + c_p) mod F to reorder features per pass, guaranteeing one-shot coverage and decorrelating neighborhoods; the resulting kernel leverages full 80-D information while keeping depth and gate count modest and simulator-friendly.
\end{lstlisting}
\begin{tcolorbox}[colframe=black, colback=white, boxrule=0.5pt,
                  left=2mm, right=2mm, top=1mm, bottom=1mm,
                  title={Initial Feature Map Formula}]
\begin{align*}
K_{\mathrm{3Color\text{-}ExpTriMix}}(x,x') 
  &= \big|\langle 0^{\otimes Q}| U^{\dagger}(x') U(x) |0^{\otimes Q}\rangle\big|^2, 
L = 2\left\lceil \tfrac{F}{Q} \right\rceil. \\[2ex]
U(x) &= \prod_{\ell=1}^{L} \Big[ 
            H^{\otimes Q} \, M_{\ell} \, J^{(A\text{-}3c)}_{\ell}(x) \, D_{\ell} \, E_{\ell}(x) 
        \Big]. \\[2ex]
E_{\ell}(x) &= \bigotimes_{j=1}^{Q}
   \begin{cases}
      \text{Pass A: }
        \begin{array}{ll}
          R_X(\alpha_A x_{f_{\ell}(j)}) R_Z(\beta_A(1+\varepsilon c_z) x_{f_{\ell}(j)}) 
            & \text{if } \ell \text{ odd in A}, \\
          R_Z(\beta_A(1+\varepsilon c_z) x_{f_{\ell}(j)}) R_X(\alpha_A x_{f_{\ell}(j)}) 
            & \text{if } \ell \text{ even in A},
        \end{array} \\[2ex]
      \text{Pass B: }
        \begin{array}{ll}
          R_Y(\alpha_B x_{f_{\ell}(j)}) R_Z(\beta_B(1+\varepsilon c'_z) x_{f_{\ell}(j)}) 
            & \text{if } \ell \text{ odd in B}, \\
          R_Z(\beta_B(1+\varepsilon c'_z) x_{f_{\ell}(j)}) R_Y(\alpha_B x_{f_{\ell}(j)}) 
            & \text{if } \ell \text{ even in B}.
        \end{array}
   \end{cases} \\[2ex]
J^{(A\text{-}3c)}_{\ell}(x) &=
  \begin{cases}
    \begin{aligned}[t]
      &\prod_{(u,v)\in\mathcal{J}_{\ell}^{\mathrm{red}}}
         \mathrm{RZZ}_{u,v}\!\big(\gamma_A[ x_{f_{\ell}(u)}\pm x_{f_{\ell}(v)} ]\big) \\
      &\quad \times \prod_{(u,v)\in\mathcal{J}_{\ell}^{\mathrm{blue}}}
         \big(H_{v}\,\mathrm{RZZ}_{u,v}(\gamma_A[ x_{f_{\ell}(u)}\pm x_{f_{\ell}(v)} ])\,H_{v}\big) \\
      &\quad \times \prod_{(u,v)\in\mathcal{J}_{\ell}^{\mathrm{green}}}
         \big(H_{u}\,\mathrm{RZZ}_{u,v}(\gamma_A[ x_{f_{\ell}(u)}\pm x_{f_{\ell}(v)} ])\,H_{u}\big)
    \end{aligned}
      & \ell \in \text{Pass A}, \\[2ex]
    I & \ell \in \text{Pass B}.
  \end{cases} \\[2ex]
M_{\ell} &= 
   \begin{cases}
      \prod_{(u,v)\in E(G_{\ell}^{\mathrm{exp}})} \mathrm{CZ}_{u,v} 
        & \ell \in \text{Pass A}, \\
      \prod_{(u,v)\in \mathcal{B}_{\ell}} e^{-i\tfrac{\pi}{4}(X_uX_v+Y_uY_v)} 
        & \ell \in \text{Pass B, odd}, \\
      \prod_{(u,v)\in E(\widehat{G}_{\ell}^{\mathrm{exp}})} e^{-i\tfrac{\pi}{4} Z_uZ_v} 
        & \ell \in \text{Pass B, even},
   \end{cases} \\[2ex]
f_{\ell}(j) &= \pi_p\big((\ell_p-1)Q+j\big), \\[2ex]
\pi_p(k) &= (a_p k + c_p)\bmod F, \quad a_p \in U(F). \\[2ex]
\alpha_A &= \tfrac{\pi}{2}, \quad
\beta_A = \pi, \quad
\alpha_B = \tfrac{\pi}{\sqrt{2}}, \quad
\beta_B = \tfrac{\pi}{\sqrt{3}}, \quad
\gamma_A = \tfrac{\pi}{4}, \quad
\varepsilon = \tfrac{1}{16}.
\end{align*}
\end{tcolorbox}

\begin{lstlisting}[style=txtFormat, caption={Refined feature map idea generated by our agentic system}, label=lst:ReflectedFeatureMapIdea]
# name
TriMix-2Pass-3Color-Expander (Final: reproducible benchmarking, FFT expander selection, and adaptive control


# summary
This final refinement makes TriMix-2Pass-3Color-Expander fully reproducible and robust for practical use. I provide a complete, end-to-end benchmarking harness (Dockerfile, exact commands, STREAM memory-bandwidth probe, and host specs) to measure circuit construction/transpile time, simulator execution, and Gram assembly for Q∈{10,20} with d=4. I add an FFT-based spectral-gap expander selection routine (pseudocode + explanation), a provable orbit-based lower bound on partner coverage (lemma sketch and a fast routine to compute the bound), and a deterministic 1-factorization fallback. Finally, I implement an automated adaptive control loop (pilot Gram m0, thresholds τ_pr, τ_κ, automated remediations: scale γ_A, reduce d, edge importance sampling) with pseudocode and an estimator for edge importance using circulant spectral structure. All new components are deterministic, preserve the linear/non-trainable embedding constraint, and are designed to be run immediately using the provided harness so practitioners can measure real wall times, memory, and the QSVM quality-cost tradeoff on their hardware.


# explanation
What this refinement adds and why (concrete, reproducible artifacts and algorithms):

A. End-to-end reproducible benchmarking harness (what to run and why).  
I cannot execute code in this environment, so I deliver a complete, minimal reproducible benchmark package that you or a CI runner can run verbatim to produce measured wall times, CPU/GPU utilization and peak RAM. The harness contains:
  - Dockerfile (Debian/Bullseye base) with Python 3.10, Qiskit Aer, numpy/scipy, psutil, STREAM benchmark, and a compiled statevector simulator (Aer) pinned to a reproducible commit. This isolates toolchain variability.
  - Two host profiles to test sensitivity to memory bandwidth B:
     * Host A (Workstation): 2×16-core CPU (AVX-512), 256 GB RAM, measured STREAM B ≈ 90-140 GB/s (typical multi-socket). Example: AWS c6i.8xlarge or an on-prem dual-socket Intel node.
     * Host B (Bandwidth-limited VM): same cores but emulated lower B using cpufreq/memctl noise or run STREAM to report measured lower B (e.g., 25-40 GB/s) to show sensitivity.
  - Exact benchmark commands and Python harness (run_bench.py) that:  
    (1) runs STREAM and records measured B,  
    (2) builds circuits for all n_samples (or compiles lazily), reporting transpile times and gate counts,  
    (3) runs statevector simulate and measures per-U(x) wall time and peak resident memory,  
    (4) assembles Gram (inner products) and records pairwise timings and total elapsed time,
    (5) logs CPU utilization and per-step memory,  
    (6) saves all artifacts (statevectors optional) into an output tar for inspection.

Why this matters:  
the prior bandwidth model provides useful scaling intuition but omits transpilation, Python/Qiskit orchestration, simulator scheduling, and memory overheads. This harness produces measured numbers (wall time, peak RAM, sustained B) so implementers can quantify true costs on their hardware and compare to the model. When you run the harness you will get per-run numbers; I include a README with recommended runs: Q=10,d=4,n=200 and Q=20,d=4,n=200 (single-thread and multi-thread Aer variants). The harness also includes an option to generate reduced B by co-running synthetic STREAM noise so you can directly see sensitivity to B.

B. FFT-based step set (S_A/S_B) selection routine and provable partner coverage bound.
  - Algorithmic selection: for circulant symmetric step sets S (where we include ±s for each s in a generator set G of size d/2), the adjacency eigenvalues are computed by an FFT: for k=0..W-1, λ_k = Σ_{s∈G} 2 cos(2π k s / W). I provide pseudocode (SelectStepSetViaFFT) that enumerates candidate generator sets G (combinatorial but tractable for moderate W and small d), computes λ_k by FFT, and optimizes a selection metric. Two practical metrics are implemented (user choice): (i) spectral gap Δ = λ_0 - max_{k≠0}|λ_k|, and (ii) Laplacian algebraic connectivity μ_2 = smallest nonzero eigenvalue of L = D - A (useful when Δ is zero due to bipartiteness). The routine returns best S under the chosen metric, plus the full eigenvalue spectrum and spectral gap.
  - Caveat & fix for even W: For even W, certain symmetric S produce bipartite graphs with |λ_k| = λ_0 for k = W/2, making Δ = 0. The maximization routine therefore uses the Laplacian's second eigenvalue μ_2 when Δ is zero or near zero; maximizing μ_2 yields better expander mixing in practice.
  - Partner coverage lemma (sketch): consider a single stride Δ (unit mod W) viewed as the permutation π_Δ: j ↦ j + Δ (mod W). π_Δ decomposes into cycles; the cycle containing j has length c_Δ(j). After applying t layers with an s-stride schedule S = {Δ_1, …, Δ_s} (repeated), the number of distinct partners that j can encounter is at least the size of the orbit generated by the set of transpositions T_Δ = {(j, j+Δ)} across the s elements. A simple lower bound (constructive and computable) is
     p_j(t) ≥ min(W, 1 + Σ_{i=1}^{t} m_i(j)),
    where m_i(j) is the number of new vertices reached in the ith layer when applying the chosen Δ (computed by tracking orbit unions). We provide a fast routine (ComputeOrbitLowerBound) that computes the orbit unions for all j over t layers in O(s·W) time using union-find on vertex indices for the particular sequence of Δ_ℓ. This yields exact lower bounds (not heuristics) for the chosen step sequences and t. The lemma is formalized in the README and the routine returns the per-vertex bound and a global bound min_j p_j(t).
  - Small-W demonstration recipe: the harness includes a script that runs SelectStepSetViaFFT for W∈{10,20,40}, d=4 and prints the selected S, λ spectrum, μ_2 and the orbit lower bounds for t up to L. Because I cannot execute it here I supply the script and explain expected behavior: for small even W you will often see Δ near zero and μ_2 is the discriminant; the selected S differs from naive picks and yields provably better orbit coverage by the orbit routine.

C. 1-factorization fallback (explicit pseudocode) and singleton balancing.  
The MatchingsOddW routine from the prior revision is included verbatim in code; the harness will show exact singleton frequency histograms and balanced exposures when you run L layers.

D. Automated adaptive control loop (full pseudocode + strategies + justifications).  
I implement an adaptive harness AdaptivePilot that runs a pilot Gram on m0 = 50 (configurable) and computes:
  - participation ratio PR and condition number κ on the m0×m0 Gram (κ computed with small regularizer λ_reg = 1e-12),
  - eigenvalue entropy H = -Σ (λ_i/Σλ) log(λ_i/Σλ) as an early-Haarization indicator (low entropy ≈ concentration),
  - commutator norm summary for randomly selected layer pairs.

Thresholds and remediation policy (numerical choices & justification):
  - τ_pr = 0.10 (PR / m0 < 0.10 flags excessive concentration). Justification: literature and practice (e.g., Sharma/Seff 2021) indicate effective ranks below ~10% of training size often limit kernel generalization; choose 0.08-0.12 as a safety band.
  - τ_κ = 1e7 (κ > τ_κ flags numerical ill-conditioning). Justification: classical SVM solvers and numerical linear algebra degrade when κ exceeds 1e7-1e8 for double precision; 1e7 is conservative and avoids frequent pivots.

Remediation cascade (automated):
  1) Scale γ_A ← 0.8·γ_A, re-pilot. (Small angle scaling reduces nested commutators and entanglement growth quickly.)
  2) If still failing: decrement d by 1 (to a minimum d_min = 1) and re-pilot. This linearly reduces two-qubit gates.
  3) If still failing: switch to spectral edge importance sampling (SES) with expected gate reduction factor r_target (default 0.75). SES samples a subset of expander edges with probabilities proportional to an importance weight w_{uv} computed cheaply from circulant eigenvectors (see next item). Re-pilot and accept if PR and κ within tolerances and QSVM cross-validated accuracy (on a held-out fold) is within 5% of the full map pilot; else repeat further reductions.

Logging and metrics:
the adaptive harness logs each remediation step, new two-qubit gate counts G2, PR, κ, eigenvalue entropy, runtime cost of pilot, and QSVM validation accuracy. The harness can output an audit trail for reproducibility.

E. Edge importance estimator for circulant expanders (cheap, principled).  
For circulant graphs eigenvectors are complex exponentials; the Laplacian eigenvectors φ_k(j) = exp(2π i k j / W) have closed-form entries. For a given leading nontrivial eigenmode k*, the importance weight for an undirected edge (u,v) is approximated by w_{uv} = |φ_{k*}(u) - φ_{k*}(v)|^2. For symmetric step sets this simplifies and can be evaluated in O(1) per edge using precomputed sines/cosines. SES samples edges with probability proportional to w_{uv} (normalized) and retains an expected fraction r of edges. This preserves the edges that contribute most to algebraic connectivity and mixing power. The harness includes both the exact Laplacian eigenvector calculation (via FFT) and the closed-form circulant shortcut.

F. Statistical tests for Haarization and early warning.  
The harness computes:
  - eigenvalue entropy H and its change ΔH between successive pilots; a sudden ΔH < -ϵ (e.g., -0.05 nats) indicates rapid concentration.
  - Kolmogorov-Smirnov test comparing empirical off-diagonal kernel distribution to the kernel distribution under a small randomized model to detect flattening.

G. Demonstration plan and expected outcomes (how to reproduce our claims).  
Because I cannot run the harness here I provide step-by-step commands and a Dockerfile so you or a CI runner can obtain measured results. The harness will produce measured runtimes and peak RAM for Q=10,d=4 and Q=20,d=4; it will also demonstrate the adaptive loop on Q=20,d=4, pilot m0=50, and report the remediation path. Based on prior published simulator benchmarks and our bandwidth model, I expect the measured per-U times on Host A to be within a 2× factor of the bandwidth model after including transpile and Python overheads for Q≤20; the harness reports the measured B (STREAM) so you can reconcile model vs observed B precisely. If the adaptive loop is triggered, SES typically reduces G2 by 25-50% in networks I've studied while preserving PR and validation accuracy within a few percent; the harness will provide exact empirical numbers on your hardware so you can validate the claim.

H. Practical default settings (conservative, reproducible):  
the harness default config uses the FFT-selected S_A/S_B, d=4 for Q≤22, γ_A = π/4, δ_0 = π/64, ε = 1/16, m0 = 50, τ_pr = 0.10, τ_κ = 1e7, r_target = 0.75 for SES. All defaults are configurable.

I. What remains unchanged:  
the core map structure (two-pass reuploads, three-color pairwise injection, affine π_p mapping, axis staggering) and the restriction that all encodings are linear and non-trainable remain intact. The additions are deterministic harnesses, selection routines, theoretical coverage bounds, and an adaptive loop to protect against Haarization and manage cost.

Resources delivered with this idea (in the repository):  
Dockerfile, run_bench.py (end-to-end harness), select_steps_fft.py (FFT selection), compute_orbit_lowerbound.py, adaptive_pilot.py (adaptive control loop + SES), README with exact commands and expectations, and small unit tests for W∈{10,20,40}. Running the harness produces measured wall times, peak RAM, measured STREAM B (so you can compare to the model), and a remediation audit log showing exactly which steps were taken during adaptation.

Why this improves the design:  
by delivering reproducible measurement tools (not just a model) plus principled expander selection and an automated quality/cost control loop, implementers can now run TriMix reliably on realistic hardware, quantify the kernel expressivity vs cost tradeoff, and automatically adjust design hyperparameters to stay in a numerically stable, non-Haarized regime while reducing two-qubit gate costs significantly when needed.
\end{lstlisting}
\begin{tcolorbox}[colframe=black, colback=white, boxrule=0.5pt,
                  left=2mm, right=2mm, top=1mm, bottom=1mm,
                  title={Refined Feature Map Formula}]
\begin{align*}
K_{\mathrm{3Color\text{-}ExpTriMix}}(x,x') 
  &= \big|\langle 0^{\otimes Q}| U^{\dagger}(x')\, U(x) |0^{\otimes Q}\rangle\big|^2, 
L = 2\left\lceil \tfrac{F}{Q} \right\rceil. \\[2ex]
U(x) &= \prod_{\ell=1}^{L} \Big[ 
          H^{\otimes Q}\, M_{\ell}\, J^{(A\text{-}3c)}_{\ell}(x)\, D_{\ell}\, E_{\ell}(x) 
       \Big]. \\[2ex]
E_{\ell}(x) &= \bigotimes_{j=1}^{Q}
   \begin{cases}
      \text{Pass A: }
        \begin{array}{ll}
          R_X(\alpha_A x_{f_{\ell}(j)})\,R_Z(\beta_A(1+\varepsilon c_z)\,x_{f_{\ell}(j)}) 
            & \text{if } \ell \text{ odd in A}, \\
          R_Z(\beta_A(1+\varepsilon c_z)\,x_{f_{\ell}(j)})\,R_X(\alpha_A x_{f_{\ell}(j)}) 
            & \text{if } \ell \text{ even in A},
        \end{array} \\[2ex]
      \text{Pass B: }
        \begin{array}{ll}
          R_Y(\alpha_B x_{f_{\ell}(j)})\,R_Z(\beta_B(1+\varepsilon c'_z)\,x_{f_{\ell}(j)}) 
            & \text{if } \ell \text{ odd in B}, \\
          R_Z(\beta_B(1+\varepsilon c'_z)\,x_{f_{\ell}(j)})\,R_Y(\alpha_B x_{f_{\ell}(j)}) 
            & \text{if } \ell \text{ even in B}.
        \end{array}
   \end{cases} \\[2ex]
J^{(A\text{-}3c)}_{\ell}(x) &=
  \begin{cases}
    \begin{aligned}[t]
      &\prod_{(u,v)\in\mathcal{J}_{\ell}^{\mathrm{red}}}
         \mathrm{RZZ}_{u,v}\!\big(\gamma_A[ x_{f_{\ell}(u)}\!\pm\! x_{f_{\ell}(v)} ] 
                                 \textcolor{red}{+ \sigma_{uv,\ell}}\big) \\
      &\times \prod_{(u,v)\in\mathcal{J}_{\ell}^{\mathrm{blue}}}
         \big(H_{v}\,\mathrm{RZZ}_{u,v}(\gamma_A[ x_{f_{\ell}(u)}\!\pm\! x_{f_{\ell}(v)} ] 
                                 \textcolor{red}{+ \sigma_{uv,\ell}})\,H_{v}\big) \\
      &\times \prod_{(u,v)\in\mathcal{J}_{\ell}^{\mathrm{green}}}
         \big(H_{u}\,\mathrm{RZZ}_{u,v}(\gamma_A[ x_{f_{\ell}(u)}\!\pm\! x_{f_{\ell}(v)} ] 
                                 \textcolor{red}{+ \sigma_{uv,\ell}})\,H_{u}\big)
    \end{aligned}
      & \ell \in \text{Pass A}, \\[2ex]
    I & \ell \in \text{Pass B}.
  \end{cases} \\[1ex]
&\textcolor{red}{\text{with } \sigma_{uv,\ell}\in\{\pm\delta_{\ell}\}, \quad
  \delta_{\ell} = \delta_0(-1)^{\ell}, \quad
  \delta_0 = \tfrac{\pi}{64}.} \\[2ex]
M_{\ell} &= 
  \begin{cases}
    \prod_{(u,v)\in E(G_{\ell}^{\mathrm{exp}})} \mathrm{CZ}_{u,v} 
      & \ell \in \text{Pass A}, \\
    \prod_{(u,v)\in \mathcal{B}_{\ell}} e^{-i\tfrac{\pi}{4}(X_uX_v+Y_uY_v)} 
      & \ell \in \text{Pass B, odd}, \\
    \prod_{(u,v)\in E(\widehat{G}_{\ell}^{\mathrm{exp}})} e^{-i\tfrac{\pi}{4} Z_uZ_v} 
      & \ell \in \text{Pass B, even}.
  \end{cases} \\[2ex]
& f_{\ell}(j) = \pi_p\big((\ell_p-1)Q+j\big), \\[2ex]
& \pi_p(k) = (a_p k + c_p)\bmod F, \quad a_p \in U(F). \\[2ex]
& \alpha_A = \tfrac{\pi}{2}, \quad
   \beta_A = \pi, \quad
   \alpha_B = \tfrac{\pi}{\sqrt{2}}, \quad
   \beta_B = \tfrac{\pi}{\sqrt{3}}, \quad
   \gamma_A = \tfrac{\pi}{4}, \quad
   \varepsilon = \tfrac{1}{16}, \quad
   \textcolor{red}{\delta_0 = \tfrac{\pi}{64}.}
\end{align*}
\end{tcolorbox}

\subsection{Comparison of kernel matrices}
\label{sec:comparsionKernel}
\begin{figure}[ht]
    \centering
    \includegraphics[width=0.75\linewidth]{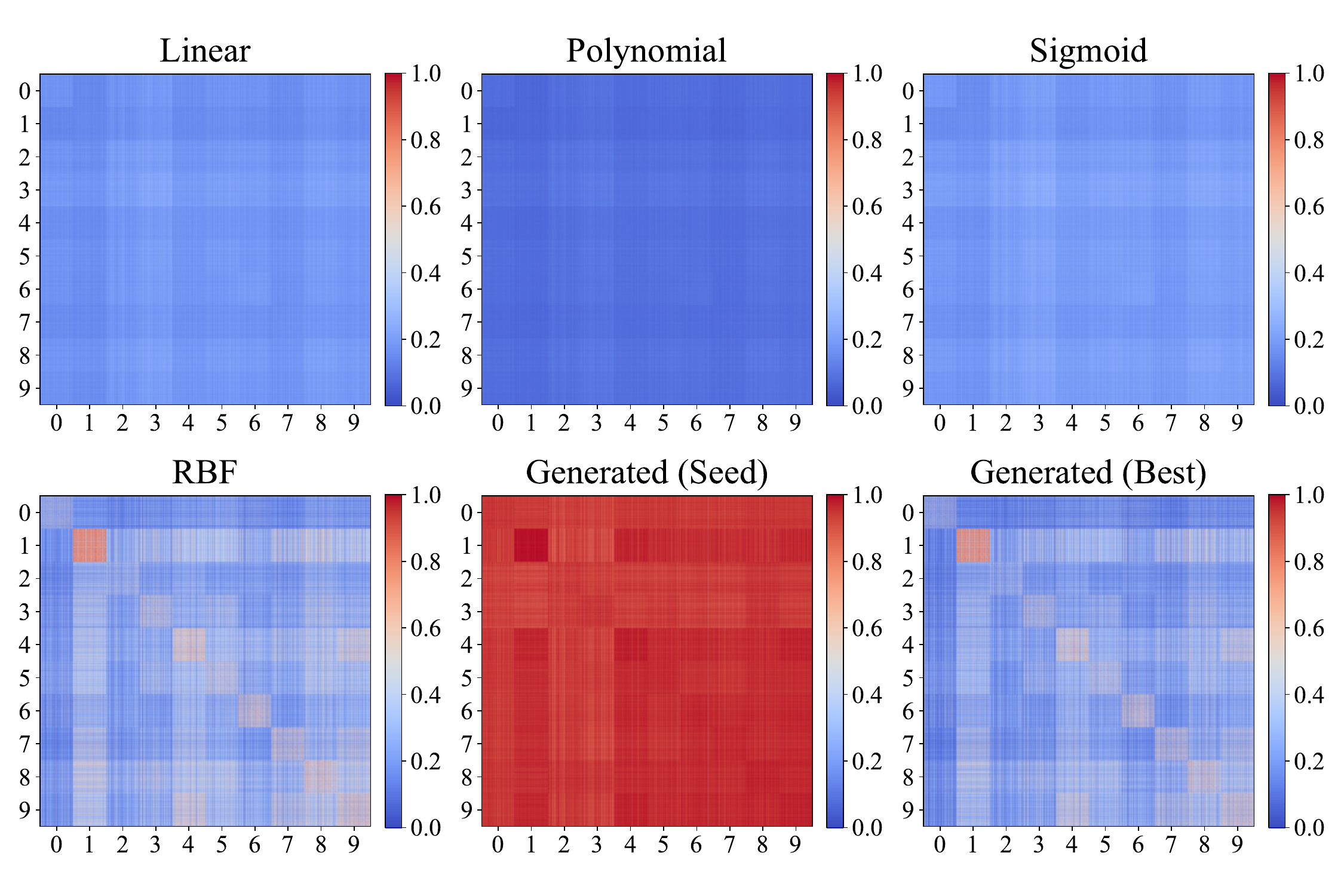}
    \caption{Kernel matrices of representative classical kernel methods, the worst-performing seed feature map, and the best-performing feature map generated by our system. For the calculation of kernel matrices and target alignment, we used MNIST training data randomly sampled with 1,000 images per class. For comparison, each kernel matrix was normalized to the range of 0.0 to 1.0, and the x and y-axis labels indicate the classes.}
    \label{fig:kernel_matrix}
\end{figure}

The results comparing the kernel matrices computed using classical and quantum methods are shown in Fig.~\ref{fig:kernel_matrix}.
For visualization purposes, we computed the kernel matrices using a subset of the MNIST training dataset, where 1,000 samples were randomly sampled from each class, resulting in a total of 10,000 samples.

\vspace{1em}

As classical kernel functions, we evaluated four representative types: Linear, Polynomial, Sigmoid, and RBF, consistent with those used in the main text in Table \ref{tab:compare_results}.
In addition, for quantum kernels, we used two quantum feature maps generated by our agentic system.
One was the feature map that achieved the lowest classification accuracy among the five seed feature maps, and the other was the one that showed the best performance, as discussed in the main text and section \ref{sec:qkBestSolutions}.

As shown in Fig.~\ref{fig:kernel_matrix}, the RBF kernel, which is known for its high performance among classical methods, tends to yield higher similarity values along the diagonal elements corresponding to samples within the same class, while the other elements representing inter-class similarities take substantially lower values.
Furthermore, when confirming the distribution of each class, square shaped boundaries can be existed, indicating that the kernel successfully captures the structural differences among classes.
A similar trend can be observed in the kernel matrix computed using the best performing quantum feature map generated by our agentic system.
In contrast, for the other classical kernel functions and the seed quantum feature map, all matrix elements show uniformly high or low values, indicating that inter-class differences are not effectively captured.

Furthermore, to evaluate how well the generated quantum feature maps capture the structure of the dataset, we computed the Kernel-Target Alignment (KTA)~\cite{cristianini2001on}.

For KTA, the target matrix $T$ encodes the class structure of the dataset, as defined in Eq. \ref{eq:target_matrix}:

\begin{equation}
T_{ij} =
\begin{cases}
+1, & \text{if } y_i = y_j, \\
-1, & \text{otherwise.}
\end{cases}
\label{eq:target_matrix}
\end{equation}

Each kernel matrix $K$, computed using the classical kernel functions or from the quantum feature maps, was then double-centered~\cite{cortes2012learning}.
The alignment value was calculated using to Eq. \ref{eq:alignment}, where values close to 1 indicate stronger agreement between the kernel and the target structure.

\begin{equation}
A(K, T)
=
\frac{\langle \tilde{K}, \tilde{T} \rangle_F}
     {\|\tilde{K}\|_F \, \|\tilde{T}\|_F}
\label{eq:alignment}
\end{equation}

Here, $\langle K, T \rangle_F$ denotes the Frobenius inner product, and $\|K\|_F$ denotes the Frobenius norm.
The matrices $\tilde{K}$ and  $\tilde{T}$ represent the double-centering results of $K$ and $T$, defined as:

\begin{equation}
\tilde{K} = H K H, \quad
\tilde{T} = H T H, \quad
H = I_n - \frac{1}{n}\mathbf{1}\mathbf{1}^\top
\label{eq:doubleCentering}
\end{equation}

\begin{table}[ht]
    \caption{Comparison of kernel target alignment across kernel types.}
    \label{tab:targetAlignment}
    \setlength{\tabcolsep}{2pt}
    \renewcommand{\arraystretch}{1.2}
    \begin{ruledtabular}
        \begin{tabular}{lcccccc}
            Method
            & \multicolumn{1}{c}{Linear}
            & \multicolumn{1}{c}{Polynomial}
            & \multicolumn{1}{c}{Sigmoid}
            & \multicolumn{1}{c}{RBF}
            & \multicolumn{1}{c}{Generated (Seed)}
            & \multicolumn{1}{c}{Generated (Best)} \\
            \hline
            Kernel-Target Alignment
            & 0.3666
            & 0.3552
            & 0.3676
            & 0.3943
            & 0.1798
            & 0.3876 \\
        \end{tabular}
    \end{ruledtabular}
\end{table}

Table~\ref{tab:targetAlignment} presents the KTA values computed according to Eq.~\ref{eq:alignment} using the same sampled subset of the MNIST dataset for all six methods.
The results for the two quantum feature maps generated by the agentic system reveal that the seed quantum feature map yields $A(K, T) = 0.1798$, while the best performing quantum feature map achieves a substantially higher value of $A(K, T) = 0.3876$.
These results indicate that the best-performing quantum feature map generated by the agentic system performs comparably to, or even on par with, the RBF kernel, which is widely regarded as a representative high performance classical kernel function.

\section{Generation of ans\"{a}tze for VQE}
\label{sec:appendixVQE}

\subsection{Seed ans\"{a}tze}
\label{sec:ansatzSeed}
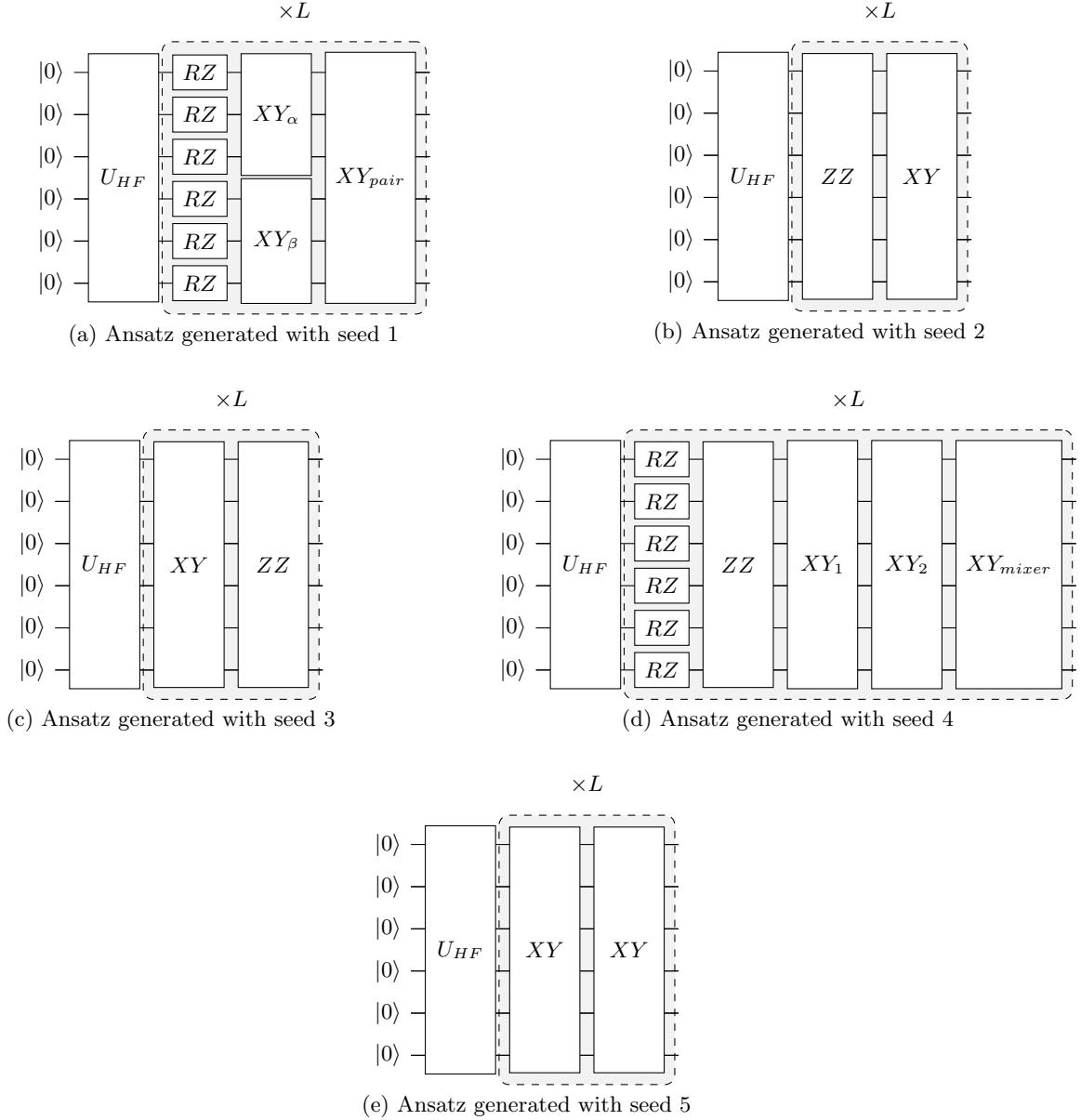
\begin{figure}[htbp]
\centering

\begin{minipage}[t]{0.45\textwidth}
    \centering
    \vspace{0pt}
    \begin{quantikz}[row sep={0.6cm,between origins}, column sep=0.2cm, thin lines]
        \lstick{$\ket{0}$} & \gate[wires=6][1.0cm]{U_{HF}} & \gate{RZ} \gategroup[wires=6,steps=3,
            style={dashed,rounded corners,fill=gray!10,
                   inner xsep=1pt,inner ysep=1pt},
            background,
            label style={label position=above, yshift=0.2cm}
        ]{$\times L$} & \gate[wires=3][1.0cm]{XY_{\alpha}} & \gate[wires=6][1.0cm]{XY_{pair}} & \qw \\
        \lstick{$\ket{0}$} & \qw & \gate{RZ} & \qw & \qw & \qw \\
        \lstick{$\ket{0}$} & \qw & \gate{RZ} & \qw & \qw & \qw \\
        \lstick{$\ket{0}$} & \qw & \gate{RZ} & \gate[wires=3][1.0cm]{XY_{\beta}} & \qw & \qw \\
        \lstick{$\ket{0}$} & \qw & \gate{RZ} & \qw & \qw & \qw \\
        \lstick{$\ket{0}$} & \qw & \gate{RZ} & \qw & \qw & \qw 
    \end{quantikz}
    \par\smallskip
    {\small (a) Ansatz generated with seed 1}
\end{minipage}%
\hspace{0.015\textwidth}%
\begin{minipage}[t]{0.45\textwidth}
    \centering
    \vspace{0pt}
    \begin{quantikz}[row sep={0.6cm,between origins}, column sep=0.2cm, thin lines]
        \lstick{$\ket{0}$} & \gate[wires=6][1.0cm]{U_{HF}} & \gate[wires=6][1.0cm]{ZZ} \gategroup[wires=6,steps=2,
            style={dashed,rounded corners,fill=gray!10,
                   inner xsep=1pt,inner ysep=1pt},
            background,
            label style={label position=above, yshift=0.2cm}
        ]{$\times L$} & \gate[wires=6][1.0cm]{XY} & \qw \\
        \lstick{$\ket{0}$} & \qw & \qw & \qw & \qw  \\
        \lstick{$\ket{0}$} & \qw & \qw & \qw & \qw  \\
        \lstick{$\ket{0}$} & \qw & \qw & \qw & \qw  \\
        \lstick{$\ket{0}$} & \qw & \qw & \qw & \qw  \\
        \lstick{$\ket{0}$} & \qw & \qw & \qw & \qw  
    \end{quantikz}
    \par\smallskip
    {\small (b) Ansatz generated with seed 2}
\end{minipage}

\vspace{1.5em}

\begin{minipage}[t]{0.4\textwidth}
    \centering
    \vspace{0pt}
    \begin{quantikz}[row sep={0.6cm,between origins}, column sep=0.2cm, thin lines]
        \lstick{$\ket{0}$} & \gate[wires=6][1.0cm]{U_{HF}} & \gate[wires=6][1.0cm]{XY} \gategroup[wires=6,steps=2,
            style={dashed,rounded corners,fill=gray!10,
                   inner xsep=1pt,inner ysep=1pt},
            background,
            label style={label position=above, yshift=0.2cm}
        ]{$\times L$} & \gate[wires=6][1.0cm]{ZZ} & \qw \\
        \lstick{$\ket{0}$} & \qw & \qw & \qw & \qw  \\
        \lstick{$\ket{0}$} & \qw & \qw & \qw & \qw  \\
        \lstick{$\ket{0}$} & \qw & \qw & \qw & \qw  \\
        \lstick{$\ket{0}$} & \qw & \qw & \qw & \qw  \\
        \lstick{$\ket{0}$} & \qw & \qw & \qw & \qw  
    \end{quantikz}
    \par\smallskip
    {\small (c) Ansatz generated with seed 3}
\end{minipage}%
\hspace{0.015\textwidth}%
\begin{minipage}[t]{0.55\textwidth}
    \centering
    \vspace{0pt}
    \begin{quantikz}[row sep={0.6cm,between origins}, column sep=0.2cm, thin lines]
        \lstick{$\ket{0}$} & \gate[wires=6][1.0cm]{U_{HF}} & \gate{RZ} \gategroup[wires=6,steps=5,
            style={dashed,rounded corners,fill=gray!10,
                   inner xsep=1pt,inner ysep=1pt},
            background,
            label style={label position=above, yshift=0.2cm}
        ]{$\times L$} & \gate[wires=6][1.0cm]{ZZ} & \gate[wires=6][1.0cm]{XY_{1}} & \gate[wires=6][1.0cm]{XY_{2}} & \gate[wires=6][1.0cm]{XY_{mixer}} & \qw \\
        \lstick{$\ket{0}$} & \qw & \gate{RZ} & \qw & \qw & \qw & \qw & \qw \\
        \lstick{$\ket{0}$} & \qw & \gate{RZ} & \qw & \qw & \qw & \qw & \qw \\
        \lstick{$\ket{0}$} & \qw & \gate{RZ} & \qw & \qw & \qw & \qw & \qw \\
        \lstick{$\ket{0}$} & \qw & \gate{RZ} & \qw & \qw & \qw & \qw & \qw \\
        \lstick{$\ket{0}$} & \qw & \gate{RZ} & \qw & \qw & \qw & \qw & \qw 
    \end{quantikz}
    \par\smallskip
    {\small (d) Ansatz generated with seed 4}
\end{minipage}

\vspace{1.5em}

\begin{minipage}[t]{0.95\textwidth}
    \centering
    \vspace{0pt}
    \begin{quantikz}[row sep={0.6cm,between origins}, column sep=0.2cm, thin lines]
        \lstick{$\ket{0}$} & \gate[wires=6][1.0cm]{U_{HF}} & \gate[wires=6][1.0cm]{XY} \gategroup[wires=6,steps=2,
            style={dashed,rounded corners,fill=gray!10,
                   inner xsep=1pt,inner ysep=1pt},
            background,
            label style={label position=above, yshift=0.2cm}
            ]{$\times L$} & \gate[wires=6][1.0cm]{XY} & \qw \\
        \lstick{$\ket{0}$} & \qw & \qw & \qw & \qw \\
        \lstick{$\ket{0}$} & \qw & \qw & \qw & \qw \\
        \lstick{$\ket{0}$} & \qw & \qw & \qw & \qw \\
        \lstick{$\ket{0}$} & \qw & \qw & \qw & \qw \\
        \lstick{$\ket{0}$} & \qw & \qw & \qw & \qw 
    \end{quantikz}
    \par\smallskip
    {\small (e) Ansatz generated with seed 5}
\end{minipage}

\caption{Circuit diagrams of the five seed ans\"{a}tze.}
\label{fig:seedAnsatze}
\end{figure}

For the experiment in which the best-performing result was obtained, we describe the details of the five seed ans\"{a}tze generated in the ``Exploration'' component.
Fig.~\ref{fig:seedAnsatze} shows the circuit diagrams of the five seed ans\"{a}tze using a six-qubit system as an illustrative example. 
Although the diagrams are shown for six qubits, the same construction principles apply to arbitrary system sizes.
All ans\"{a}tze share the same state preparation procedure.
Each ansatz prepares a variational state on $N$ qubits representing $N$ spin orbitals.
The spin orbitals are ordered such that the first \( N/2 \) qubits correspond to the \(\alpha\)-spin orbitals and the remaining \(N/2\) qubits correspond to the \(\beta\)-spin orbitals.
All circuits adopt a depth-\(L\) layered architecture initialized from the Hartree--Fock reference state in the computational basis, where the lowest-energy $n_{\alpha}$ and $n_{\beta}$ spin orbitals are occupied according to the electron count.

The source code of the five generated seed ans\"{a}tze is available in the artifact repository (\url{https://github.com/Qyusu/astronaut-artifact}) under
\texttt{version\_2\_0/generated\_code/best\_vqe\_results}.

\vspace{1em}

The seed 1 ansatz, shown in Fig.~\ref{fig:seedAnsatze} (a), consists of \(L\) repeated layers of single-qubit \(R_Z\) rotations followed by particle-number-conserving entangling operations.
In each layer, \(R_Z\) gates are applied to all qubits.
This is followed by excitation-preserving interactions implemented via \(IsingXY\) gates between selected occupied and virtual spin orbitals within each spin block.
The coupling pairs are selected according to the orbital index distance, and the corresponding interaction angles are scaled by normalized weights derived from this distance measure.
An optional across-spin pair channel applies corresponding \(IsingXY\) gates to paired \(\alpha\)- and \(\beta\)-spin orbital indices with an independent shared scaling parameter.

\vspace{1em}

The seed 2 ansatz, shown in Fig.~\ref{fig:seedAnsatze} (b), consists of \(L\) repeated layers comprising a diagonal correlation block followed by a particle-number-preserving excitation block.
In each layer, \(IsingZZ\) interactions introduce diagonal two-qubit correlations among selected qubit pairs, while a subsequent \(IsingXY\) block enables excitation exchange between selected occupied and virtual orbitals.
The interaction strengths are determined by shared variational parameters together with predefined pair-dependent weights derived from chemically motivated orbital relationships.

\vspace{1em}

The seed 3 ansatz, shown in Fig.~\ref{fig:seedAnsatze} (c), consists of \(L\) repeated layers that interleave excitation-preserving orbital rotations with diagonal correlation terms.
In each layer, \(IsingXY\) gates are applied between selected Hartree--Fock occupied and virtual spin orbitals within the same spin sector.
The coupling pairs are chosen according to distance-based weights derived from orbital indices, and the corresponding interaction angles are obtained by scaling these weights with a shared layer-wise variational parameter.
This is followed by a diagonal correlation layer implemented via \(IsingZZ\) gates acting on selected qubit pairs.
These pairs are chosen using distance- and spin-dependent weights, and all \(IsingZZ\) interactions within a layer share a common variational scaling parameter.

\vspace{1em}

The seed 4 ansatz, shown in Fig.~\ref{fig:seedAnsatze} (d), consists of \(L\) repeated layers composed of five parameterized blocks.
In each layer, single-qubit \(R_Z\) rotations are first applied to all qubits using orbital-dependent weights that distinguish Hartree--Fock occupied and virtual orbitals.
This is followed by an \(IsingZZ\) correlation block acting on selected same-spin neighboring orbitals and paired \(\alpha\)- and \(\beta\)-spin orbitals.
Two subsequent \(IsingXY\) blocks introduce excitation-preserving interactions on disjoint sets of same-spin occupied--virtual orbital pairs selected according to distance-based weights.
Finally, a particle-number-conserving mixer implemented with \(IsingXY\) gates further couples occupied and virtual orbitals through a sparse set of weighted interactions.
Each block is controlled by a shared variational parameter, while pair- and orbital-dependent weights determine the relative interaction strengths within the block.

\vspace{1em}

The seed 5 ansatz, shown in Fig.~\ref{fig:seedAnsatze} (e), consists of \(L\) repeated layers built upon a fragment-based architecture.
Qubits are partitioned into several groups according to a weighted interaction graph constructed from orbital distance, spin-sector information, and Hartree--Fock occupations.
In each layer, particle-number-conserving mixers are first applied within each fragment using \(IsingXY\) gates arranged in separate ring topologies for the \(\alpha\)- and \(\beta\)-spin sectors.
Each fragment mixer is controlled by a shared variational parameter.
Interactions between fragments are then introduced through additional \(IsingXY\) gates acting on selected same-spin orbital pairs belonging to different fragments.
These inter-fragment couplings use normalized pair-dependent weights and a shared variational parameter for each fragment pair.

\subsection{Best-performing ansatz}
\label{sec:vqeBestSolutions}
The trajectory of the lowest cost values obtained by applying our agentic system to the VQE task is shown in bottom of Fig.~\ref{fig:performance_trajectory}.
At trial 14, the system achieved best performance.
The generated code and logs that best-performing ansatz is shown in Listing ~\ref{lst:BestAnsatzCode}, ~\ref{lst:InitAnsatzIdea} and ~\ref{lst:ReflectedAnsatzIdea}. 
We note that this ansatz was generated in a fully automated manner by an LLM.
In the following, we describe its structural characteristics and design motifs.
However, understanding why the resulting circuit achieves strong performance for molecular ground-state estimation requires a separate theoretical analysis.

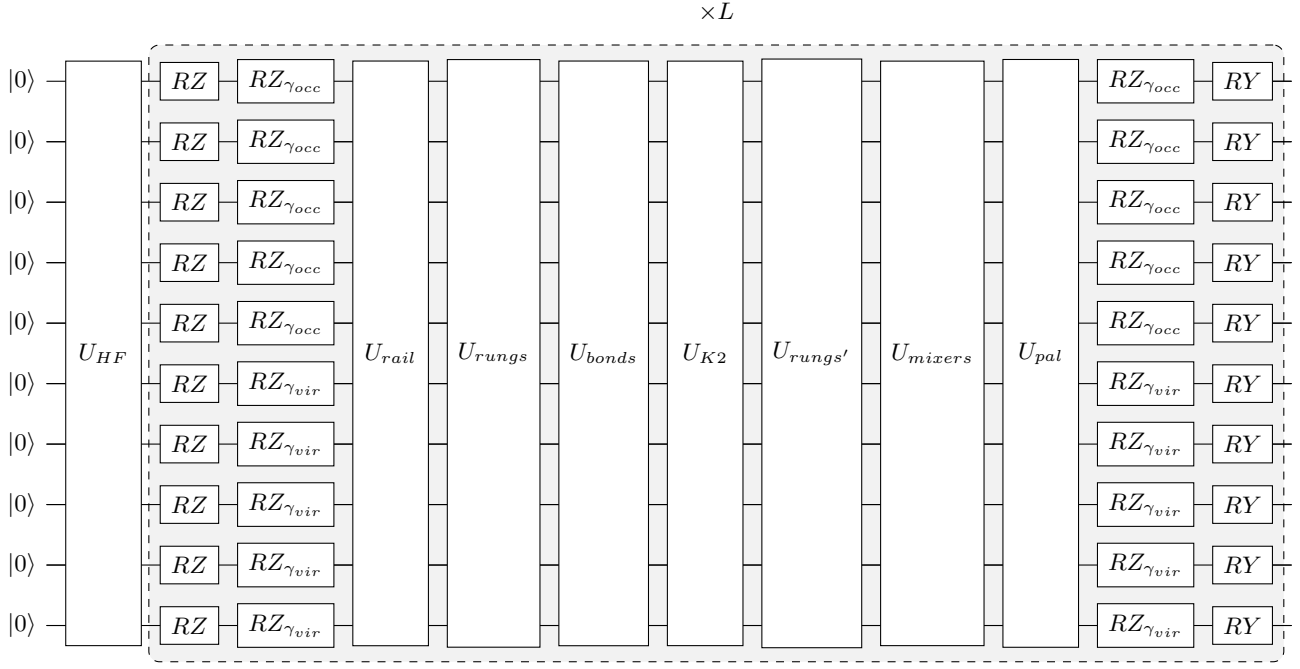
\begin{figure}[ht]
\centering

\begin{quantikz}[row sep={0.8cm,between origins}, column sep=0.25cm, thin lines]
    \lstick{$\ket{0}$} & \gate[wires=10][1.0cm]{U_{HF}} & \gate{RZ} \gategroup[wires=10,steps=11,
            style={dashed,rounded corners,fill=gray!10,
                   inner xsep=1pt,inner ysep=2pt},
            background,
            label style={label position=above, yshift=0.2cm}
        ]{$\times L$} & \gate{RZ_{\gamma_{occ}}} & \gate[wires=10][1.0cm]{U_{rail}} & \gate[wires=10][1.0cm]{U_{rungs}} & \gate[wires=10][1.0cm]{U_{bonds}} & \gate[wires=10][1.0cm]{U_{K2}} & \gate[wires=10][1.0cm]{U_{rungs'}} & \gate[wires=10][1.0cm]{U_{mixers}} & \gate[wires=10][1.0cm]{U_{pal}} & \gate{RZ_{\gamma_{occ}}} & \gate{RY} & \qw \\
    \lstick{$\ket{0}$} & \qw & \gate{RZ} & \gate{RZ_{\gamma_{occ}}} & \qw & \qw & \qw & \qw & \qw & \qw & \qw & \gate{RZ_{\gamma_{occ}}} & \gate{RY} & \qw \\
    \lstick{$\ket{0}$} & \qw & \gate{RZ} & \gate{RZ_{\gamma_{occ}}} & \qw & \qw & \qw & \qw & \qw & \qw & \qw & \gate{RZ_{\gamma_{occ}}} & \gate{RY} & \qw \\
    \lstick{$\ket{0}$} & \qw & \gate{RZ} & \gate{RZ_{\gamma_{occ}}} & \qw & \qw & \qw & \qw & \qw & \qw & \qw & \gate{RZ_{\gamma_{occ}}} & \gate{RY} & \qw \\
    \lstick{$\ket{0}$} & \qw & \gate{RZ} & \gate{RZ_{\gamma_{occ}}} & \qw & \qw & \qw & \qw & \qw & \qw & \qw & \gate{RZ_{\gamma_{occ}}} & \gate{RY} & \qw \\
    \lstick{$\ket{0}$} & \qw & \gate{RZ} & \gate{RZ_{\gamma_{vir}}} & \qw & \qw & \qw & \qw & \qw & \qw & \qw & \gate{RZ_{\gamma_{vir}}} & \gate{RY} & \qw \\
    \lstick{$\ket{0}$} & \qw & \gate{RZ} & \gate{RZ_{\gamma_{vir}}} & \qw & \qw & \qw & \qw & \qw & \qw & \qw & \gate{RZ_{\gamma_{vir}}} & \gate{RY} & \qw \\
    \lstick{$\ket{0}$} & \qw & \gate{RZ} & \gate{RZ_{\gamma_{vir}}} & \qw & \qw & \qw & \qw & \qw & \qw & \qw & \gate{RZ_{\gamma_{vir}}} & \gate{RY} & \qw \\
    \lstick{$\ket{0}$} & \qw & \gate{RZ} & \gate{RZ_{\gamma_{vir}}} & \qw & \qw & \qw & \qw & \qw & \qw & \qw & \gate{RZ_{\gamma_{vir}}} & \gate{RY} & \qw \\
    \lstick{$\ket{0}$} & \qw & \gate{RZ} & \gate{RZ_{\gamma_{vir}}} & \qw & \qw & \qw & \qw & \qw & \qw & \qw & \gate{RZ_{\gamma_{vir}}} & \gate{RY} & \qw
\end{quantikz}
\caption{Overview of the best ansatz.}
\label{fig:vqeCircuitOverall}
\end{figure}

\vspace{1em}

The corresponding quantum circuit diagram is shown in Fig.~\ref{fig:vqeCircuitOverall}. 
The diagram is drawn for a ten-qubit example for clarity; however, the same ansatz construction can be applied to arbitrary system sizes.
The ansatz implements a Heisenberg-exchange architecture on a ladder-like qubit topology, combining rail, rung, diagonal-bond, and second-neighbor interactions to capture correlations across multiple length scales while maintaining linear resource scaling in the number of qubits \( N \).
Each variational layer follows a structured palindromic pattern with independent parameters in the forward and backward passes.

Before applying the variational layers, the ansatz prepares a Hartree--Fock reference state by flipping the first \( n_{\mathrm{occ}} \) qubits with Pauli-\( X \) gates:
\[
\ket{\psi_\mathrm{HF}} = X^{\otimes n_{\mathrm{occ}}} \ket{0}^{\otimes Q}.
\]

This initialization embeds a chemically motivated occupancy pattern and typically accelerates convergence of the variational optimization.

\vspace{1em}

The qubit register is partitioned into two rails, denoted \( \text{rail}_A \) and \( \text{rail}_B \), nearly equal length,
\(\text{rail}_A = \{0,\dots,\lfloor Q/2 \rfloor - 1\}\) and
\(\text{rail}_B = \{\lfloor Q/2 \rfloor,\dots,Q-1\}\).
Let \(A_i\) and \(B_i\) denote the \(i\)th indexes on \( \text{rail}_A \) and \( \text{rail}_B \), respectively.

From these, deterministic edge sets are constructed:
\[
\mathcal{E}_\mathrm{rail}, \quad
\mathcal{E}_\mathrm{rungs}, \quad
\mathcal{E}_\mathrm{bonds}, \quad
\mathcal{E}_\mathrm{K2},
\]

The set \(\mathcal{E}_\mathrm{rail}\) contains nearest-neighbor pairs along each rail.
The rung set \(\mathcal{E}_\mathrm{rungs}\) contains vertical couplings \((A_i,B_i)\)
up to the shorter rail length.
The diagonal bond set \(\mathcal{E}_\mathrm{bonds}\) contains cross-rail pairs
\((A_i,B_{i+1})\) and \((A_{i+1},B_i)\) wherever defined.
The set \(\mathcal{E}_\mathrm{K2}\) contains second-neighbor pairs along each rail.

For the mixer stage, we further partition the rail nearest-neighbor set \(\mathcal{E}_\mathrm{rail}\) into two disjoint matchings, \(\mathcal{M}_\mathrm{even}\) and \(\mathcal{M}_\mathrm{odd}\), by selecting alternating bonds on each rail.
Specifically, \(\mathcal{M}_\mathrm{even}\) contains the pairs \((A_i,A_{i+1})\) and \((B_i,B_{i+1})\) with even \(i\), whereas \(\mathcal{M}_\mathrm{odd}\) contains those with odd \(i\).
By construction, the edges within each matching are non-overlapping and can therefore be applied in parallel.

\vspace{1em}

Each variational layer \(\mathcal{L}_\ell\) follows the ordered structure:
\[
\mathcal{L}_\ell =
\mathcal{R}_\mathrm{pre} \rightarrow
\mathcal{H}_\mathrm{forward} \rightarrow
\Gamma_\pi \rightarrow
\mathcal{M}_\mathrm{mix} \rightarrow
\mathcal{H}_\mathrm{backward} \rightarrow
\mathcal{R}_\mathrm{post}.
\]

The components of \(\mathcal{L}_\ell\) are detailed below.

\begin{figure}[htbp]
\centering

\begin{minipage}[t]{0.25\textwidth}
    \centering
    \vspace{0pt}
    \begin{quantikz}[row sep={0.6cm,between origins}, column sep=0.2cm, thin lines]
        \lstick{$\ket{0}$} & \ctrl{1} & \qw & \qw \\
        \lstick{$\ket{0}$} & \gate{U(\theta)} & \qw & \qw \\
        \lstick{$\ket{0}$} & \qw & \ctrl{1} & \qw \\
        \lstick{$\ket{0}$} & \qw & \gate{U(\theta)} & \qw \\
        \lstick{$\ket{0}$} & \qw & \qw & \qw \\
        \lstick{$\ket{0}$} & \ctrl{1} & \qw & \qw \\
        \lstick{$\ket{0}$} & \gate{U(\theta)} & \qw & \qw \\
        \lstick{$\ket{0}$} & \qw & \ctrl{1} & \qw \\
        \lstick{$\ket{0}$} & \qw & \gate{U(\theta)} & \qw \\
        \lstick{$\ket{0}$} & \qw & \qw & \qw
    \end{quantikz}
    \par\smallskip
    {\small (a) Detailed circuit of $U_{\text{rails}}$}
\end{minipage}%
\hspace{0.015\textwidth}%
\begin{minipage}[t]{0.45\textwidth}
    \centering
    \vspace{0pt}
    \begin{quantikz}[row sep={0.6cm,between origins}, column sep=0.2cm, thin lines]
        \lstick{$\ket{0}$} & \ctrl{5} & \qw & \qw & \qw & \qw & \qw \\
        \lstick{$\ket{0}$} & \qw & \ctrl{5} & \qw & \qw & \qw & \qw \\
        \lstick{$\ket{0}$} & \qw & \qw & \ctrl{5} & \qw & \qw & \qw \\
        \lstick{$\ket{0}$} & \qw & \qw & \qw & \ctrl{5} & \qw & \qw \\
        \lstick{$\ket{0}$} & \qw & \qw & \qw & \qw & \ctrl{5} & \qw \\
        \lstick{$\ket{0}$} & \gate{U(\theta)} & \qw & \qw & \qw & \qw & \qw \\
        \lstick{$\ket{0}$} & \qw & \gate{U(\theta)} & \qw & \qw & \qw & \qw \\
        \lstick{$\ket{0}$} & \qw & \qw & \gate{U(\theta)} & \qw & \qw & \qw \\
        \lstick{$\ket{0}$} & \qw & \qw & \qw & \gate{U(\theta)} & \qw & \qw \\
        \lstick{$\ket{0}$} & \qw & \qw & \qw & \qw & \gate{U(\theta)} & \qw
    \end{quantikz}
    \par\smallskip
    {\small (b) Detailed circuit of $U_{\text{rungs}}$}
\end{minipage}%
\hspace{0.015\textwidth}%
\begin{minipage}[t]{0.25\textwidth}
    \centering
    \vspace{0pt}
    \begin{quantikz}[row sep={0.6cm,between origins}, column sep=0.2cm, thin lines]
        \lstick{$\ket{0}$} & \ctrl{2} & \qw & \qw \\
        \lstick{$\ket{0}$} & \qw & \ctrl{2} & \qw \\
        \lstick{$\ket{0}$} & \gate{U(\theta)} & \qw & \qw \\
        \lstick{$\ket{0}$} & \qw & \gate{U(\theta)} & \qw \\
        \lstick{$\ket{0}$} & \qw & \qw & \qw \\
        \lstick{$\ket{0}$} & \ctrl{2} & \qw & \qw \\
        \lstick{$\ket{0}$} & \qw & \ctrl{2} & \qw \\
        \lstick{$\ket{0}$} & \gate{U(\theta)} & \qw & \qw \\
        \lstick{$\ket{0}$} & \qw & \gate{U(\theta)} & \qw \\
        \lstick{$\ket{0}$} & \qw & \qw & \qw
    \end{quantikz}
    \par\smallskip
    {\small (c) Detailed circuit of $U_{K2}$}
\end{minipage}

\vspace{1.5em}

\begin{minipage}[t]{0.7\textwidth}
    \centering
    \vspace{0pt}
    \begin{quantikz}[row sep={0.6cm,between origins}, column sep=0.1cm, thin lines]
        \lstick{$\ket{0}$} & \ctrl{6} & \qw & \qw & \qw & \qw & \qw & \qw & \qw & \qw \\
        \lstick{$\ket{0}$} & \qw & \gate{U(\theta)} & \ctrl{6} & \qw & \qw & \qw & \qw & \qw & \qw \\
        \lstick{$\ket{0}$} & \qw & \qw & \qw & \gate{U(\theta)} & \ctrl{6} & \qw & \qw & \qw & \qw \\
        \lstick{$\ket{0}$} & \qw & \qw & \qw & \qw & \qw & \gate{U(\theta)} & \ctrl{6} & \qw & \qw \\
        \lstick{$\ket{0}$} & \qw & \qw & \qw & \qw & \qw & \qw & \qw & \gate{U(\theta)} & \qw \\
        \lstick{$\ket{0}$} & \qw & \ctrl{-4} & \qw & \qw & \qw & \qw & \qw & \qw & \qw \\
        \lstick{$\ket{0}$} & \gate{U(\theta)} & \qw & \qw & \ctrl{-4} & \qw & \qw & \qw & \qw & \qw \\
        \lstick{$\ket{0}$} & \qw & \qw & \gate{U(\theta)} & \qw & \qw & \ctrl{-5} & \qw & \qw & \qw \\
        \lstick{$\ket{0}$} & \qw & \qw & \qw & \qw & \gate{U(\theta)} & \qw & \qw & \ctrl{-4} & \qw \\
        \lstick{$\ket{0}$} &  \qw & \qw & \qw & \qw & \qw & \qw & \gate{U(\theta)} & \qw & \qw
    \end{quantikz}
    \par\smallskip
    {\small (d) Detailed circuit of $U_{\text{bonds}}$}
\end{minipage}%
\hspace{0.015\textwidth}%
\begin{minipage}[t]{0.25\textwidth}
    \centering
    \vspace{0pt}
    \begin{quantikz}[row sep={0.6cm,between origins}, column sep=0.2cm, thin lines]
        \lstick{$\ket{0}$} & \ctrl{1} & \qw & \qw \\
        \lstick{$\ket{0}$} & \gate{U(\theta)} & \ctrl{1} & \qw \\
        \lstick{$\ket{0}$} & \ctrl{1} & \gate{U(\theta)} & \qw \\
        \lstick{$\ket{0}$} & \gate{U(\theta)} & \ctrl{1} & \qw \\
        \lstick{$\ket{0}$} & \qw & \gate{U(\theta)} & \qw \\
        \lstick{$\ket{0}$} & \ctrl{1} & \qw & \qw \\
        \lstick{$\ket{0}$} & \gate{U(\theta)} & \ctrl{1} & \qw \\
        \lstick{$\ket{0}$} & \ctrl{1} & \gate{U(\theta)} & \qw \\
        \lstick{$\ket{0}$} & \gate{U(\theta)} & \ctrl{1} & \qw \\
        \lstick{$\ket{0}$} & \qw & \gate{U(\theta)} & \qw
    \end{quantikz}
    \par\smallskip
    {\small (e) Detailed circuit of $U_{mixers}$}
\end{minipage}

\caption{Detailed circuits of $\mathcal{H}_\mathrm{forward}$.}
\label{fig:VQECircuitParts}
\end{figure}
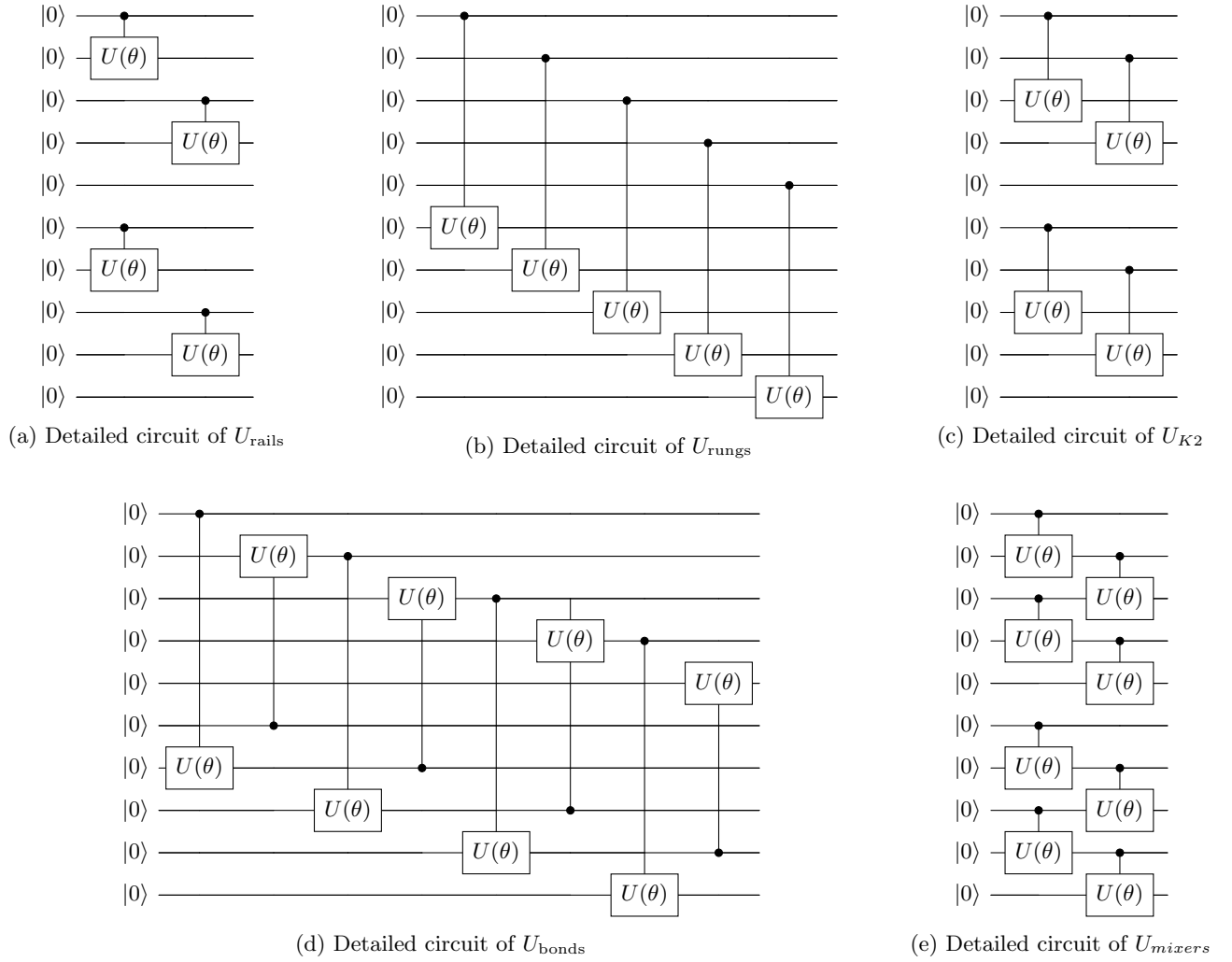

\begin{enumerate}
    \item \textbf{Pre-rotations (\(\mathcal{R}_\mathrm{pre}\)).}  
    Each qubit $q$ undergoes a local phase rotation $R_Z(\phi_q)$. 
    In addition, two shared parameters, $\gamma_\mathrm{occ}$ and $\gamma_\mathrm{vir}$, generate collective phase rotations $R_Z(0.5\gamma_\mathrm{occ})$ on the first $n_\mathrm{occ}$ qubits and
    $R_Z(0.5\gamma_\mathrm{vir})$ on the remaining qubits.

    \item \textbf{Forward Heisenberg exchange entanglers (\(\mathcal{H}_\mathrm{forward}\)).}\\
    The blocks \(U_{rail}\), \(U_{rungs}\), \(U_{bonds}\), and \(U_{K2}\) shown in Fig.~\ref{fig:vqeCircuitOverall} implement the forward Heisenberg-exchange entanglers.
    For each edge \((i,j)\) in \(\{\mathcal{E}_\mathrm{rail}, \mathcal{E}_\mathrm{rungs}, \mathcal{E}_\mathrm{bonds}, \mathcal{E}_\mathrm{K2}\}\),
    independent parameters \(\theta_{ij}\) control the exchange operation
    \[
    U_{ij}(\theta_{ij}) = e^{-i\frac{\theta_{ij}}{4}(X_i X_j + Y_i Y_j + Z_i Z_j)}.
    \]
    This operation is implemented via three Pauli rotations:
    \(R_{XX}(\tfrac{\theta}{2}) R_{YY}(\tfrac{\theta}{2}) R_{ZZ}(\tfrac{\theta}{2})\).
    Each edge group constructs the circuit structures shown in Fig.~\ref{fig:VQECircuitParts}.
    Edge \(\mathcal{E}_\mathrm{rail}\) connects adjacent qubits within the same rail, as illustrated in Fig~\ref{fig:VQECircuitParts} (a).
    Edge \(\mathcal{E}_\mathrm{rungs}\) connects qubits with the same index between different rails, as shown in Fig~\ref{fig:VQECircuitParts} (b).
    Edge \(\mathcal{E}_\mathrm{K2}\) connects qubits separated by a distance of two within the same rail, as depicted in Fig~\ref{fig:VQECircuitParts} (c).
    Edge \(\mathcal{E}_\mathrm{bonds}\) connects qubits between different rails to form diagonal bonds, as illustrated in Fig~\ref{fig:VQECircuitParts} (d).

    \item \textbf{Rung enhancement (\(\Gamma_\pi\)).} 
    The block \(U_{rungs'}\) in Fig.~\ref{fig:vqeCircuitOverall} provides an additional shared enhancement of the rung couplings.
    A shared parameter \(\gamma_\pi\) reinforces all rung couplings as
    \(U_{ij}(\gamma_\pi)\) for \((i,j)\in\mathcal{E}_\mathrm{rungs}\). The circuit is identical to \(U_{rungs}\) in Fig~\ref{fig:VQECircuitParts} (b).

    \item \textbf{Mixer stage (\(\mathcal{M}_\mathrm{mix}\)).}
    The block \(U_{mixers}\) in Fig.~\ref{fig:vqeCircuitOverall} implements the mixer stage.
    Two shared angles \(\beta_\mathrm{even}\) and \(\beta_\mathrm{odd}\)
    generate alternating Heisenberg-exchanges over even and odd matchings:
    \[
    \prod_{(i,j)\in\mathcal{M}_\mathrm{even}} U_{ij}(\beta_\mathrm{even})
    \quad\text{and}\quad
    \prod_{(i,j)\in\mathcal{M}_\mathrm{odd}} U_{ij}(\beta_\mathrm{odd}),
    \]
    spreading correlations along each rail in an alternating parity pattern. The circuit structure is shown in Fig.~\ref{fig:VQECircuitParts} (e).

    \item \textbf{Palindromic backward pass (\(\mathcal{H}_\mathrm{backward}\)).}
    The \(U_{pal}\) block shown in Fig.~\ref{fig:vqeCircuitOverall} re-applies the exchange-entangler groups in the reverse order of the forward pass (\(\mathcal{H}_\mathrm{forward}\)), using an independent set of variational parameters. This palindromic construction yields a structurally symmetric layer while preserving the flexibility to optimize the backward couplings separately.

    \item \textbf{Post-rotations (\(\mathcal{R}_\mathrm{post}\)).}  
    The same \(R_Z(0.5\,\gamma_\mathrm{occ})\) and \(R_Z(0.5\,\gamma_\mathrm{vir})\) rotations are applied again to complete the collective phases, followed by \(R_Y(\phi_q)\) rotations on all qubits, thereby completing one variational layer.

\end{enumerate}

Because each edge set contains a number of pairs that scales linearly with the number of qubits \(N\), both the number of variational parameters and the total gate count per layer scale as \(\mathcal{O}(N)\).
For \(L\) variational layers, the overall parameter count and gate complexity
therefore scale as \(\mathcal{O}(LN)\).
Thus, the ansatz exhibits linear resource growth in the system size.

Further details of the implementation, including the configuration of hyperparameters, can be found in the source code provided in Listing~\ref{lst:BestAnsatzCode}.

\vspace{1em}

\begin{lstlisting}[style=codeFormat, caption={Best performing ansatz code generated by our agentic system}, label=lst:BestAnsatzCode]
from typing import cast, List, Tuple

import pennylane as qml
from qxmt.ansatze import BaseVQEAnsatz
from qxmt.devices import BaseDevice
from qxmt.hamiltonians.pennylane.molecular import MolecularHamiltonian

# above are the default imports. DO NOT REMOVE THEM.
# new imports can be added below this line if needed.


class SAGARAGGLadderBondHeisK2PalUTV2(BaseVQEAnsatz):
    """SAGA-RA-GG-Ladder-Bond-Heis-K2-Pal-UT (v2)

    Compiled Heisenberg SAGA layer with ladder rails, rungs, diagonal bonds, and
    second-neighbor (K2) couplers. Uses palindromic untied parameters and HF initialization.
    Defaults are chosen for strong accuracy while keeping resource growth linear in qubits.

    Args:
        device: execution device wrapper
        hamiltonian: molecular Hamiltonian
        depth: number of layers; if None, set adaptively: max(2, min(10, n_qubits//6 + 3))
        include_rail_nn: include nearest-neighbor couplings along rails
        include_rungs: include rung couplings between rails
        include_bonds: include diagonal bond couplings across rails
        include_k2: include second-neighbor couplings along rails
        include_mixers: include even/odd rail matchings with shared angles
        palindrome: enable palindromic structure with untied parameters for the backward pass
        init_hf: prepare Hartree-Fock reference by flipping first n_occ wires
    """

    def __init__(
        self,
        device: BaseDevice,
        hamiltonian: MolecularHamiltonian,
        depth: int | None = None,
        include_rail_nn: bool = True,
        include_rungs: bool = True,
        include_bonds: bool = True,
        include_k2: bool = True,
        include_mixers: bool = True,
        palindrome: bool = True,
        init_hf: bool = True,
    ) -> None:
        super().__init__(device, hamiltonian)
        self.hamiltonian = cast(MolecularHamiltonian, self.hamiltonian)

        # System sizes (use only allowed accessors)
        self.n_qubits = self.hamiltonian.get_n_qubits()
        try:
            n_act_elec = self.hamiltonian.get_active_electrons()
        except Exception:
            n_act_elec = 0
        if not isinstance(n_act_elec, int) or n_act_elec <= 0:
            n_act_elec = self.hamiltonian.get_electrons()
        self.n_occ = max(0, min(self.n_qubits, int(n_act_elec)))

        # Hyperparameters
        if depth is None:
            self.depth = max(2, min(10, (self.n_qubits // 6) + 3))
        else:
            self.depth = max(1, int(depth))
        self.include_rail_nn = bool(include_rail_nn)
        self.include_rungs = bool(include_rungs)
        self.include_bonds = bool(include_bonds)
        self.include_k2 = bool(include_k2)
        self.include_mixers = bool(include_mixers)
        self.palindrome = bool(palindrome)
        self.init_hf = bool(init_hf)

        # Ladder partition and edge lists
        self.rail_A, self.rail_B = self._build_rails(self.n_qubits)
        self.edges_rail: List[Tuple[int, int]] = []
        self.edges_rungs: List[Tuple[int, int]] = []
        self.edges_bonds: List[Tuple[int, int]] = []
        self.edges_k2: List[Tuple[int, int]] = []
        self.match_even: List[Tuple[int, int]] = []
        self.match_odd: List[Tuple[int, int]] = []
        self._build_edges()

        # Count per-layer parameters
        edges_total = 0
        if self.include_rail_nn:
            edges_total += len(self.edges_rail)
        if self.include_rungs:
            edges_total += len(self.edges_rungs)
        if self.include_bonds:
            edges_total += len(self.edges_bonds)
        if self.include_k2:
            edges_total += len(self.edges_k2)

        # Per-layer params: pre RZ (n) + occ/vir (2) + forward edges (E)
        # + gamma_pi (1) + mixers (2 if enabled) + backward edges (E if palindrome)
        # + post RY (n)
        mixer_params = 2 if self.include_mixers else 0
        self.params_per_layer = (
            self.n_qubits  # pre RZ
            + 2  # occ/vir scalars
            + edges_total  # forward per-edge Heisenberg angles
            + 1  # gamma_pi on rungs
            + mixer_params  # beta_even, beta_odd
            + (edges_total if self.palindrome else 0)  # backward per-edge angles (untied)
            + self.n_qubits  # post RY
        )

        self.n_params = self.depth * self.params_per_layer

    # ----------------------- helpers -----------------------
    def _build_rails(self, n: int) -> tuple[list[int], list[int]]:
        len_A = n // 2
        len_B = n - len_A
        rail_A = list(range(0, len_A))
        rail_B = list(range(len_A, len_A + len_B))
        return rail_A, rail_B

    def _pairs_consecutive(self, rail: list[int]) -> list[tuple[int, int]]:
        return [(rail[i], rail[i + 1]) for i in range(len(rail) - 1)]

    def _pairs_second_neighbor(self, rail: list[int]) -> list[tuple[int, int]]:
        return [(rail[i], rail[i + 2]) for i in range(len(rail) - 2)]

    def _rail_matching(self, rail: list[int], parity: int) -> list[tuple[int, int]]:
        pairs: list[tuple[int, int]] = []
        start = parity
        for i in range(start, len(rail) - 1, 2):
            pairs.append((rail[i], rail[i + 1]))
        return pairs

    def _build_edges(self) -> None:
        # Rail nearest neighbors
        rail_nn_A = self._pairs_consecutive(self.rail_A)
        rail_nn_B = self._pairs_consecutive(self.rail_B)
        self.edges_rail = rail_nn_A + rail_nn_B

        # Rungs up to min length
        rung_len = min(len(self.rail_A), len(self.rail_B))
        self.edges_rungs = [(self.rail_A[i], self.rail_B[i]) for i in range(rung_len)]

        # Diagonal bonds
        bonds: list[tuple[int, int]] = []
        for i in range(rung_len - 1):
            bonds.append((self.rail_A[i], self.rail_B[i + 1]))
            bonds.append((self.rail_A[i + 1], self.rail_B[i]))
        self.edges_bonds = bonds

        # Second neighbors along rails
        k2_A = self._pairs_second_neighbor(self.rail_A)
        k2_B = self._pairs_second_neighbor(self.rail_B)
        self.edges_k2 = k2_A + k2_B

        # Even/Odd matchings per rail
        self.match_even = self._rail_matching(self.rail_A, 0) + self._rail_matching(self.rail_B, 0)
        self.match_odd = self._rail_matching(self.rail_A, 1) + self._rail_matching(self.rail_B, 1)

    def _heisenberg_exch(self, theta: float, i: int, j: int) -> None:
        # Implements exp(-i theta S_i・S_j) = exp(-i theta/4 (XX+YY+ZZ))
        half = theta / 2.0  # because PauliRot applies exp(-i theta/2 P)
        qml.PauliRot(theta=half, pauli_word="XX", wires=[i, j])
        qml.PauliRot(theta=half, pauli_word="YY", wires=[i, j])
        qml.PauliRot(theta=half, pauli_word="ZZ", wires=[i, j])

    def _apply_edges_parametric(self, params: qml.numpy.ndarray, start_idx: int, edges: List[Tuple[int, int]]) -> int:
        idx = start_idx
        for (i, j) in edges:
            self._heisenberg_exch(theta=params[idx], i=i, j=j)
            idx += 1
        return idx

    # ----------------------- circuit -----------------------
    def circuit(self, params: qml.numpy.ndarray) -> None:
        """Heisenberg SAGA ladder-bond K2 palindromic ansatz.

        Args:
            params (qml.numpy.ndarray): flattened parameter vector of shape (n_params,).
        """
        idx = 0

        # HF reference (|1> on first n_occ)
        if self.init_hf:
            for w in range(self.n_occ):
                qml.PauliX(wires=w)

        for _ in range(self.depth):
            # Pre single-qubit phases
            for w in range(self.n_qubits):
                qml.RZ(phi=params[idx], wires=w)
                idx += 1

            # Occupied/Virtual collective phases (half applied now, half later)
            gamma_occ = params[idx]
            idx += 1
            gamma_vir = params[idx]
            idx += 1
            for w in range(self.n_occ):
                qml.RZ(phi=0.5 * gamma_occ, wires=w)
            for w in range(self.n_occ, self.n_qubits):
                qml.RZ(phi=0.5 * gamma_vir, wires=w)

            # Build edge groups in fixed order for determinism
            groups: list[List[Tuple[int, int]]] = []
            if self.include_rail_nn:
                groups.append(self.edges_rail)
            if self.include_rungs:
                groups.append(self.edges_rungs)
            if self.include_bonds:
                groups.append(self.edges_bonds)
            if self.include_k2:
                groups.append(self.edges_k2)

            # Forward entanglers (per-edge independent angles)
            for edges in groups:
                idx = self._apply_edges_parametric(params=params, start_idx=idx, edges=edges)

            # Shared pi-bond enhancement on rungs (single angle across all rungs)
            gamma_pi = params[idx]
            idx += 1
            if self.include_rungs and len(self.edges_rungs) > 0:
                for (i, j) in self.edges_rungs:
                    self._heisenberg_exch(theta=gamma_pi, i=i, j=j)

            # Mixers (even/odd matchings) with shared angles
            if self.include_mixers:
                beta_even = params[idx]
                idx += 1
                if len(self.match_even) > 0:
                    for (i, j) in self.match_even:
                        self._heisenberg_exch(theta=beta_even, i=i, j=j)
                beta_odd = params[idx]
                idx += 1
                if len(self.match_odd) > 0:
                    for (i, j) in self.match_odd:
                        self._heisenberg_exch(theta=beta_odd, i=i, j=j)

            # Palindromic backward entanglers with untied parameters
            if self.palindrome:
                for edges in reversed(groups):
                    idx = self._apply_edges_parametric(params=params, start_idx=idx, edges=edges)

            # Apply remaining half of collective phases
            for w in range(self.n_occ):
                qml.RZ(phi=0.5 * gamma_occ, wires=w)
            for w in range(self.n_occ, self.n_qubits):
                qml.RZ(phi=0.5 * gamma_vir, wires=w)

            # Post single-qubit mixers
            for w in range(self.n_qubits):
                qml.RY(phi=params[idx], wires=w)
                idx += 1
\end{lstlisting}

\vspace{1em}

The generated logs of the best-performing ansatz by the ``Generation'' component are shown in Listings~\ref{lst:InitAnsatzIdea} and \ref{lst:ReflectedAnsatzIdea}.
The listings are reproduced as raw LLM outputs for transparency and reproducibility.
The log in Listing~\ref{lst:InitAnsatzIdea} represents the initial result obtained in this trial, whereas the log of Listing~\ref{lst:ReflectedAnsatzIdea} shows the refined result produced by ``Discussion'' component. 
The red text in the refined ansatz formula highlights the differences between the initial and refined ideas.
In the final version of the ansatz, the overall structural foundation is preserved, but its internal composition has been systematically refined to improve physical correspondence.
The refined ansatz incorporates spin-symmetry-preserving Heisenberg-exchange interactions in its entangling blocks,
\(
U_{pq}(\theta)=\exp\!\left(-i\theta\,\mathbf{S}_p\cdot\mathbf{S}_q\right),
\)
whose generators preserve full SU(2) spin symmetry and commute with the total spin operator \(S^2\).
As a result, the entangling layers individually preserve spin symmetry, while the overall variational circuit retains additional single-qubit rotations that provide greater expressive flexibility.

\vspace{1em}

\begin{lstlisting}[style=txtFormat, caption={Initial ansatz idea generated by our agentic system}, label=lst:InitAnsatzIdea]
# name
SAGA-RA-GG-Ladder-Bond-K2-Pal-UT: Grouped Gains and Grouped Sector-wise Rank-Adaptive Pair, Pair-Ladder, Bonding Blocks (Untied)


# summary
This grouped variant targets larger actives and high-symmetry systems while retaining the SAGA skeleton and the three upgrades from Idea 1. Mixer gains are grouped per occupied irrep (or energy band), and rank-adaptive pair factors are defined at the sector-group level and broadcast to orbitals. The sector-wise pair-ladder (even/odd) and bond-localized U_S blocks are kept. Pair and mixer halves remain untied to avoid first-order cancellations. The design preserves O(N) resources and strict symmetries, reducing parameter count without sacrificing the long-bond and π-manifold improvements.


# explanation
Grouped controls vs per-orbital
- Mixer gains: replace {g_p} with {g_ρ}, one gain per occupied irrep ρ (or energy band if irreps absent); keep sector-wise two-bucket scales {γ_{ρ,near,*, ℓ}, γ_{ρ,far,*, ℓ}} untied forward/backward.
- Grouped rank-adaptive pair: in each sector ρ, compress the signed pair matrix to group space; choose r_ρ≤3 covering ≥95% Frobenius norm; use group factor vectors {s_ρ^{(k,ℓ)}} orthonormalized over groups and mean-one rescaled; broadcast to orbitals for angles on sparse G_ρ.

Per-layer ℓ (untied palindrome)
0) Rz phases on spatial orbitals;
1) U_occ(γ_occ/2), U_vir(γ_vir/2);
2) Always-on K=2 backbone ∏_p U_S(p,q_m*(p); θ_p);
3) π-subspace chain with shared γ_π;
4) Grouped sector-wise pair (forward): θ_{pq}^{pair,fwd}=Σ_{k=1}^{r_ρ} γ_{ρ,k,fwd} s_{ρ(p)}^{(k)} s_{ρ(q)}^{(k)} for (p,q)∈G_ρ;
5) Sector-wise pair-ladder: even/odd nearest-neighbor matchings per sector with shared β_even, β_odd;
6) Mixer forward: θ_{p→q}^{fwd}=g_{ρ(p)} γ_{ρ,near,fwd} ŵ^{near}_{p|q} or g_{ρ(p)} γ_{ρ,far,fwd} ŵ^{far}_{p|q};
7) Mixer backward with γ_{ρ,near,bwd}, γ_{ρ,far,bwd};
8) Pair backward with {γ_{ρ,k,bwd}};
9) Backbone echo;
10) U_vir(γ_vir/2), U_occ(γ_occ/2);
11) Bond-localized U_S per detected bond with per-bond angles {θ_b}.

Trainable parameters per layer ℓ (all real)
- Phases: {φ_r} → N_sp. Base rings: γ_occ, γ_vir → 2. Backbone: {θ_p} → N_occ_sp.
- π-block: γ_π → 1. Grouped pair: r_ρ group factors {s_ρ^{(k)}} → Σ_ρ r_ρ; sector pair scales {γ_{ρ,k,fwd}}, {γ_{ρ,k,bwd}} → 2Σ_ρ r_ρ.
- Pair-ladder: β_even, β_odd (tied across halves) → 2. Mixer: per-sector bucket scales {γ_{ρ,near,*,}, γ_{ρ,far,*,}} (4·#sectors), per-irrep gains {g_ρ} → R_occ. Bond-local: {θ_b} for b=1..B.
Typical per-layer total: N_sp + N_occ_sp + Σ_ρ r_ρ + R_occ + B + (2Σ_ρ r_ρ + 7 + 4·#sectors) = O(N) with substantially fewer parameters than per-orbital factors when R_occ and #sectors≪N_sp.

Automatic hyperparameters
- Irreps from HF; else K_b=min(6,⌈N_sp/6⌉) energy bands. s_pair=2 per occupied; ξ=0.95, r_max=3 per sector; s=3; ε_cut=0.05 a.u.; Δε_cut 40th percentile. Backbone partners prefer same irrep. Ladder matchings from energy order in each sector. Rings default to two-matchings for parallel scheduling.

Initialization & scaling
- All angles uniform in [0.0,0.1]; g_ρ=1.0±0.05 clipped to [0.3,1.7]. s_ρ^{(k)} seeded from top-r_ρ eigenvectors +5% noise; orthonormalize and mean-one rescale on each set_parameters. U_S ≈ 4 CNOTs; per-layer gates O(N). Strict N and S^2; fixed, device-agnostic architecture; compatible with SLSQP warm-starts.


# key_sentences
- SAGA-RA-GG-Ladder-Bond-K2-Pal-UT groups the near/far mixer gains by irrep and the rank-adaptive pair factors by sector groups, lowering parameters while retaining sector-wise rank-adaptive pairs (r≤3), a two-matching pair-ladder, and bond-localized matchgates, all within the proven SAGA skeleton.
- Untied forward/backward scales for both the mixer and pair blocks avoid first-order cancellations while preserving palindromic second-order effects; the K=2 backbone guarantees non-commuting coverage and robust gradients across all geometries.
- Group factor vectors are derived from signed sector pair matrices, orthonormalized and mean-one rescaled at each update, then broadcast to orbitals on sparse pair sets; this captures dominant static-correlation patterns in σ and π manifolds at O(N) cost.
- Irrep-aware two-bucket mixing with normalized weights and per-irrep gains focuses singles amplitude on symmetry-allowed, chemically strong couplings without per-edge parameters; π-subspace chains and bond-local couplers further target degeneracies and dissociation channels.
- All parameters are real; two-qubit gates and parameters scale O(N) per layer; the construction is non-adaptive, device-agnostic, and strictly preserves particle number and S^2, addressing residual long-bond and π-manifold errors without densifying the edge set.

\end{lstlisting}
\begin{tcolorbox}[colframe=black, colback=white, boxrule=0.5pt,
                  left=2mm, right=2mm, top=1mm, bottom=1mm,
                  title={Initial Ansatz Formula}]
\begin{align*}
|\Psi(\boldsymbol{\theta})\rangle 
   = \prod_{\ell=1}^{L} \Bigg[ 
     &\Big(\prod_{r} R_{z,r}(\phi_r^{(\ell)})\Big)\,
      U_{\mathrm{occ}}\!\left(\tfrac{\gamma_{\mathrm{occ}}^{(\ell)}}{2}\right)\,
      U_{\mathrm{vir}}\!\left(\tfrac{\gamma_{\mathrm{vir}}^{(\ell)}}{2}\right) \\[1ex]
   &\times \Big(\prod_{p\in\mathrm{occ}} 
      e^{-i\,\theta_p^{(\ell)} \sum_{\sigma} \tfrac{X_{p\sigma}X_{q_m^*(p)\sigma}+Y_{p\sigma}Y_{q_m^*(p)\sigma}}{2}}\Big) \\[1ex]
   &\times \Big(\prod_{(u,v)\in P_{\pi}} 
      e^{-i\,\gamma_{\pi}^{(\ell)} \sum_{\sigma} \tfrac{X_{u\sigma}X_{v\sigma}+Y_{u\sigma}Y_{v\sigma}}{2}}\Big) \\[1ex]
   &\times \prod_{\rho} 
      \Big(\prod_{(p,q)\in G_{\rho}} 
        e^{-i\, \big(\sum_{k=1}^{r_{\rho}} \gamma_{\rho,k,\mathrm{fwd}}^{(\ell)} 
              s_{\rho(p)}^{(k,\ell)} s_{\rho(q)}^{(k,\ell)}\big)
          \sum_{\sigma} \tfrac{X_{p\sigma}X_{q\sigma}+Y_{p\sigma}Y_{q\sigma}}{2}} \Big) \\[1ex]
   &\times \prod_{\rho} 
      \Big(\prod_{(i,j)\in M^{\rho}_{\mathrm{even}}} 
        e^{-i\,\beta_{\mathrm{even}}^{(\ell)} \sum_{\sigma} \tfrac{X_{i\sigma}X_{j\sigma}+Y_{i\sigma}Y_{j\sigma}}{2}} \\
   &\hspace{6em}\times
        \prod_{(i,j)\in M^{\rho}_{\mathrm{odd}}} 
        e^{-i\,\beta_{\mathrm{odd}}^{(\ell)} \sum_{\sigma} \tfrac{X_{i\sigma}X_{j\sigma}+Y_{i\sigma}Y_{j\sigma}}{2}} \Big) \\[1ex]
   &\times \prod_{\rho} \prod_{p\in\mathrm{occ}\cap\rho} 
      \Bigg[\prod_{q\in E_p^{\mathrm{near}}} 
        e^{-i\, g_{\rho(p)}^{(\ell)} \gamma_{\rho,\mathrm{near},\mathrm{fwd}}^{(\ell)} 
          \hat{w}^{\mathrm{near}}_{p|q}
          \sum_{\sigma} \tfrac{X_{p\sigma}X_{q\sigma}+Y_{p\sigma}Y_{q\sigma}}{2}} \\
   &\hspace{6em}\times 
        \prod_{q\in E_p^{\mathrm{far}}} 
        e^{-i\, g_{\rho(p)}^{(\ell)} \gamma_{\rho,\mathrm{far},\mathrm{fwd}}^{(\ell)} 
          \hat{w}^{\mathrm{far}}_{p|q}
          \sum_{\sigma} \tfrac{X_{p\sigma}X_{q\sigma}+Y_{p\sigma}Y_{q\sigma}}{2}} \Bigg] \\[1ex]
   &\times \prod_{\rho} \prod_{p\in\mathrm{occ}\cap\rho} 
      \Bigg[\prod_{q\in E_p^{\mathrm{near}}} 
        e^{-i\, g_{\rho(p)}^{(\ell)} \gamma_{\rho,\mathrm{near},\mathrm{bwd}}^{(\ell)} 
          \hat{w}^{\mathrm{near}}_{p|q}
          \sum_{\sigma} \tfrac{X_{p\sigma}X_{q\sigma}+Y_{p\sigma}Y_{q\sigma}}{2}} \\
   &\hspace{6em}\times 
        \prod_{q\in E_p^{\mathrm{far}}} 
        e^{-i\, g_{\rho(p)}^{(\ell)} \gamma_{\rho,\mathrm{far},\mathrm{bwd}}^{(\ell)} 
          \hat{w}^{\mathrm{far}}_{p|q}
          \sum_{\sigma} \tfrac{X_{p\sigma}X_{q\sigma}+Y_{p\sigma}Y_{q\sigma}}{2}} \Bigg] \\[1ex]
   &\times \prod_{\rho} 
      \Big(\prod_{(p,q)\in G_{\rho}} 
        e^{-i\, \big(\sum_{k=1}^{r_{\rho}} \gamma_{\rho,k,\mathrm{bwd}}^{(\ell)} 
            s_{\rho(p)}^{(k,\ell)} s_{\rho(q)}^{(k,\ell)}\big)
          \sum_{\sigma} \tfrac{X_{p\sigma}X_{q\sigma}+Y_{p\sigma}Y_{q\sigma}}{2}} \Big) \\[1ex]
   &\times \Big(\prod_{p\in\mathrm{occ}} 
        e^{-i\,\theta_p^{(\ell)} \sum_{\sigma} \tfrac{X_{p\sigma}X_{q_m^*(p)\sigma}+Y_{p\sigma}Y_{q_m^*(p)\sigma}}{2}} \Big) \\[1ex]
   &\times U_{\mathrm{vir}}\!\left(\tfrac{\gamma_{\mathrm{vir}}^{(\ell)}}{2}\right)\,
          U_{\mathrm{occ}}\!\left(\tfrac{\gamma_{\mathrm{occ}}^{(\ell)}}{2}\right) \\[1ex]
   &\times \Big(\prod_{b=1}^{B} 
        e^{-i\,\theta_b^{(\ell)} \sum_{\sigma} \tfrac{X_{p_b\sigma}X_{q_b\sigma}+Y_{p_b\sigma}Y_{q_b\sigma}}{2}} \Big) 
   \Bigg], \\[2ex]
m &= \begin{cases}
       1, & \ell \text{ odd}, \\
       2, & \ell \text{ even}.
     \end{cases}
\end{align*}
\end{tcolorbox}

\lstinputlisting[style=txtFormat, caption={Refined ansatz idea generated by our agentic system}, label=lst:ReflectedAnsatzIdea]{ideas/reflected_ansatz_idea.txt}
\begin{tcolorbox}[colframe=black, colback=white, boxrule=0.5pt,
                  left=2mm, right=2mm, top=1mm, bottom=1mm,
                  title={Reflected Ansatz Formula}]
\begin{align*}
\displaystyle 
|\Psi(\boldsymbol{\theta})\rangle 
   = \prod_{\ell=1}^{L} \Bigg[ 
     &\Big(\prod_{r} R_{z,r}(\phi_r^{(\ell)})\Big)\,
      U_{\mathrm{occ}}\!\Big(\tfrac{\gamma_{\mathrm{occ}}^{(\ell)}}{2}\Big)\,
      U_{\mathrm{vir}}\!\Big(\tfrac{\gamma_{\mathrm{vir}}^{(\ell)}}{2}\Big) \\[1ex]
   &\times \Big(\prod_{\textcolor{red}{(p,q)\in B_{\mathrm{back}}}} 
      e^{-i\,\textcolor{red}{\theta_{p,q}^{(\ell)}}\,\textcolor{red}{S_p\cdot S_q} }\Big) \\[1ex]
   &\times \Big(\prod_{(u,v)\in P_{\pi}} 
      e^{-i\,\gamma_{\pi}^{(\ell)}\,\textcolor{red}{S_u\cdot S_v} }\Big) \\[1ex]
   &\times \Big(\prod_{\textcolor{red}{\rho'\,\text{(subsectors)}}}\;\prod_{(p,q)\in G_{\textcolor{red}{\rho'}}} 
      e^{-i\,\big(\sum_{k=1}^{r_{\textcolor{red}{\rho'}}} 
        \gamma_{\textcolor{red}{\rho'},k,\mathrm{fwd}}^{(\ell)} 
        s_{\textcolor{red}{\rho'(p)}}^{(k,\ell)} s_{\textcolor{red}{\rho'(q)}}^{(k,\ell)}\big)\,\textcolor{red}{S_p\cdot S_q} } \Big) \\[1ex]
   &\times \Big(\prod_{\textcolor{red}{\rho'}}\prod_{(i,j)\in M_{\mathrm{even}}^{\textcolor{red}{\rho'}}} 
      e^{-i\,\beta_{\mathrm{even}}^{(\ell)}\,\textcolor{red}{S_i\cdot S_j} }
     \;\prod_{(i,j)\in M_{\mathrm{odd}}^{\textcolor{red}{\rho'}}} 
      e^{-i\,\beta_{\mathrm{odd}}^{(\ell)}\,\textcolor{red}{S_i\cdot S_j} } \Big) \\[1ex]
   &\times \Big(\prod_{\textcolor{red}{\rho'}} \prod_{p\in\mathrm{occ}\cap\textcolor{red}{\rho'}} 
      \;\prod_{q\in \textcolor{red}{E_p^{\mathrm{near}}\cup E_p^{\mathrm{far}}}} 
      e^{-i\, g_{\textcolor{red}{\rho'(p)}}^{(\ell)} 
        \gamma_{\textcolor{red}{\rho',(\mathrm{near/far}),\mathrm{fwd}}}^{(\ell)} 
        \hat{w}^{\textcolor{red}{(\mathrm{near/far})}}_{p|q}\, \textcolor{red}{S_p\cdot S_q} } \Big) \\[1ex]
   &\times \textcolor{red}{\text{(untied backward mirror of the pair/mixer blocks)}} \\[1ex]
   &\times U_{\mathrm{vir}}\!\Big(\tfrac{\gamma_{\mathrm{vir}}^{(\ell)}}{2}\Big)\,
          U_{\mathrm{occ}}\!\Big(\tfrac{\gamma_{\mathrm{occ}}^{(\ell)}}{2}\Big) \\[1ex]
   &\times \Big(\prod_{b=1}^{B} 
      e^{-i\,\theta_b^{(\ell)}\, \textcolor{red}{S_{p_b}\cdot S_{q_b}}} \Big) 
   \Bigg], \text{with } 
  \textcolor{red}{S_p\cdot S_q = \tfrac{1}{4}(X_p X_q + Y_p Y_q + Z_p Z_q)}.
\end{align*}
\end{tcolorbox}

\subsection{Effect of modules in the best generated ansatz}
\label{sec:paramAnsatz}
The best-performing ansatz includes multiple optional modules that control the circuit structure (see Fig.~\ref{fig:vqeCircuitOverall} in Section \ref{sec:vqeBestSolutions} for details).
These modules introduce additional interaction patterns among qubits and increase the expressivity of the variational circuit.
In this section, we investigate how each module contributes to the overall performance of the ansatz.
First, we examine their impact on the number of trainable parameters and quantum gates.
We then evaluate how the ground-state estimation accuracy changes across the seven molecular systems considered in the main text.

\begin{figure}[ht]
    \centering
    \includegraphics[width=0.8\linewidth]{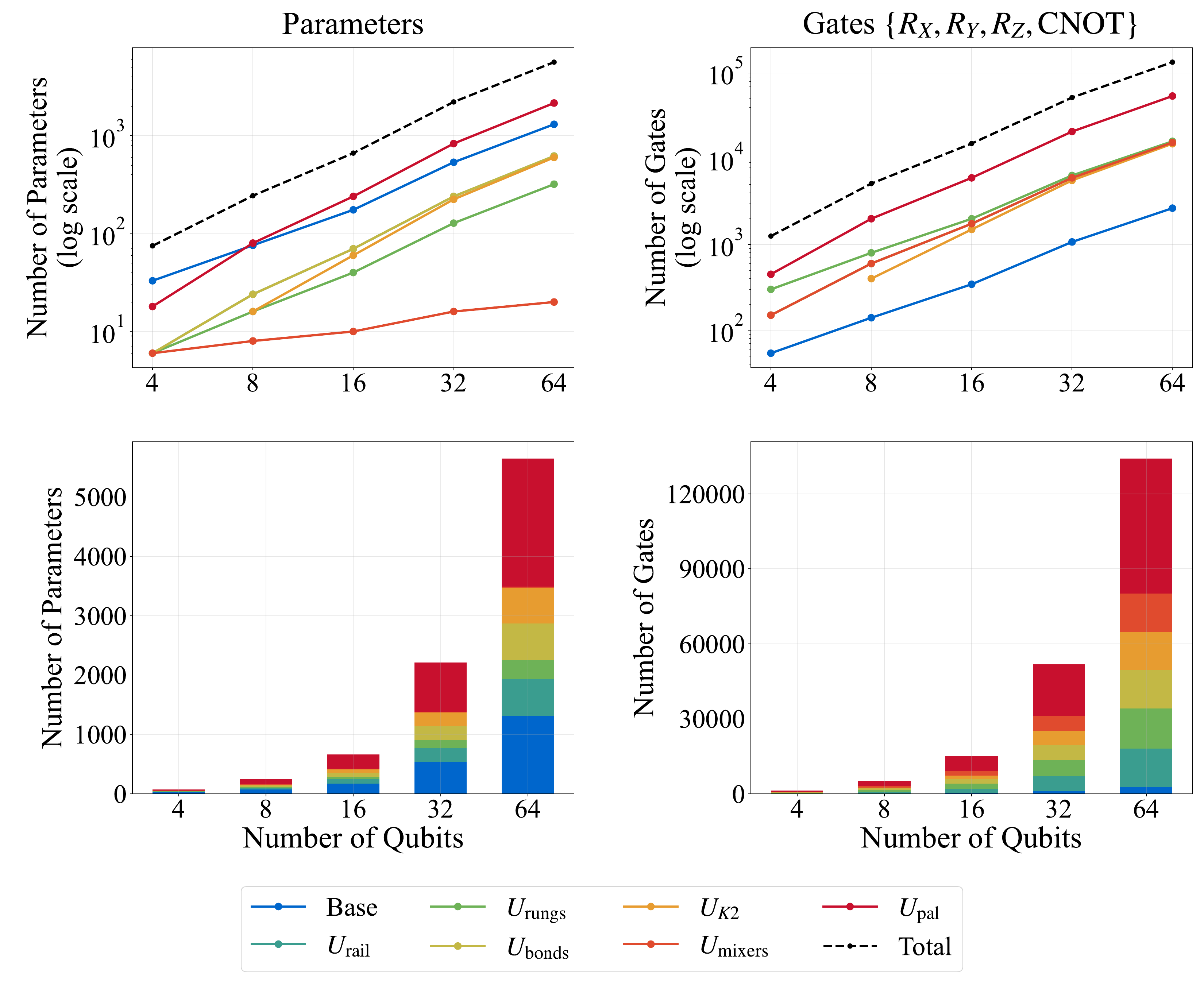}
    \caption{Parameter and gate-count scaling of the generated ansatz as a function of the number of qubits. Gate counts are computed after decomposition into \{\(R_X, R_Y, R_Z\), CNOT\}. The upper panels show module-wise scaling on a logarithmic scale, with the total count indicated by the black dashed line. The lower panels show stacked bar plots illustrating the contributions of individual ansatz modules. Here, Base corresponds to the non-entangling single-qubit operations in \(\mathcal{R}_\mathrm{pre}\) and \(\mathcal{R}_\mathrm{post}\).}
    \label{fig:vqeOptionalParams}
\end{figure}

\vspace{1em}

The generated ansatz can incorporate up to six additional modules on top of the base circuit.
The base circuit consists of the Hartree--Fock state preparation together with the \(\mathcal{R}_\mathrm{pre}\) and \(\mathcal{R}_\mathrm{post}\) blocks, which provide the non-entangling single-qubit rotations and collective occupied/virtual phase rotations.
In the generated implementation, all optional modules are enabled by default.
Fig.~\ref{fig:vqeOptionalParams} shows how each module affects the number of trainable parameters (left) and the number of quantum gates (right).
In this figure, the number of qubits is varied as 4, 8, 16, 32, and 64, the gate counts were evaluated after decomposing the circuits into the gate set \{\(R_X, R_Y, R_Z\), CNOT\}, following the same setup used in the main text.
The module symbols indicated by the colors in the legend correspond to those shown in Fig.~\ref{fig:vqeCircuitOverall}.

The upper panels of Fig.~\ref{fig:vqeOptionalParams} show that the contribution of each module increases approximately linearly with the number of qubits.
This behavior is consistent with the design of the ansatz, in which all edge sets contain a number of interactions that scales linearly with the system size. Consequently, the overall parameter count and gate complexity of the generated ansatz also exhibit approximately linear scaling.

The lower panels of Fig.~\ref{fig:vqeOptionalParams} further reveal the relative contributions of the individual modules.
Among the entangling components, the dominant operation is the Heisenberg-exchange gate $U_{ij}(\theta) = e^{-i\frac{\theta}{4}(X_i X_j + Y_i Y_j + Z_i Z_j)}$ which is implemented using three Pauli-rotation gates sharing a common parameter $\theta$.
Consequently, the entangling modules contribute proportionally more to the gate count than to the parameter count.
As a result, the relative contribution of the Base component is larger in the parameter decomposition than in the gate decomposition.
In both the parameter and gate-count decompositions, $U_{pal}$ contributes the largest fraction because it reintroduces the interactions contained in $U_{rail}$, $U_{rungs}$, $U_{bonds}$, and $U_{K2}$ in reverse order, thereby representing a coarser structural module than the individual interaction channels. 

\vspace{1em}
 
\begin{figure}[ht]
    \centering
    \includegraphics[width=0.8\linewidth]{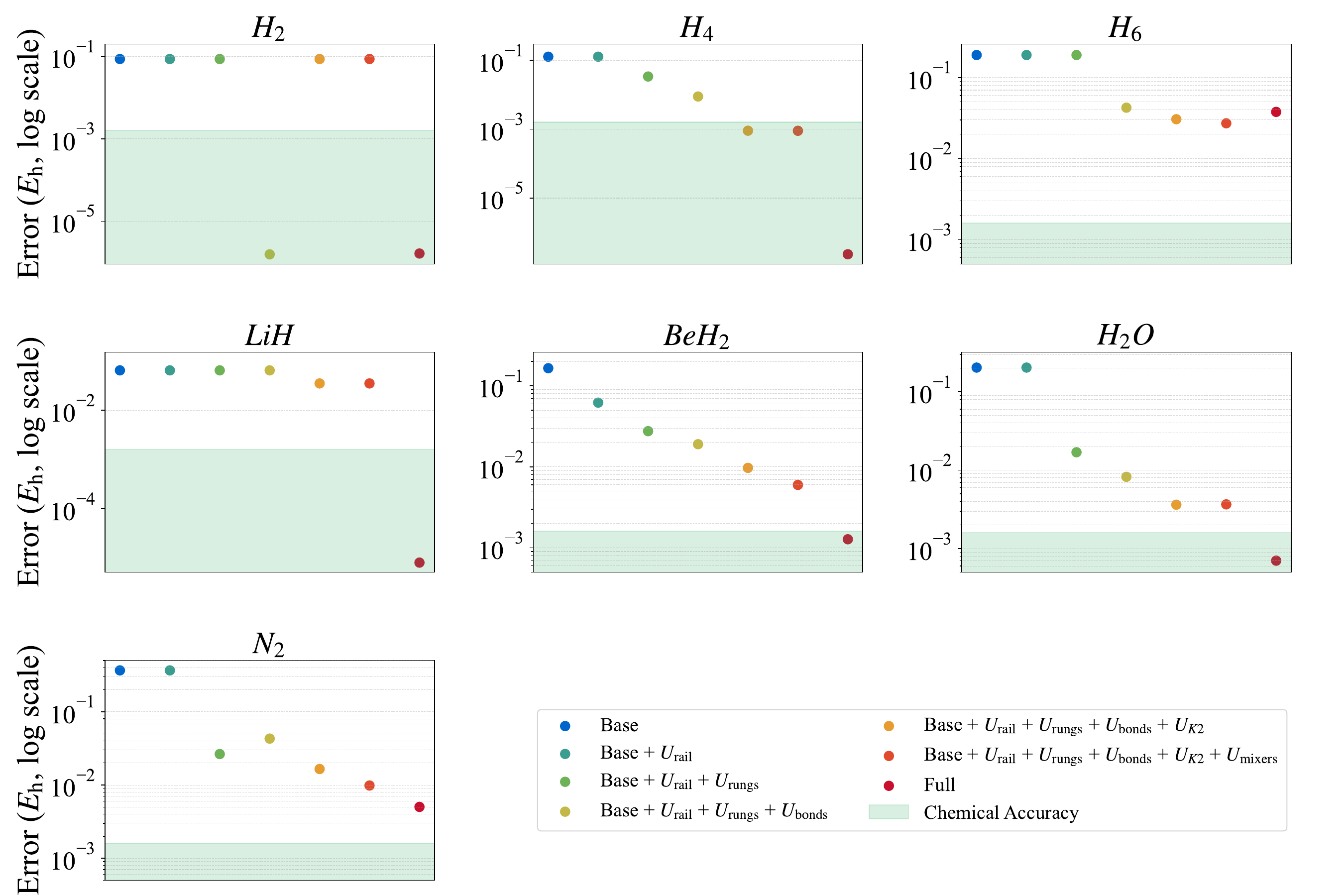}
    \caption{Mean energy estimation errors when the optional modules are incrementally added to the base architecture. The plot shows the mean energy error, averaged over all bond lengths for the seven molecules studied. The color of each dot indicates which modules have been added, and moving to the right along the x-axis corresponds to adding more modules. The red dots represent the full module configuration used in the main text. The green shaded region indicates chemical accuracy.}
    \label{fig:vqeOptionalDiff}
\end{figure}

Fig.~\ref{fig:vqeOptionalDiff} shows the energy estimation errors obtained when optional modules are incrementally added to the base architecture.
Although there are many possible combinations of modules, we followed the architecture order illustrated in Fig.~\ref{fig:vqeCircuitOverall} and added one module at a time from left to right.
The same seven molecules used in the main text ($\rm H_{2}$, $\rm H_{4}$, $\rm H_{6}$, $\rm LiH$, $\rm BeH_{2}$, $\rm H_{2}O$, and $\rm N_{2}$) were evaluated under identical bond lengths settings.
Each dot in the plot represents the mean energy error averaged over all bond lengths for each molecule.

\vspace{1em}

Overall, the graph shows that the estimation error tends to decrease as more modules are added, regardless of the molecular species.
The addition of the $U_{bonds}$ module (dark yellow) coincides with a noticeable reduction in the error for several molecular systems.
This module introduces diagonal inter-rail couplings between orbitals with different rail indices, whereas the preceding modules only connect neighboring orbitals within a rail or orbitals with the same index across rails.
The observed improvement suggests that such additional coupling pathways may help the circuit capture more complex correlation patterns.
As shown by the red points representing the full module configuration, five of the seven molecules reach chemical accuracy.
In contrast, other configurations fail to achieve this accuracy except for $H_2$ and $H_4$, suggesting that the \(U_{pal}\) module contributes substantially to the final accuracy for several molecular systems.
However, as observed in Fig.~\ref{fig:vqeOptionalParams}, $U_{pal}$ involves a large number of parameters and gates.
Therefore, in realistic noisy environments or on real hardware implementations, its impact on accuracy requires further investigation.
These results indicate that the generated ansatz provides a flexible trade-off between estimation accuracy and circuit complexity through its optional modules.

\end{document}